\documentclass[12pt]{article}

\usepackage[english]{babel}
\usepackage[utf8]{inputenc}

\usepackage[intlimits]{amsmath} 
\usepackage{amsthm,mathrsfs}    
\usepackage[comma,authoryear]{natbib}
\usepackage{url}
\usepackage{algorithm}
\usepackage{morefloats}

\numberwithin{equation}{section}
\usepackage{amssymb,amsmath}
\usepackage{subfigure}
\usepackage{bm}
\usepackage{fullpage}
\usepackage{booktabs} 
\usepackage{graphicx,graphics,epsfig}
\usepackage{float}
\usepackage{multirow}

\theoremstyle{plain}
\newtheorem{thm}{Theorem}

\newcommand{\bthm}{\begin{thm}}
\newcommand{\ethm}{\end{thm}}

\newcommand{\bpf}{\begin{proof}}
\newcommand{\epf}{\end{proof}}

\theoremstyle{definition}
\newtheorem{defn}{Definition}
\newtheorem{rem}{Remark}

\usepackage[margin=.9in,a4paper]{geometry}
\usepackage[compact]{titlesec}
\titlespacing*{\section}{0pt}{.8pt}{.8pt}
\titlespacing*{\subsection}{0pt}{.8pt}{.8pt}
\titlespacing*{\subsubsection}{0pt}{.8pt}{.8pt}
\newcommand{\bib}{\bibliography{ref-bib}\bibliographystyle{ims}}
\usepackage{deep-macro}

\begin{document}
\begin{center}
{\Large {\bf Large-Scale Mode Identification\\[.25em] and Data-Driven Sciences}}  
\\[.2in]
Subhadeep Mukhopadhyay\\
Temple University, Department of Statistical Science \\ Philadelphia, Pennsylvania, 19122, U.S.A.\\
\end{center} 
\vspace{-1em}
\begin{abstract}
Bump-hunting or mode identification is a fundamental problem that arises in almost every scientific field of data-driven discovery. Surprisingly, very few data modeling tools are available for automatic (not requiring manual case-by-case investigation), objective (not subjective), and nonparametric (not based on restrictive parametric model assumptions) mode discovery, which can scale to large data sets. This article introduces \texttt{LPMode}--an algorithm based on a new theory for detecting multimodality of a probability density. We apply LPMode to answer important research questions arising in various fields from environmental science, ecology, econometrics, analytical chemistry to astronomy and cancer genomics.
\end{abstract}
\section{Introduction}
\subsection{Goals}
Many scientific problems seek to identify modes in the true unknown probability density function $f(x)$ of a variable $X$, given i.i.d observations $X_1,\ldots,X_n$. The presence of unexpected ``bumps'' in a distribution are interpreted as interesting `new' phenomena or discoveries whose scientific explanation is a research problem.

However, in the era of big data, we have a slightly more complicated situation where modern data-driven sciences routinely gather measurements on tens of thousands or even millions of variables instead of only one variable. The goal is to learn and compare the multi-modality shape of each variables. This problem of finding structures in the form of hidden bumps arises in many data-intensive sciences. For example, cancer biologists may be interested to identify genes with multi-modal expression from a large-scale microarray database as they might serve as ideal biomarker candidates that can be useful for discovering unknown cancer subtypes or designing personalized treatment. On the other hand, applied economist may be interested to understand how the muti-modality pattern or shape of income per capita distribution evolve over time, which might provide insights into the economic polarization (or lack thereof). Due to the massive scale of these kinds of problems, it is not practical to manually investigate the modality in a case-by-case basis. As a practical requirement, we seek to develop algorithms that are \emph{automated and systematic}.  In this paper, we address the intellectual challenge of developing novel algorithm for `\emph{large-scale nonparametric mode exploration}'--a problem of outstanding interest at the present time. To the best of our knowledge, there has been \emph{no previous work} that can address this important \emph{multi-disciplinary applied data problem}.


Two different classes of bump-hunting methods are currently prevailing in the literature, which provide insights at different levels of granularity and details: (i) testing multimodality or deviation from unimodality; (ii) determining how many modes are present in a probability density function. The purpose of this paper is to present a new \emph{genre} of nonparametric mode identification technique for (iii) comprehensive mode identification: determining number of modes (along with locations), as well as standard errors or confidence intervals of the associated mode positions to assess significance and uncertainty.
\vskip.4em

\subsection{Two modeling cultures}
The vast majority of bump-hunting techniques available to date can broadly be divided into two parts based on the modeling cultures. The \emph{first line} of work is based on parametric mixture model. The Gaussian mixture model (GMM) is the most heavily studied model in this class. GMM-based parametric bump hunting methodologies are well-studied and have a large literature. For details, see \cite{day1969, fraley2002}, and references therein. The \emph{second and most popular} approach extracts modes by estimating the kernel density function, thus completely removes the parametric restrictions. The idea of using kernel density for nonparametric mode identification goes back to the seminal work of \cite{parzen1962}. This was furthered studied by \cite{silverman1981} based on the concept of ``critical bandwidths'' and bootstrapping, which is known to be highly conservative, non-robust (sensitive to outliers), and generate different answers based on various calibration techniques (e.g. bandwidth). For recent applications of kernel density based approach in mode clustering see \cite{chacon2013data,chacon2015} and \cite{chen2016mode}.
\vskip.4em

In the present paper, we shall instead focus on developing a comprehensive solution to the problem (iii) by blending the traditional parametric and nonparametric statistical modeling cultures. The modeling attitude taken here can be classified as a nonparametrically designed parametric statistical modeling culture. It provides \emph{nonparametric generalization of the (parametric) Edgeworth-like expansion} for density approximation in a way that is especially useful for bump identification. Our specialized technique  provides a fundamentally \emph{new point of view to the problem of mode-discovery}, which borrows modern nonparametric machineries from \cite{Deep14LP}.

\section{Methods}
\subsection{Skew-G density representation}  We introduce a new class of density representation scheme, called Skew-G models, which lies at the heart of our analysis. As a prototypical example we consider acidity index (more specifically acid-neutralizing capacity (ANC) -- a fundamental index of natural water acid-base status) measured in a sample of $n=155$ lakes in North-Central Wisconsin\footnote{\url{ttp://dnr.wi.gov/topic/surfacewater/assessments.html}}. Assessing water quality standards by monitoring acidity level of lakes (a higher level of acidity is a threat to biodiversity) is a matter of paramount importance as this has a direct impact on our environment. The study has two interrelated goals: to check whether the data supports normal ANC distribution and if not, then the next step is to investigate the presence of multiple distinct populations of lakes with different distributions of ANC.

\begin{figure}[htb]
\centering
\includegraphics[height=\textheight,width=.45\textwidth,keepaspectratio,trim=2cm .7cm 1cm 1.5cm]{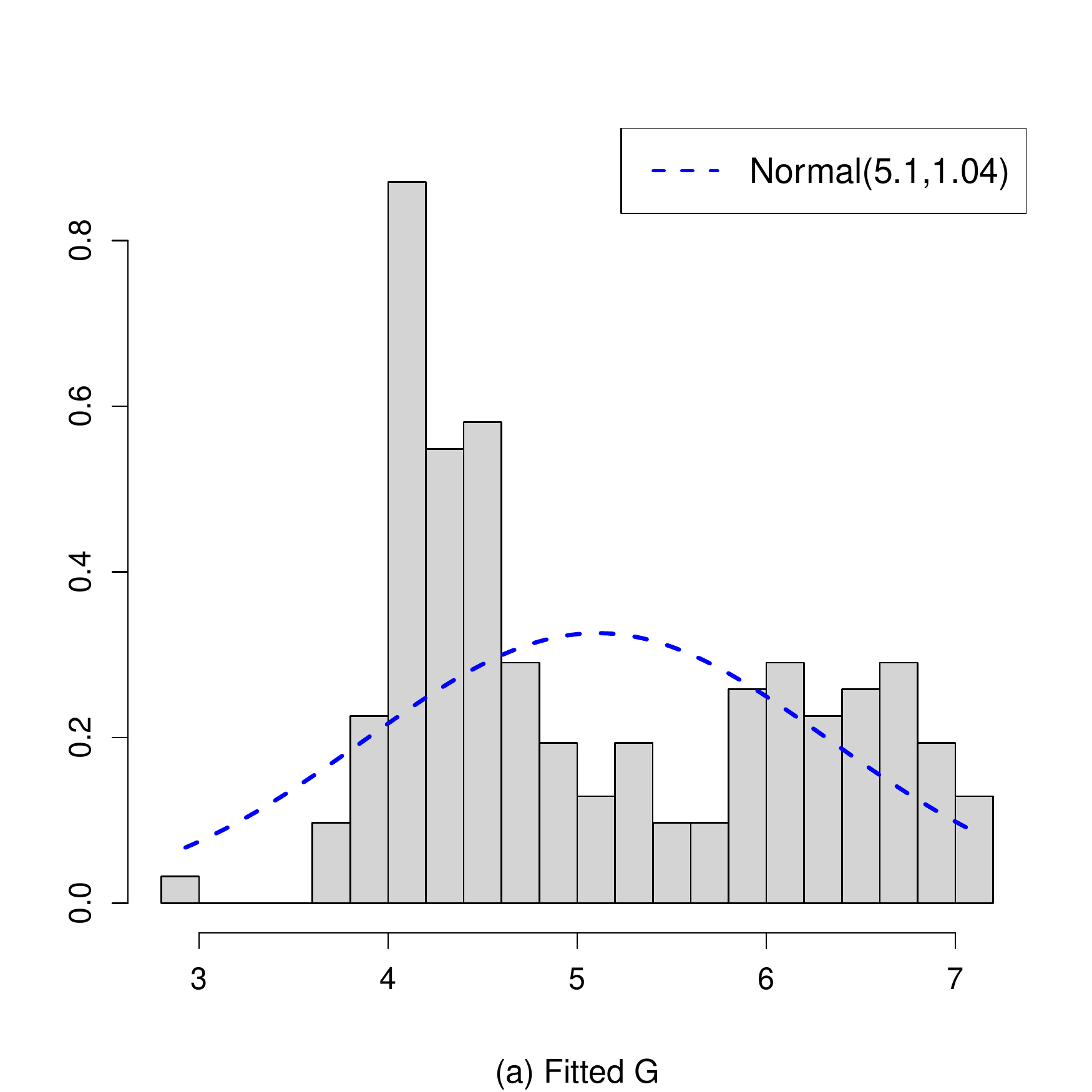}~~
\includegraphics[height=\textheight,width=.45\textwidth,keepaspectratio,trim=1.5cm .7cm 2.5cm 1.5cm]{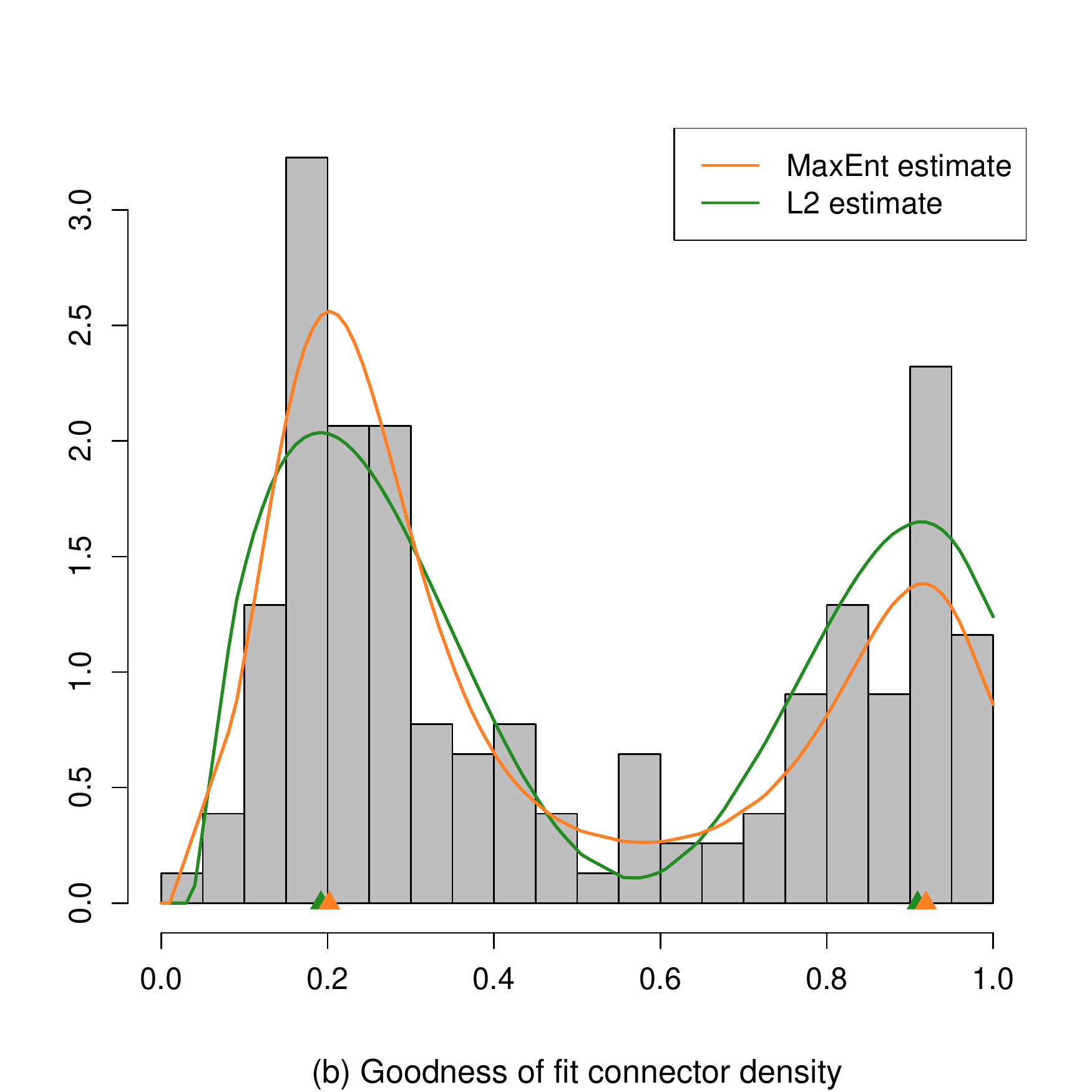}\\
\caption{(a) The normal density with mean $5.10$ and standard deviation $1.04$ (maximum-likelihood estimates) is fitted to the acidity index data. (b) The smooth bimodal connector density estimate is shown. The green curve shows the LP orthogonal series estimator and the orange curve denotes the LP maximum entropy exponential (MaxEnt) estimates; corresponding modes are shown in same color. }
\label{fig-acid2}
\end{figure}

To address this question, our first modeling goal is to understand how the true unknown density is different from the \emph{unimodal} $G$, which is ``normal'' distribution for the acidity index data. Depending on the problem, data scientists are free to choose \emph{any} suitable parametric null-model as a rough starting point to \emph{query the data}, thereby making data analysis a more interactive and automatic endeavor; see examples in Section 4. The inadequacy of the normal density model as clearly visible from Fig \ref{fig-acid2}(a)
begs the following question, which is at the heart of our approach:

\emph{What is the least or minimal nonparametric perturbation of the null parametric model $G$ is required to produce a density that best fits the data?}
\begin{defn}
Define skew-G density model, an universal representation scheme by
\beq
\label{eq:sg}
f(x)\,=\,g(x) \,\times\, d(G(x);G,F).
\eeq
Here $d(u;G,F)$ is the goodness of fit ``connector'' or ``comparison'' density defined as
\beq d(u;G,F)\,=\,\dfrac{f(Q(u;G))}{g(Q(u;G))},~~~~~~0\le u \le1, \eeq
where $Q(u;G)=\inf\{x: G(x) \ge u\}$ denotes the quantile function. The corresponding comparison distribution function is given by $D(u;G,F)=F(Q(u;G))=\int_0^u d(v;G,F)\dd v$.
\end{defn}
\begin{rem}
In the algorithmic terms, the Skew-G density modeling involves three steps: (1) start with a unimodal parametric candidate model $G$; (2)  manufacture an intermediate density $d(u;G,F)$; and (3) combine them using \eqref{eq:sg} to construct $\widehat f(x;X)$. The comparison density acts as a glue to ``connect'' the parametric unimodal null-density $G$ with the true unknown distribution $F$ to provide the best fit in a way that reveals the hidden multimodality.
\[ G ~~ \longrightarrow ~~d\big(\cdot\,;\,G,F\big) ~~\longrightarrow ~~F,~~~~~~~~~\]
For that reason, we alternatively call $d$ connector density.
\end{rem}
The skew-G density formulation simultaneously serves two purposes: (1) Exploratory goodness-of-fit assessment: The hypothesis $H_0: F = G$ can equivalently be expressed as testing $d \equiv 1$. Thus the flat uniform shape of the estimated comparison density provides a quick graphical diagnostic to test the fit of the unimodal parametric model $G$ to the true distribution $F$. Fig \ref{fig-acid2}(b) shows the nonparametric comparison density estimate (a \emph{specially designed} method will be discussed in the next section) for acidity-index data. (2) Mode identification:
The shape of $d(u;G,F)$ not only provides a tool to check the null hypothesis, but also indicates how to repair $G$ such that it adequately fits the data. Data scientists are often interested in understanding whether the assumed hypothesized model $G$ is different from the actual distribution $F$ by having \emph{extra} mode(s). Looking at the Fig \ref{fig-acid2}(b), one may anticipate that the initial clue of modal structure of a true $f$ might be hidden in the shape of the pre-density $d$. This is, indeed, the case as we demonstrate in the next section.
\begin{rem}
The comparison density function captures and exposes the modality structure of the data in a \emph{more transparent and unambiguous way} than the original density $f$, which is the \emph{key observation} behind our LPMode algorithm (see Section 3). One of the reasons for it being the added smoothness that $d(G(x))$ enjoys compare to the original density $f(x)$, which not only tackles the problem of spurious bumps but also allows efficient estimation via sparse expansion in a specialized orthogonal basis.
\end{rem}

\begin{rem}
An anonymous reviewer correctly pointed out to us that the density representation formula (Eq. 2.1 and 2.2) `is a good formulation for scientists because the term $d(G(x); G; F)$ contains the information about how the actual distribution is deviated from the approximated distribution $g$.' In fact in our formulation, the scientists can choose $G$ based on domain-knowledge (such as \cite{novikov2006}), which we incorporate in our density modeling procedure by viewing it as a `background distribution.' As a result, it allows the capability to better understand (i) how compatible is the data with the theoretically expected model? (ii) whether the \emph{unexplained part} manifests itself as a `peak'? Note that the modality of $d(u;G,F)$ (\emph{not} the original density $f(x)$) answers the last question of searching for the bump \emph{above} background, which we get as a byproduct from the proposed \texttt{LPMode} algorithm (see Section 3).
\end{rem}
\subsection{Constructing empirical orthogonal rank polynomials}
We introduce a new class of orthogonal polynomials, called \emph{LP family of rank polynomials}, and discuss the construction procedure. The word ``empirical'' in the title conveys the fact that data determine the shape of the orthogonal basis functions, which act as a key fundamental  ingredient in the nonparametric approximation of $d[G(x)]\in \sL^2(\cR)$.
\begin{defn}
The probability integral transformation with respect to the measure $G$ (or rank-$G$ transform) is defined as $G(X_F)$, where $X_F$ indicates $X \sim F$; distinguish it from the uniformly distributed rank-transform of a random variable $F(X_F)$. For notational simplicity, we suppress the subscript $F$ in $X$ from now on.
\end{defn}

\begin{defn}
Construct $T_j(X;G)\, (j=1,2,\ldots)$, an orthonormal basis for $L^2(G)$ by Gram Schmidt orthonormalization of the power of rank-$G$ transform of a random variable $\{G(X), G^2(X), $ $\ldots,G^j(X)\}$. First few polynomial bases are given below
\beas
T_0(x;G)&=&1 \\
T_1(x;G)&=& \sqrt{12} \big\{G(x)-.5\big\}\\
T_2(x;G)&=& \sqrt{5} \big\{6 G^2(x)-6G(x)+1\big\}\\
T_3(x;G)&=&\sqrt{7} \big\{20G^3(x)-30G^2(x)+12G(x)-1\big\}
\eeas
Verify these polynomials $\{T_j(x;G)\}_{j=1}^{\infty}$ are orthogonal with respect to the measure $G$
\[\big\langle T_j,T_k \big\rangle_G \,:=\,\int_{\cR} T_j(x;G) T_k(x;G) \dd G(x)\,=\,\delta_{jk}~~~\mbox{for}\, j\ne k.\]
\end{defn}
The score functions $T_j(X;G)$ can equivalently be expressed as $\Leg_j[G(X)]$, ``shifted'' \underline{L}egendre \underline{P}olynomials in $L^2[0,1]$ evaluated at rank-$G$ transform $G(X)$.
\begin{rem}
Asymptotically as $\nti$ the ``empirical'' (piecewise-constant) orthonormal LP-score functions converges to the ``smooth'' \underline{L}egendre \underline{P}olynomials under $H_0:F=G$.  To emphasize this \emph{universal limiting shape} of our empirically constructed score functions, we call it LP basis--Legendre Polynomials \emph{of ranks}. The substantial \emph{departure} of empirical LP basis functions from the Legendre polynomials is the consequence of the fact that unlike rank-transform $F(X)$, the rank-G transform random variable $G(X)$ is not uniformly distributed due to $F \ne G$, as in acid data.
\end{rem}

\begin{rem}
We perform density function approximation of $d(G(x))$ in the LP Hilbert spaces. In particular, we design $L^2$ and exponential connector density models whose sufficient statistics are LP special functions. Furthermore, we show that the orthogonal Fourier expansion coefficients can be expressed as functional of these LP bases.
\end{rem}
\subsection{Estimation and properties}
Two methods for probability density expansion are discussed by choosing the sufficient statistics to be LP basis elements, which differs us from the classical orthogonal series based methods \citep{wassermanbook,tsybakov2009}.
\begin{defn}
For a random sample $X_1,\ldots,X_n$ with the sample distribution $\widetilde F(x;X)=n^{-1}\sum_{i=1}^n$ $\I(X_i \le x)$, where $\I(\cdot)$ is the indicator function, define LP means as
\beq \LP(j;G;\widetilde F)\,=\,  \Ex\big[T_j(X;G)\,|\,\widetilde F\big].  \eeq
\end{defn}
\begin{thm} \label{thm:coeff}
The square integrable comparison density function can be represented by a convergent orthogonal series expansion in the LP Hilbert space with the associated Fourier coefficients $\LP(j;G;F)$.
\end{thm}
To establish the theorem it is enough to show that $\LP(j;G; F) = \big\langle \Leg_j, d\big\rangle_{\sL^(0,1)}$, which is shown below
\beas \LP(j;G,F)&=&\Ex[T_j(X;G)|F]\,=\,\int \Leg_j(G(x)) \dd F(x)\,=\,\int \Leg_j(u) \dd D(u;G,F). \eeas
To derive the asymptotic null-distribution of the LP-orthogonal coefficients first note that under $H_0:F=G$ we can express
\beq \label{eq:anorm}
\sqrt{n}\Big\{ \LP(j;G,\wtF) \,-\, \LP(j;G,F)\Big\} ~=~\sqrt{n}\int_{-\infty}^{\infty} T_j(x;G) \dd \big( \wtF(x) -F(x)\big),
\eeq
as functional of uniform (comparison distribution) empirical process $\mathbb{U}_n \equiv \sqrt{n} \big\{\wtD(u;G,\wtF) - u\big\}$ for $0 \le u \le 1$, which is known \citep{shorack2009} to converge weakly to a limiting Brownian bridge process $\mathbb{U}_n  \xrightarrow{w} U$. As a consequence, under $H_0$ the continuous mapping theorem yields
\beq \label{eq:ep}
\sqrt{n}\int_{-\infty}^{\infty} T_j(x;G) \dd \big( \wtF(x) -F(x)\big) \,\,\stackrel{d}{=}\,\,\int_0^1 \Leg_j(u) \dd \mathbb{U}_n ~ \xrightarrow{~d~}~\int_0^1 \Leg_j(u) \dd \mathbb{U}.\eeq
We get the following important theorem by straightforward applications of integration by parts followed by Fubini’s theorem on the last expression \eqref{eq:ep}.
\begin{thm}\label{thm:anorm}
Under $H_0$, the sample LP means have the following limiting null distribution of the, as $\nti$
\[\LP[j;G,\wtF]~\dist~\cN(0,n^{-1}),~~\text{i.i.d for all}~ j.~~\]
\end{thm}
\subsection{Model denoising} Akaike information model selection criteria (AIC) selects the significantly non-zero LP means after arranging them in the decreasing magnitude; details in Section 3. Compute LP series estimator by $\widehat d(u;G,\widetilde F)-1=\sum_j \Leg_j(u) \LP(j;G,\widetilde F)$, sum over AIC selected indices. Interestingly, the proposed AIC-based LP-Fourier coefficient selection criterion can be shown to minimize MISE estimation error \citep{hart1985choice}:
\beq \label{eq:AIC}{\rm MISE}(\widehat d_m \,\| \,d)=\Ex\big[ \int_0^1(\widehat d(u)\,-\,d(u))^2\dd u  \big]=2 \sum_{j=m+1}^{\infty} \big| \LP[j;G,F] \big|^2 + \dfrac{2}{n} \sum_{j=1}^m\Big( 1- \big| \LP[j;G,F] \big|^2 \Big),\eeq
where the first term denotes the bias component and second term denotes the variance.
\begin{rem}
Instead of Akaike’s rule, one can alternatively use Schwarz BIC rule \citep{ledwina94} as the selection criterion. In an interesting study, \cite{kallenberg2000} showed that asymptotic optimality properties continue to hold for a much more general class of penalty starting from the AIC up to even much larger penalties than the one in Schwarz’s criterion. Thus from the theoretical perspective, practitioners can use either AIC or BIC to select the `significant' LP-coefficients without much harm.
\end{rem}
For acid data the resulting Skew-G orthogonal series density estimator is given by
\[L^2 ~{\rm Estimate}:~~\widehat f(x;X) \,=\, \hat\si^{-1}\phi\big(\dfrac{x-\hat\mu}{\hat\si}\big)  \,   \Big\{ 1 +.2T_2(x;G) +.44T_3(x;G) -.48 T_4(x;G) \Big \}, \]
where $(\hat\mu, \hat\si)$ are simply the MLE estimates. Practitioners are free to use any other plug-in estimators for specifying the parameters of $G$. To ensure non-negativity of the density estimate we further enhance the $L^2$ approach by estimating maximum entropy (MaxEnt) exponential estimator of $\log d(u;G,F)= \te_0 + \sum_j \te_j \Leg_j(u)$ satisfying the following moment constraints for significant non-zero LP-means indices:
\[\mbox{Moment equality constraints:}~~~~~~ \LP[j;G,\wtF]\,=\,\int_0^1 d_{\boldsymbol \te}(u;G,F) \Leg_j(u) \dd u.~~~~~~~~~~~~~~~~~~~~~\]
The resulting maximum entropy density estimate provides the best approximation to the comparison density in the sense of information divergence. The estimated specially designed (LP) nonparametric exponential density for acid index data is given below.
\[{\rm MaxEnt\,\, Estimate}:\widehat f(x;X)= \hat\si^{-1}\phi\big(\dfrac{x-\hat\mu}{\hat\si}\big) \exp\Big\{ -.35  +.13T_2(x;G) +.60T_3(x;G) -.58 T_4(x;G)  \Big\}.\]
\begin{rem}
The LP skew-G series representation formula $f(x)=g(x)\big[1+\sum_j \LP(j;G,F)\, T_j(x;G)\big]$ can be considered as a \emph{nonparametric generalization} of Edgeworth-like expansion \citep{cox1994book} for density approximation, which modifies the normal density by multiplying with Hermite polynomials.
\end{rem}
\begin{rem}
Unlike traditional approaches where parametric models are constructed \emph{before} the sufficient statistics, our modeling philosophy starts by constructing novel data representation via LP transform; schematic description of our \emph{nonparametrically designed parametric density} estimation strategy is given below:

Data ~$\rightarrow$ ~Constructing LP nonparametric transform ~$\rightarrow$~ Sufficient statistics selection ~$\rightarrow$~ Parsimonious parametric density models.
\end{rem}

\subsection{Consistency of local mode estimates}
\begin{thm} \label{thm:kernel}
The LP-series approximated comparison density admits specialized kernel representation
\beq \label{eq:kernel} \widehat{d}_m(u;G,F)~=~n^{-1}\sum_{i=1}^n K_m(u,u_i),\eeq
where $K_m$ has the following form
\beq  \label{eq:CDop} K_m(u,u_i)~=~\dfrac{m+1}{2} \dfrac{\Big( L_{m+1}(u)L_m(u_i) - L_{m}(u)L_{m+1}(u_i)\Big)}{u-u_i}, ~~~0<u,u_i<1
\eeq
and $L_j(u)=(2j+1)^{-1/2}\Leg_j(u)$.
\end{thm}
\vskip.4em
To prove \eqref{eq:kernel}, first apply Theorem 1 to verify:
\[\widehat{d}_m(u;G,F)~=~\sum_{j=1}^m \big \langle \Leg_j, \widetilde{d} \big \rangle \Leg_j(u)~=~n^{-1}\sum_{i=1}^n K_m(u,u_i),\]
where $K_m(u,u_i)=\sum_{j=1}^m \Leg_j(u_i) \Leg_j(u)$. Finish the proof by using the Christoffel–Darboux orthogonal polynomial identity formula to show that
\[\sum_{j=1}^m \Leg_j(u_i) \Leg_j(u)~=~\dfrac{m+1}{2} \dfrac{\Big( L_{m+1}(u)L_m(u_i) - L_{m}(u)L_{m+1}(u_i)\Big)}{u-u_i}.\]

\vskip1em
\begin{rem}
(i) By virtue of this kernel representation, one can interpret $m^{-1}$ as the bandwidth parameter. (ii) Due to the simple polynomial structure, the LP-kernel satisfies the following regularity conditions with $r=0,1,2,3$: (C.1) $d^{(r)}(u;G,F)$ is uniformly continuous; (C.2) $\int_0^1 u^2 K^{(r)}(u) \dd u < \infty$; and (C.3) $\int_0^1 \big( K^{(r)}(u) \big)^2 \dd u < \infty$. (iii) As a consequence of Theorem \ref{thm:kernel} along with (C.1-C.3), we can now utilize the well-known results from kernel density estimates to prove the consistency of local modes, without reinventing the wheel.
\end{rem}
\begin{thm} \label{thm:consis}
Let $m$ be a function of $n$ satisfying $m_n \rightarrow \infty$, and $\dfrac{m_n^5}{n}\rightarrow 0$, as $n \rightarrow \infty$. Then under the regularity conditions (C.1), (C.2), and (C.3),  the Hausdorff distance
\[\mathcal{H}\big\{ \mathscr{M}_{\widehat d}, \mathscr{M}_d\big\}~ \rightarrow~ 0, ~~\text{almost surely as}~~\nti, ~~~~~\]
where $\mathscr{M}_{\widehat d}$ and $\mathscr{M}_d$ respectively denote the set of local maxima points of $\widehat d$ and $d$.
\end{thm}
The proof essentially follows by similar arguments presented in \cite{chen2016mode} in conjunction with our Theorem \ref{thm:kernel} \footnote{ However, to the best of our knowledge, the explicit theoretical arguments for Hausdorff-consistency of ensemble of `local modes' first appeared in the 1983 Ph.D. dissertation of Steven Boswell, which has not received the due credit.}.
\section{LPMode algorithm and inference}
We now present the computational steps of LPMode algorithm for nonparametric mode identification (determining locations as well as confidence intervals of the modal positions) by exploiting the duality relationship between comparison density $d$ and the true unknown density $f$. Our proposed algorithm provides a computationally efficient and scalable solution to large-scale mode-exploration problems as demonstrated in Section 4.
LPMode starts with a unimodal parametric reference or null distribution $G$ (a plausible candidate for true unknown $F$) whose parameters are estimated using  maximum-likelihood (or any other methods: method of moments, etc.)
\begin{center}
{\bf The LPMode Algorithm}
\end{center}
\medskip\hrule height .65pt
\vskip.4em

1. \emph{LP basis construction}. Construct LP system of orthonormal \emph{rank} polynomials $T_j(X;G) \in \sL^2(G)$ associated with distribution $G$ (following the recipe given in Section 2.3) satisfying $\int_{\cR} T_j(x;G) \dd G(x)=0$, and $\int_{\cR} T_j(x;G) T_k(x;G) \dd G(x)=\delta_{jk}$ for $j \ne k$. Our data-driven non-linear rank polynomials $T_j(X;G)$ act as a sufficient statistics in our modeling algorithm, which are determined by the observed data $X_1,\ldots,X_n$ and the reference density $G$.

\vskip.25em
2. \emph{Computation of LP means}. Compute LP means \[\LP(j;G;\widetilde F)\,=\,  \Ex\big[T_j(X;G)\,|\,\widetilde F\big]\,=\,\avein\, T_j(x_i;G).  \]
\vskip.25em

3. \emph{Adaptive filtering}. Identify indices $j$ for which  $\LP(j;G;\widetilde F)$ are significantly non-zero by using AIC model selection criterion applied to LP means arranged in decreasing magnitude. Choose $k$ to maximize $\AIC(k)$,
\[\AIC(k)~=~\mbox{sum of squares of first $k$ LP-means}~ – ~ 2k/n.\]

4. \emph{Estimate $L^2$ and maximum entropy comparison density}.  Compute $L^2$ estimator of $d(u;G,F)-1=\sum_j \Leg_j(u) \LP(j;G,\widetilde F)$ and maximum entropy exponential estimator $\log d(u;G,F)$ $=$ $\te_0 + \sum_j \te_j \Leg_j(u)$, sum over the significant non-zero LP-means selected in the Step (3). The intermediate goodness of fit density $d(u;G,F)$ assists to uncover the modal shape characteristics of the true density $f(x)$ by capturing the missing shape features of unimodal $G$.
\vskip.25em

5. \emph{Skew-G density estimate}. Construct estimator of $f(x)$ using LP skew-G density model \eqref{eq:sg} as a nonparametric refinement of the parametric null-model:
\[f(x)\,=\,g(x) \,\times\, d(G(x);G,F).\]

6. \emph{Identify local maxima or mode}. Identify the modes of $f(x)$ by counting the number of modes in $d(u;G,F)$. Exploit the duality relationship between comparison density $d$ and the true unknown density $f$ for mode identification. Define the sets
\beas
\mathscr{M}(\widehat d)&=& \text{points of local maxima of }~\widehat d.\\
\mathscr{M}(\widehat f) &=& \text{points of local maxima of}~ \widehat f.
\eeas
~~$\bullet$ If $|\mathscr{M}(\widehat f)| \le |\mathscr{M}(\widehat d)|$, stop and return $\mathscr{M}(\widehat f)$ as potential modes. The case `$<$' includes the possibility when the modes of $\widehat d$ might converted into shoulder (at $x$ if $\widehat f'(x)= \widehat f''(x)=0$, not a turning point) of $\widehat f(x)=g(x)\, \widehat d[G(x);G,F]$. Often this happens when $\mathscr{M}(\widehat d)$ contains the boundary points $0$ or $1$.

~~$\bullet$ Else return largest $|\mathscr{M}(\widehat d)|$ modes of $\widehat f$ based on modal-jumps $\{\widehat f(x_{i}) - \min(\widehat f(x_{i-1}), \widehat f(x_{i+1}))\}$, where $ x_{i}$ is an local-maxima of $\widehat f$.
\vskip.5em
\medskip\hrule height .65pt
\vskip1.5em
To facilitate nonparametric statistical inference, we seek to find an approximate sampling distribution of the $k$-modal positions. Comparison density offers a neat way to simulate from the ``smooth'' nonparametric density estimate $\widehat f$. We compute the standard errors (to assess the significance and uncertainty) associated with each putative mode location by the following comparison density-based accept-reject inference algorithm.
\vskip1em
\begin{center}
{\bf Comparison Density Based Inference Algorithm}
\end{center}
\medskip\hrule height .65pt
\vskip.65em

\vskip.4em
1. Apply \texttt{LPMode} algorithm on the the original i.i.d sample $X_1,\ldots,X_n$ to get the modal positions and the parameters of the estimated comparison density.
 \vskip.4em

2. Generate random variable $Y$ from the parametric distribution $G$; generate $U$ distributed as ${\rm Uniform}[0,1]$ (independent from $Y$).
 \vskip.4em
3. Accept and set $X^*=Y$ if
  \[\widehat d\big[G(y);G,F\big] \,> \,U \, \max_u \{\widehat d(u)\},\]
  otherwise discard $Y$ and go back to the sampling step (2).
  \vskip.2em
4. Repeat the process until we have simulated sample of size $n$ to produce an i.i.d sample $\{x^{*}_1\, \ldots, x^{*}_n\}$.
 \vskip.4em

5. Draw $B$  independent sets of samples each with size $n$; Repeat the steps 1-4 for each datasets and compute the $k$-modes. Return the standard errors and confidence intervals of the corresponding mode locations that captures the variability.
\vskip.55em
\medskip\hrule height .65pt
\vskip1em
As one of the referee pointed out,  our comparison density based inference algorithm is reminiscent of the smoothed nonparametric bootstrap approach. The only difference is that instead of classical kernel density based bootstrap smoothing \citep{silverman1987}, here we advocate LP Skew-G density based approach. Nonetheless,  it seems to be an interesting connection for further research.
\section{Applications}
Scientists across many disciplines are interested in quantifying modality in distributions.

\subsection{Econometrics}
We consider the dynamics of the cross-country distribution of GDP per worker over a span of $50$ years (between $1959$ to $2008$). The data is taken from The Penn World Table (PWT), version $7.0$ and the output is measured in $2005$ International dollars. We seek to investigate the evolving multi-modal structure of the GDP per worker distribution and the associated questions (e.g., when the multi-modality first emerges), which have significant economic importance.

\begin{figure*}[!thb]
\centering
\includegraphics[height=.3\textheight,width=.45\textwidth,trim=2cm 1cm 2cm 1cm]{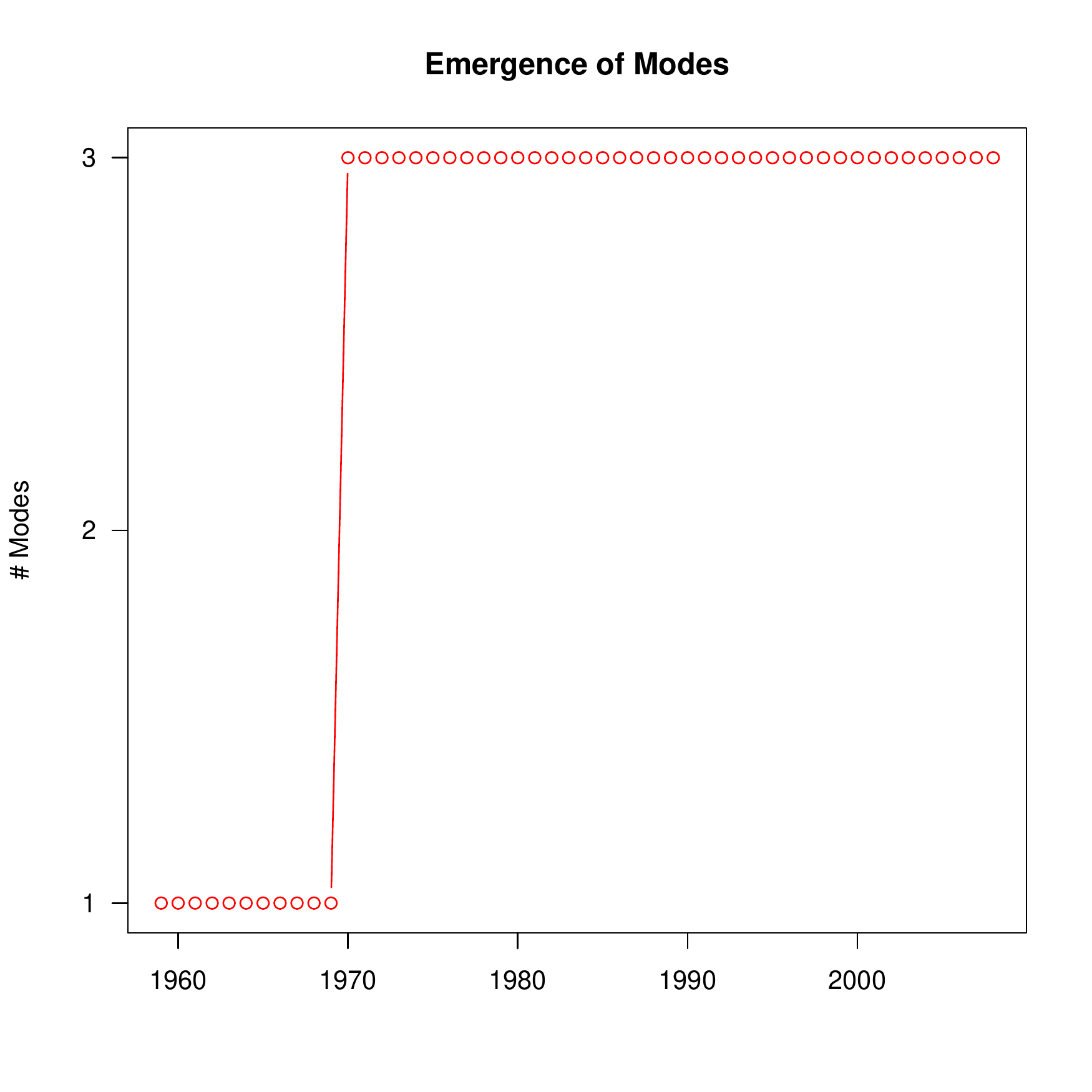}
\caption{How the multi-modal structure of the cross-country GDP per worker distribution evolving over time from $1959$ to $2008$. A sharp transition from unimodal to tri-modal shape occurs at the year $1970$.}
\label{fig:gdp-mode}
\end{figure*}

\begin{figure*}[!thb]
\centering
\vspace{1em}
\includegraphics[height=.4\textheight,width=.45\textwidth,keepaspectratio,trim=2cm .5cm .5cm 0cm]{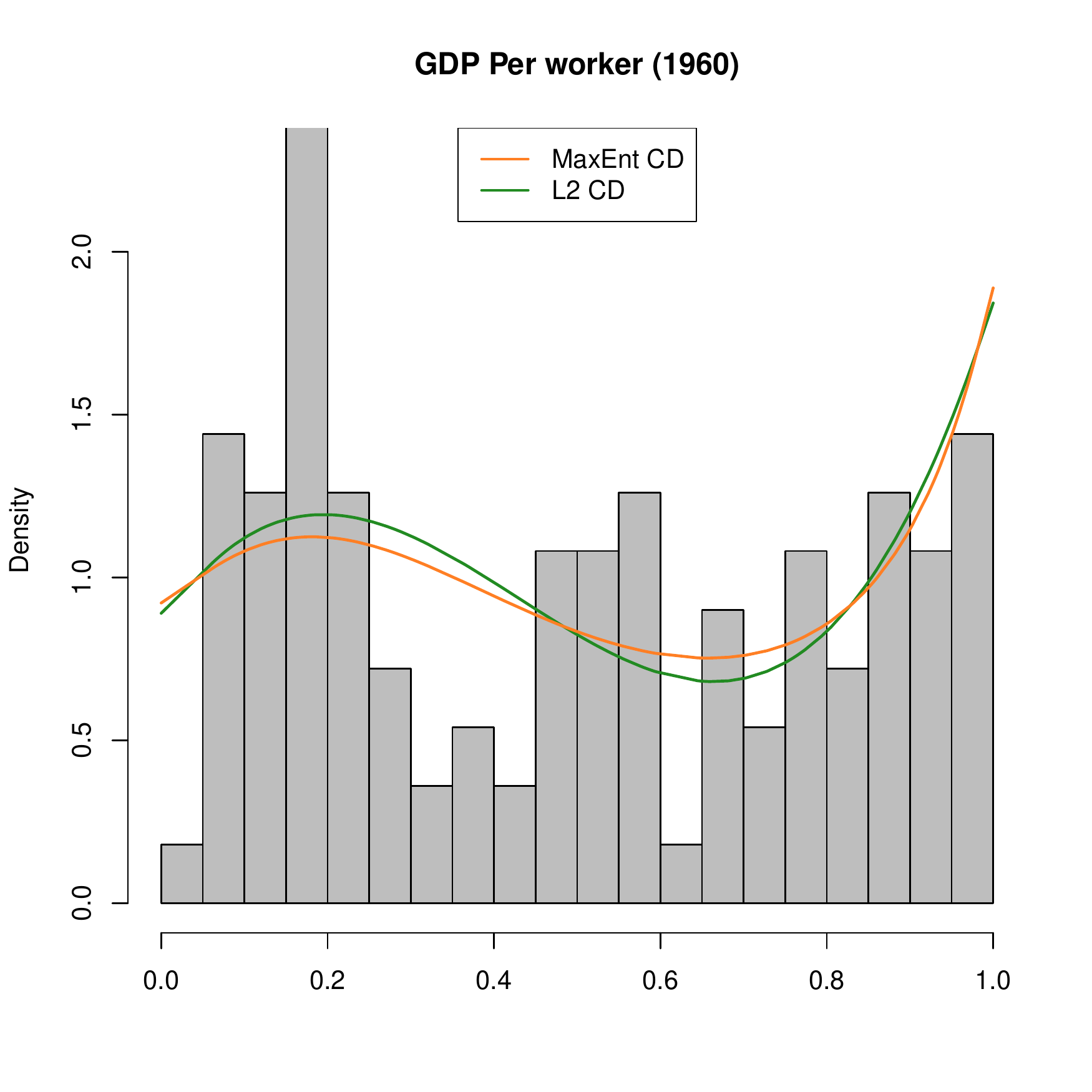}~
\includegraphics[height=.4\textheight,width=.45\textwidth,keepaspectratio,trim=.5cm .5cm 2cm 0cm]{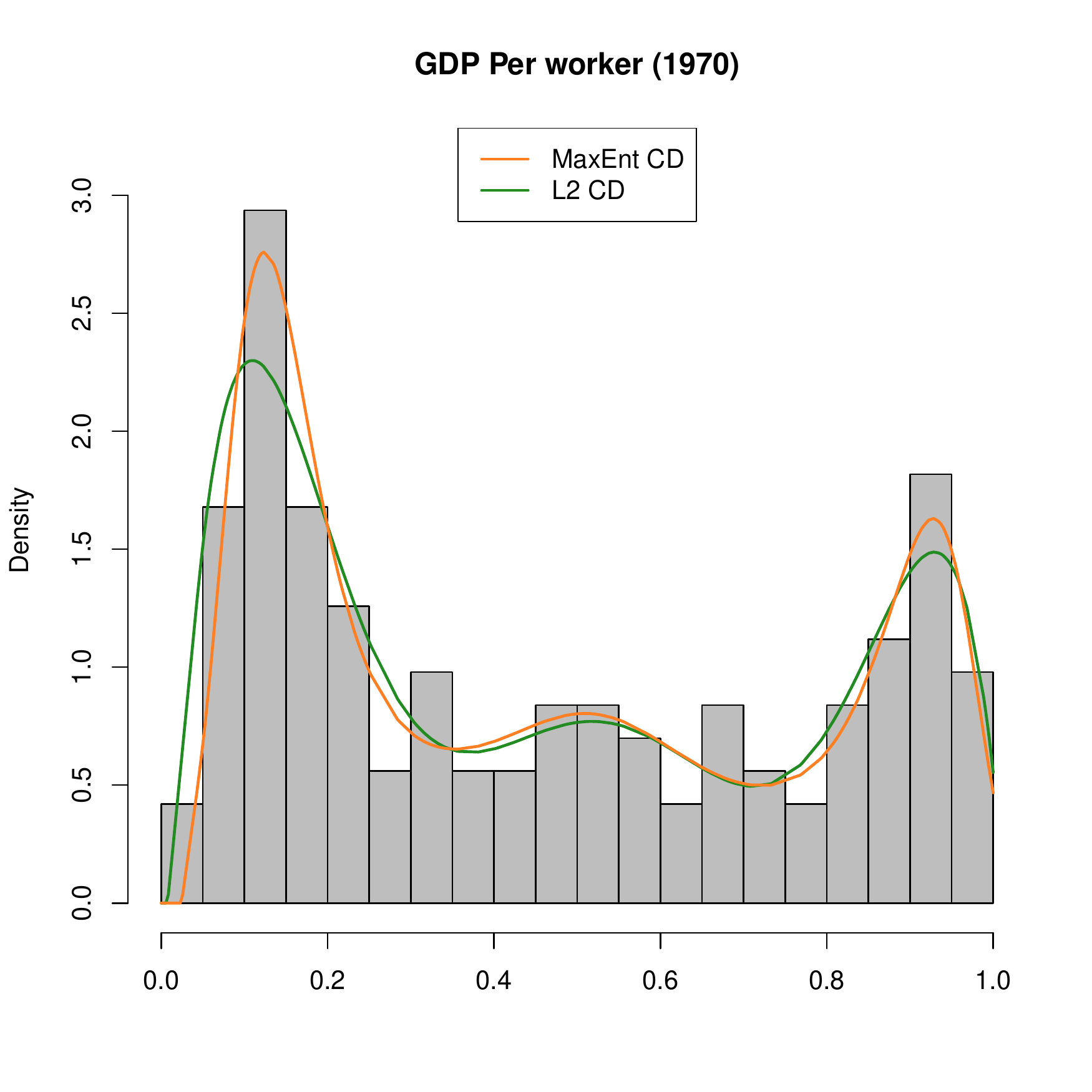}\\
\includegraphics[height=.4\textheight,width=.45\textwidth,keepaspectratio,trim=2cm .5cm .5cm 0cm]{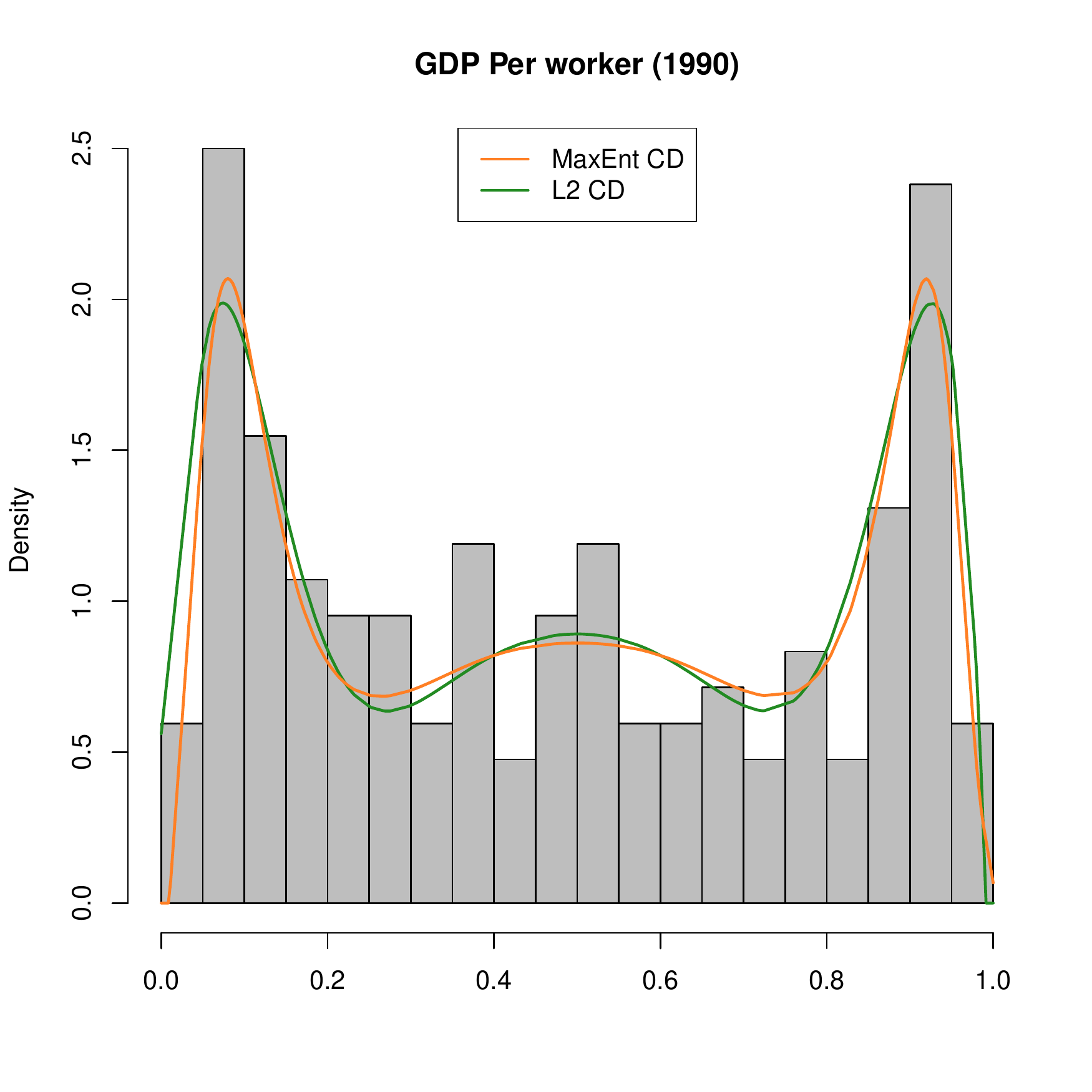}~
\includegraphics[height=.4\textheight,width=.45\textwidth,keepaspectratio,trim=.5cm .5cm 2cm 0cm]{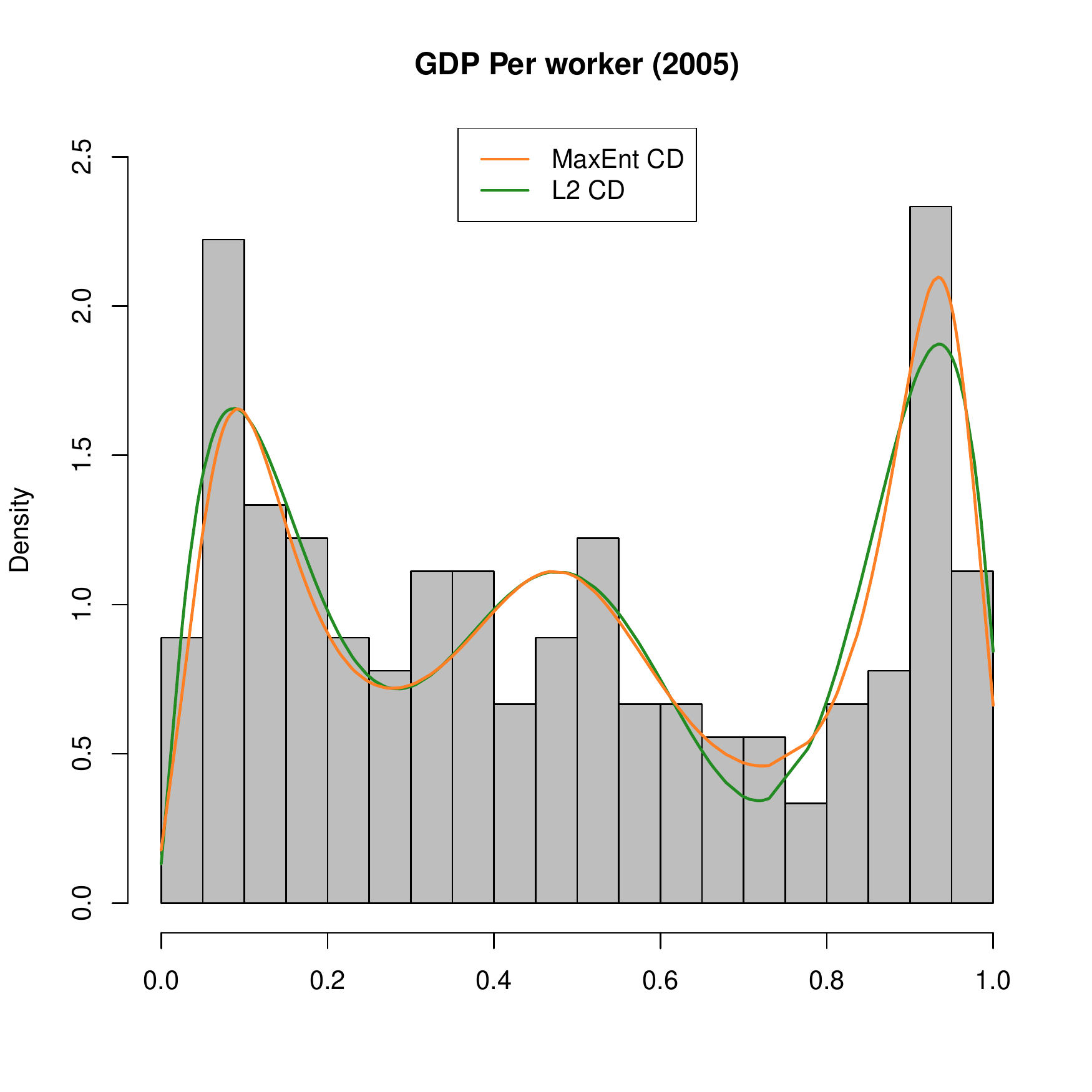}
\vspace{-1em}
\caption{The estimated comparison densities for the years $1960, 1970, 1990$ and $2005$ of the cross-country distribution GDP per worker.} \label{fig:cd}
\end{figure*}

\begin{figure*}[!thb]
\centering
\vspace{1em}
\includegraphics[height=.45\textheight,width=.48\textwidth,keepaspectratio,trim=2cm .5cm .5cm 0cm]{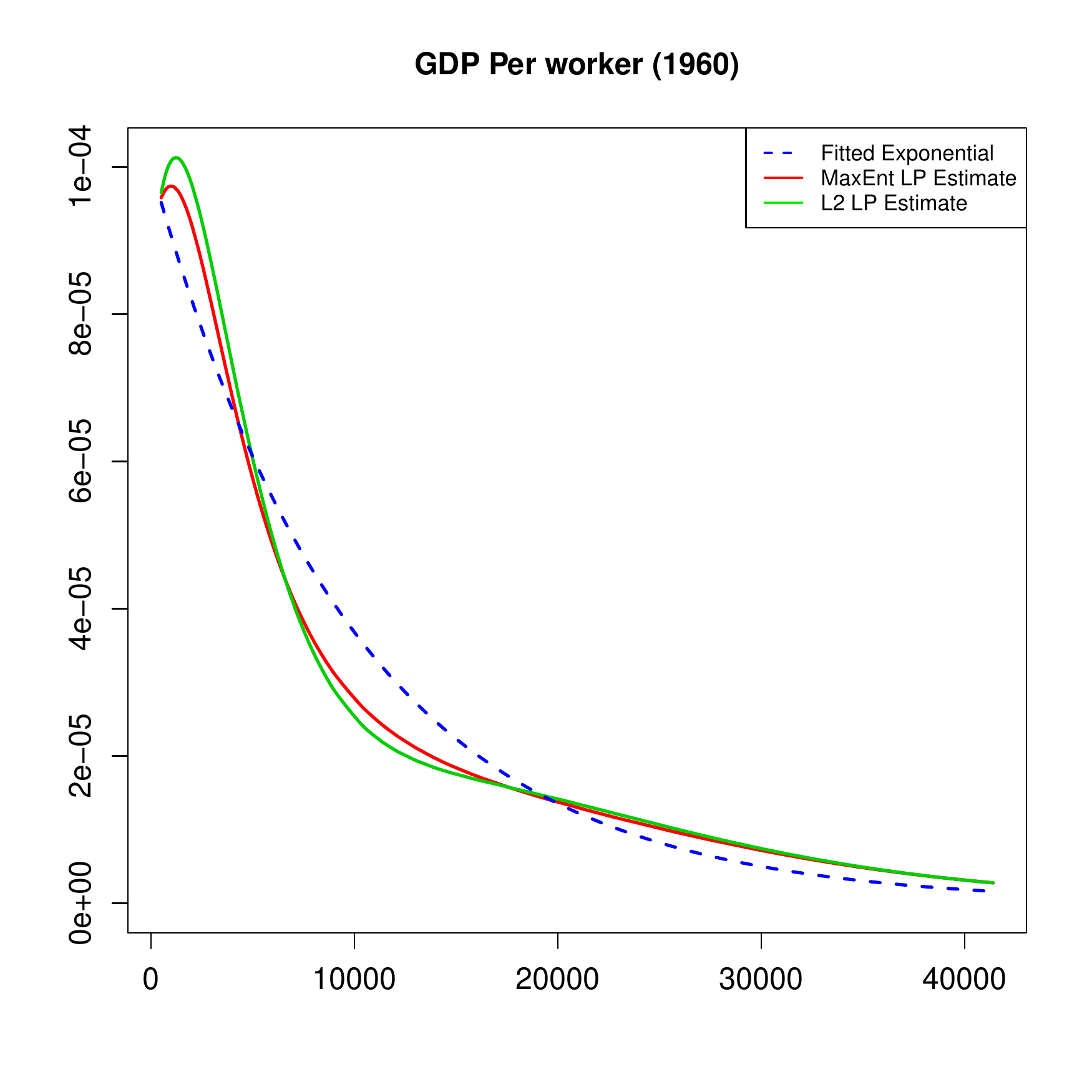}
\includegraphics[height=.45\textheight,width=.48\textwidth,keepaspectratio,trim=.5cm .5cm 2cm 0cm]{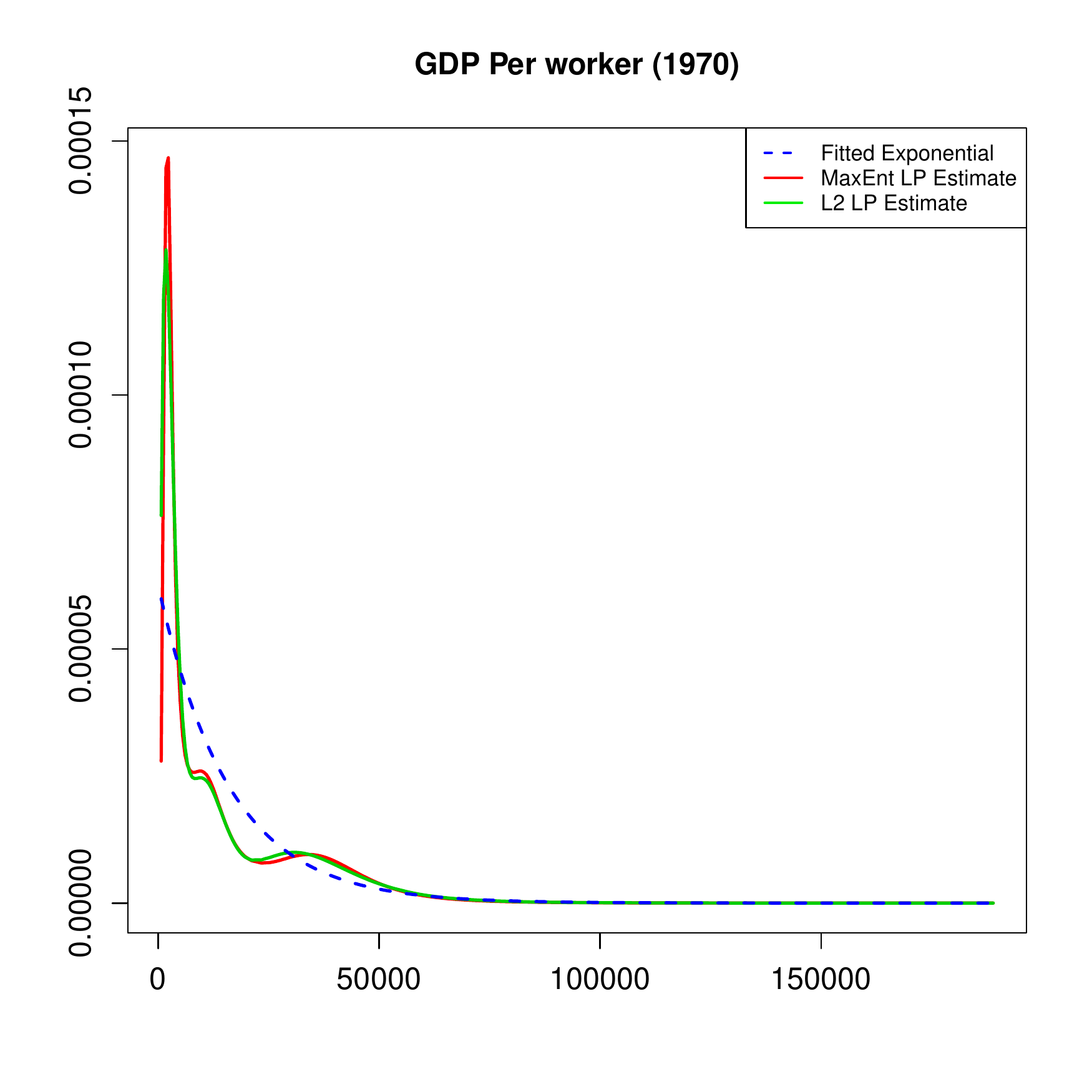}\\
\includegraphics[height=.45\textheight,width=.48\textwidth,keepaspectratio,trim=2cm .5cm .5cm 0cm]{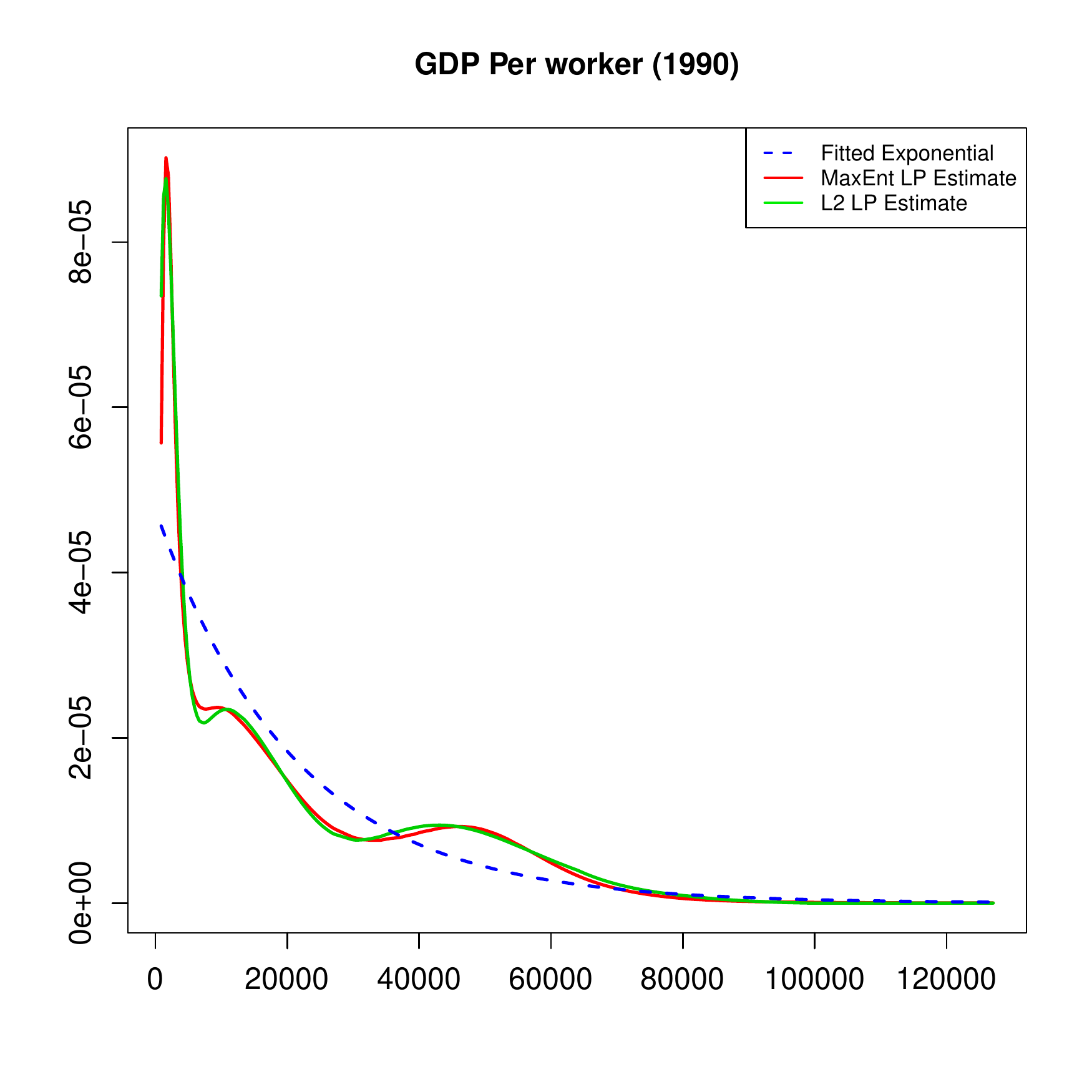}
\includegraphics[height=.45\textheight,width=.48\textwidth,keepaspectratio,trim=.5cm .5cm 2cm 0cm]{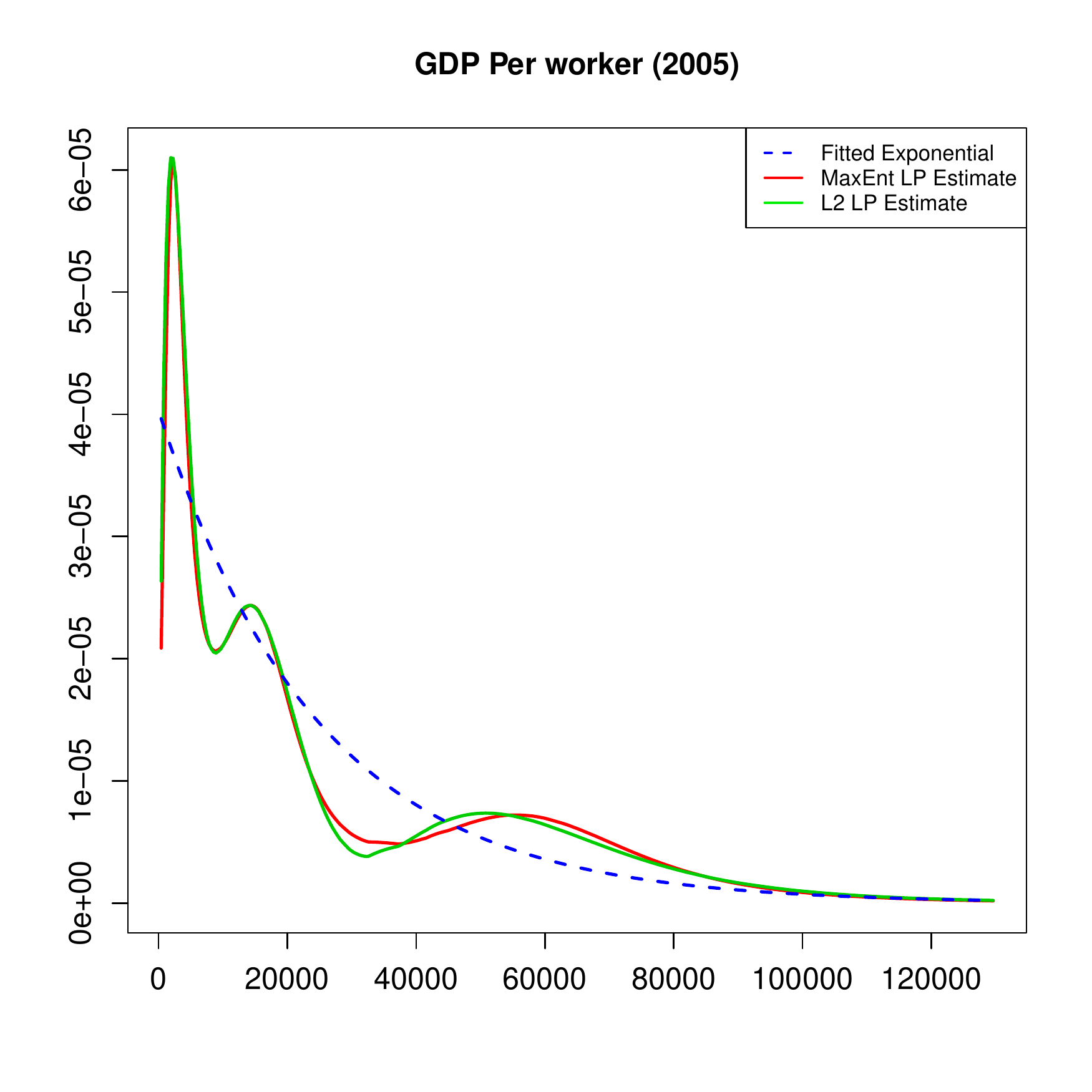}
\caption{The estimated LP skew-G estimated (along with the baseline exponential distribution) densities for the years $1960, 1970, 1990$ and $2005$ of the cross-country GDP per worker distributions.} \label{fig:density}
\end{figure*}

To achieve that we need an automatic bump-hunting algorithm that can learn and compare the multi-modality shape of the GDP per worker distributions at $50$ different points in time without requiring to inspect them manually case-by-case basis. Our \texttt{LPMode} is specially designed to tackle this kind of automated large-scale study. The cross-country GDP per worker distributions have a common shape: a peak around zero and a long exponential-like tail (see Fig \ref{fig:density}). Thus we select $G$ to be exponential, a common choice for all the time points. Our \texttt{LPMode} algorithm scans through all the $50$ distributions and provides (a) the number and locations of modes over time; and (b) nonparametrically corrected parametric specification of the cross-country GDP per worker distribution for all the years. Fig \ref{fig:gdp-mode} shows the dynamics of modes, which finds the sudden prominent emergence from unimodal distribution to tri-modal happens in the year $1970$. This aspect of mode phase-transition in the cross-country GDP per worker distributions was first empirically discovered by \cite{quah1993}, which led to a host of articles on this topic including \cite{henderson2008}. Economic scientists traditionally use Silverman's test for kernel density-based multimodality and have noted three well-known difficulties: (a) it is highly conservative, (b) it has problems of spurious modes in the tails of nonparametric distributions, and (c) it can generate different answers based on various calibration techniques (bandwidth and also calibration of Silverman's test). Thus, the authors were forced to \emph{manually investigate} the modal shapes at certain selected years and found that between $1975$--$80$ the unimodal distribution switches to bimodal shape. This is in contrast to our finding, which indicates the distribution shifted to three-peaked shape (instead of bimodal) in the year $1970$ (which marks the emergence of middle-income countries).\footnote{It is interesting to note that if we plot the histogram the tri-modality is not clear, the reason being the extreme observations `mask the extra middle bump'; we have to be cautious before we infer modality structure of a distribution simply by looking at the histogram - this could lead to misleading results.}

\begin{figure*}[!thb]
\centering
\includegraphics[height=.29\textheight,width=.6\textwidth,trim=2cm .5cm 2cm 0cm]{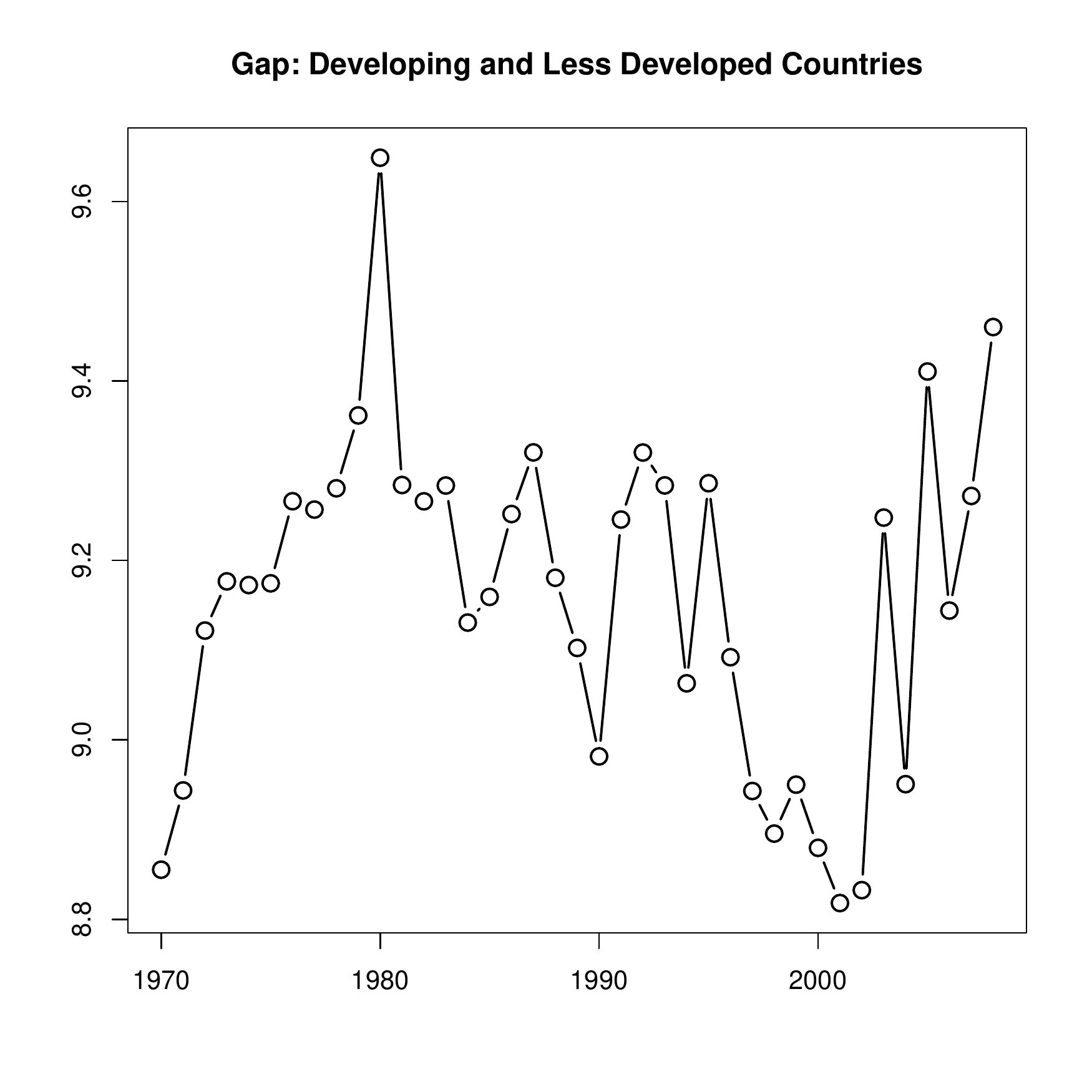}\\
\includegraphics[height=.29\textheight,width=.6\textwidth,trim=2cm .5cm 2cm 0cm]{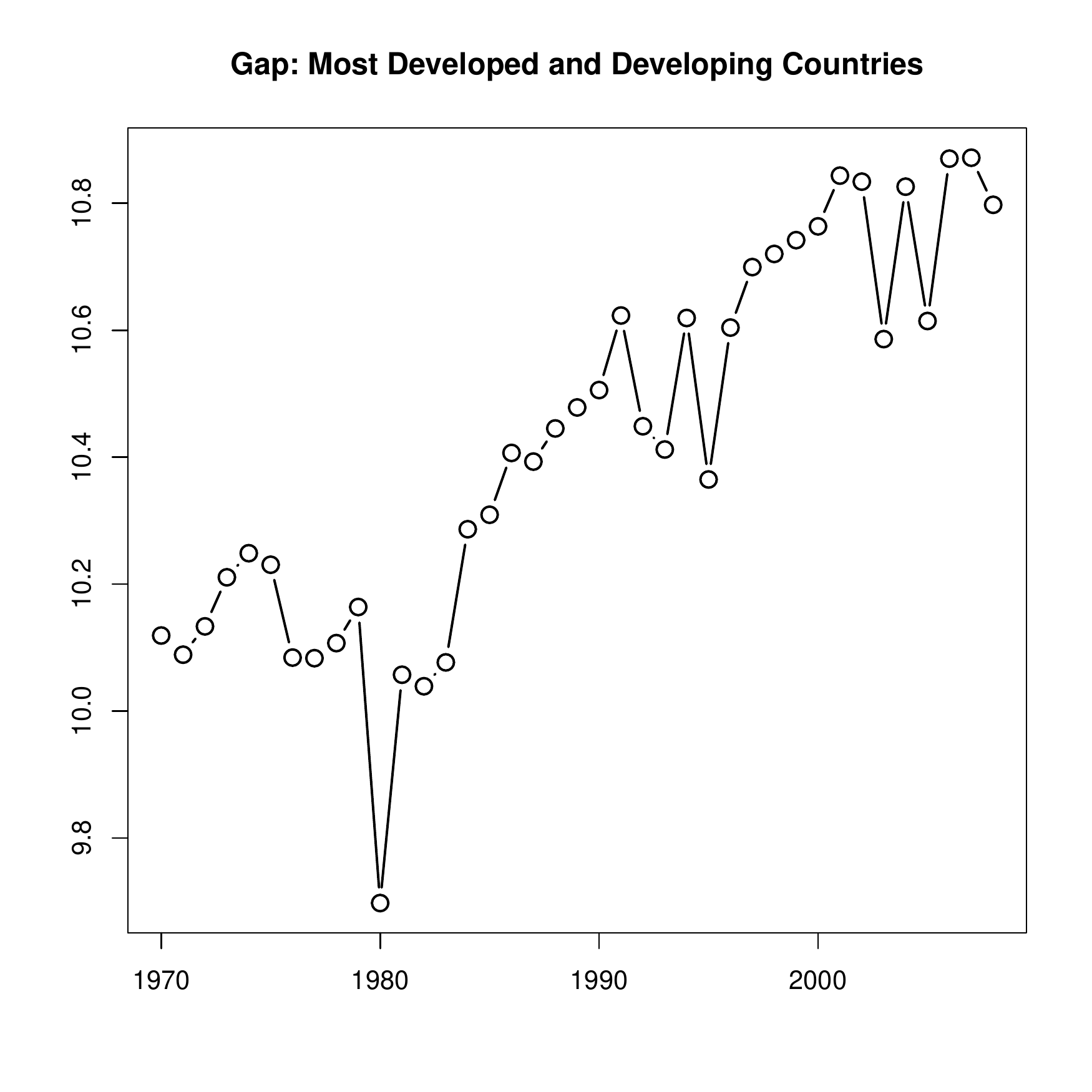}\\
\includegraphics[height=.29\textheight,width=.6\textwidth,trim=2cm 1cm 2cm 1cm]{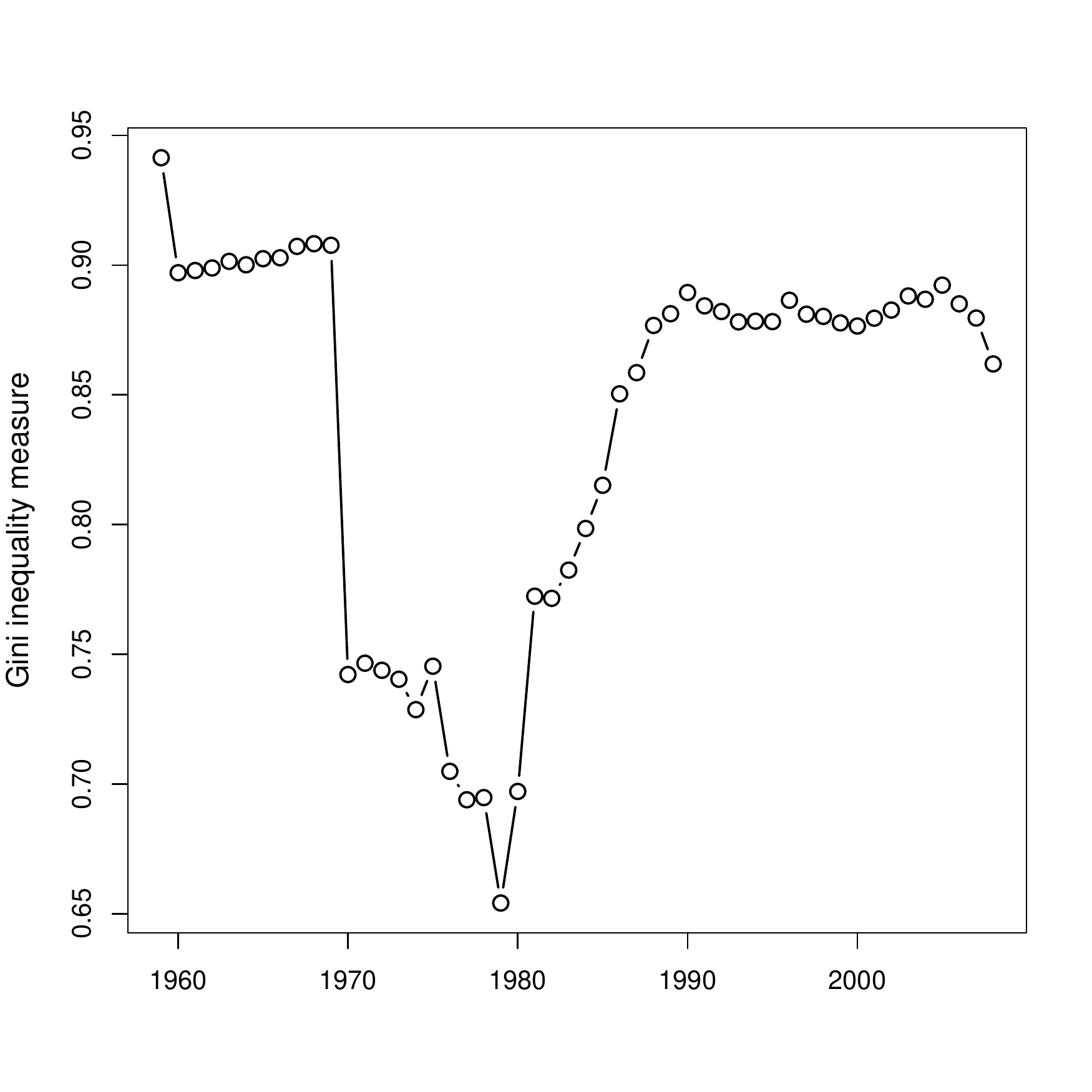}
\caption{The dynamics of the gap between the developing, less- and most-developed countries to understand the inequality in the world economy; The gap is computed as the log of the difference between the corresponding representative modes at each year. The last row shows the Gini inequality measure over time.}
\label{fig:gap}
\end{figure*}

\begin{figure*}[!thb]
\centering
\includegraphics[height=.27\textheight,width=.7\textwidth,trim=2cm .5cm 2cm 1cm]{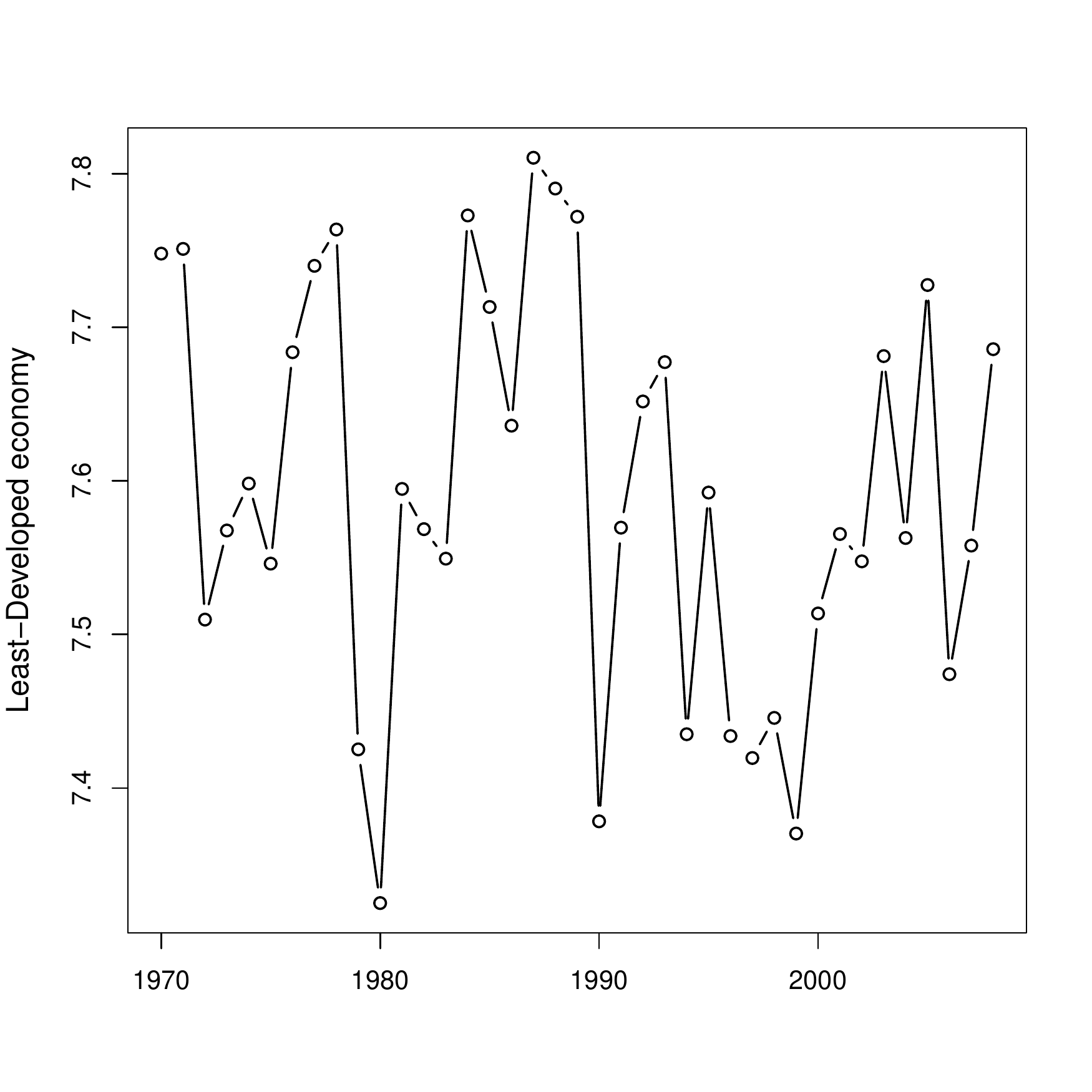}\\[.2em]
\includegraphics[height=.27\textheight,width=.7\textwidth,trim=2cm .5cm 2cm 1cm]{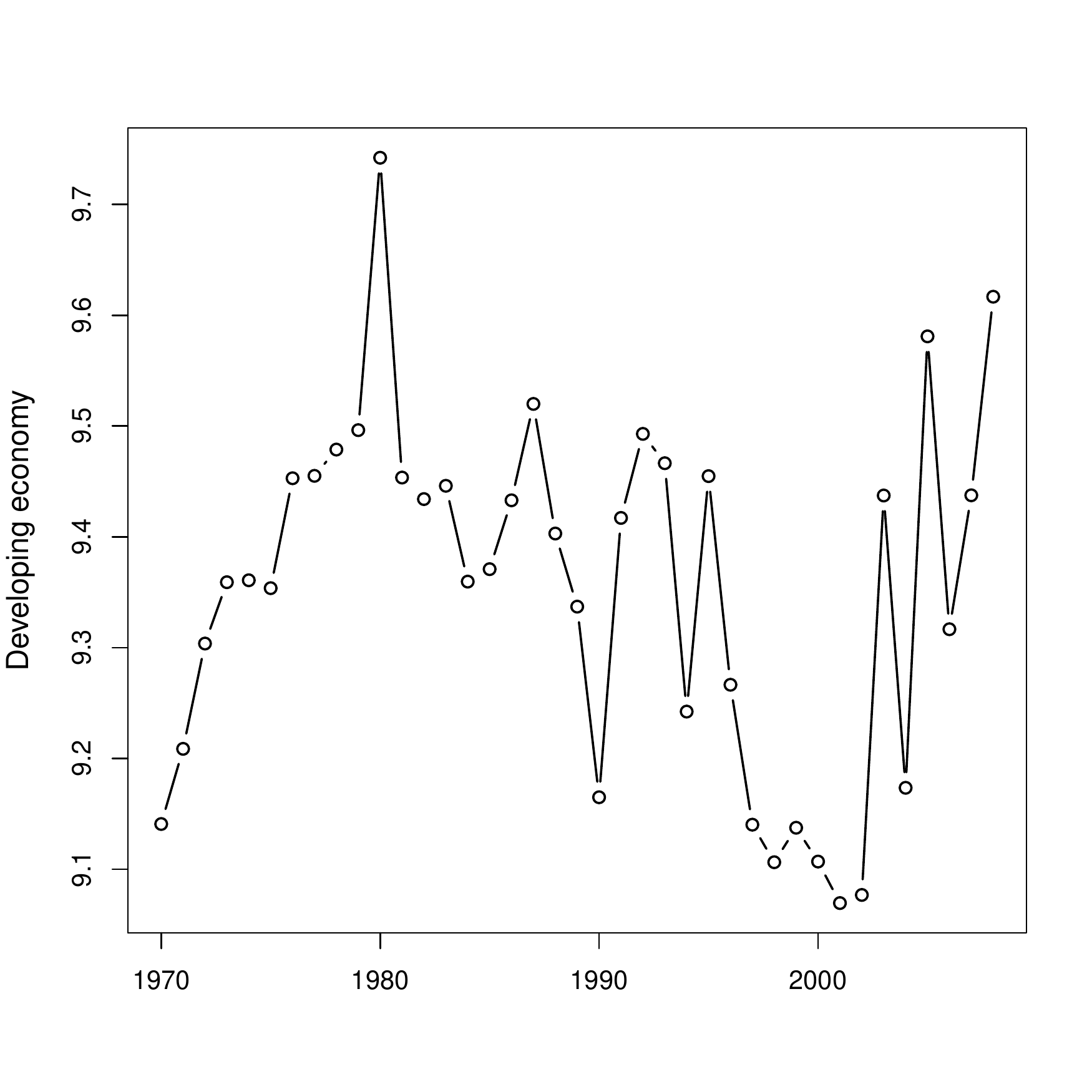}\\[.2em]
\includegraphics[height=.27\textheight,width=.7\textwidth,trim=2cm 1.5cm 2cm 1cm]{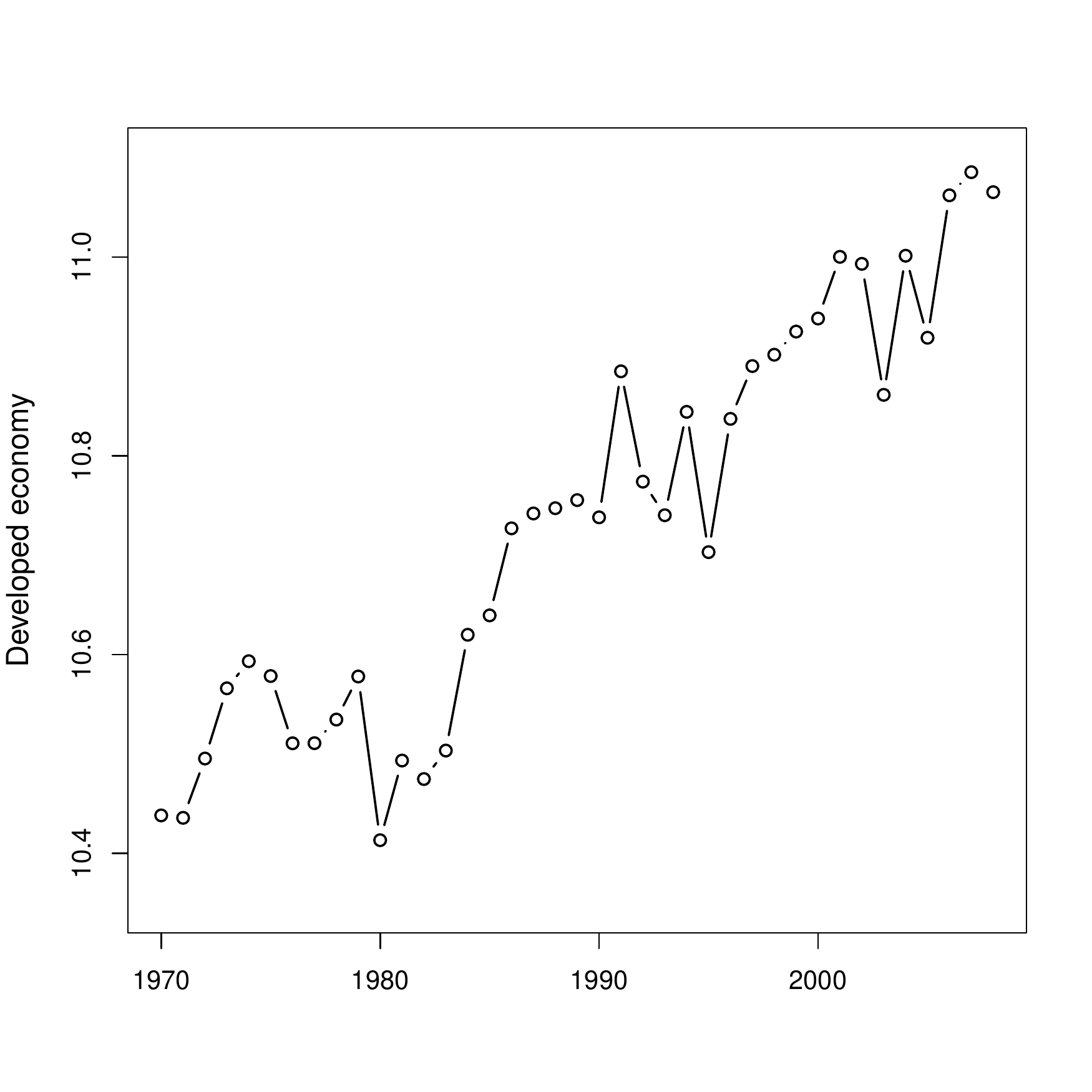}\\[.2em]
\caption{How the modes of each three classes evolved over time. We have plotted the log of the modal positions over time. Goal is to understand the dynamics of GDP per worker growth rate for each of the three classes over $50$ years time span.}
\label{fig:gdp-ts}
\end{figure*}

The three modes respectively denote the low-, middle-, and high-income countries.  Figs \ref{fig:cd} plots the comparison densities estimates for the years $1960, 1970, 1990$ and $2005$. The presence of three modes is most clearly visible from the plot of comparison densities. The evidence of three modes has recently been noted by \cite{pittau2010}, although the year of inception of tri-modality differs from ours. They have used parametric Gaussian mixture model for nine hand-picked years between $1960$ and $2000$. They noted that in contrast to the commonly held bimodality view ``the statistical tests that we use indicate the presence of three component densities in each of the nine years that we examine over this period.'' The gaps between (a) developing and less-developed countries, and (b) developing and most-developed nations are shown in Fig \ref{fig:gap}. The gap between low- and middle-income countries reduced substantially from early 1980s until the late-1990s; thereafter the gap widens, hinting at polarization between these two classes. On the other hand, recent trends suggest that developing economies have increased their rate of convergence in GDP per worker with developed economies (see Fig \ref{fig:gdp-ts}), which is manifested in the tri-modal density shape. The bottom panel of Fig \ref{fig:gap} shows the dynamics of Gini measure, computed by taking correlation of $X$ and $F(X;X)$ for each time point, which measures the degrees of inequality and polarization. Gini measure has dropped from $0.94$ in $1959$ to $0.65$ in $1980$, increased almost monotonically between $1980$ to $1990$ and then stagnated for next decade at the value $.86$, which is practically the same as that of $1960$.

\subsection{Cancer genomics}
We study the bump-hunting based cancer biomarker discovery and disease prediction using three high-dimensional microarray databases summarized in Table \ref{tab:cg}.

\begin{figure*}[t]
\centering
\vspace{1em}
\includegraphics[height=\textheight,width=.31\textwidth,keepaspectratio,trim=2cm 1cm .5cm 2cm]{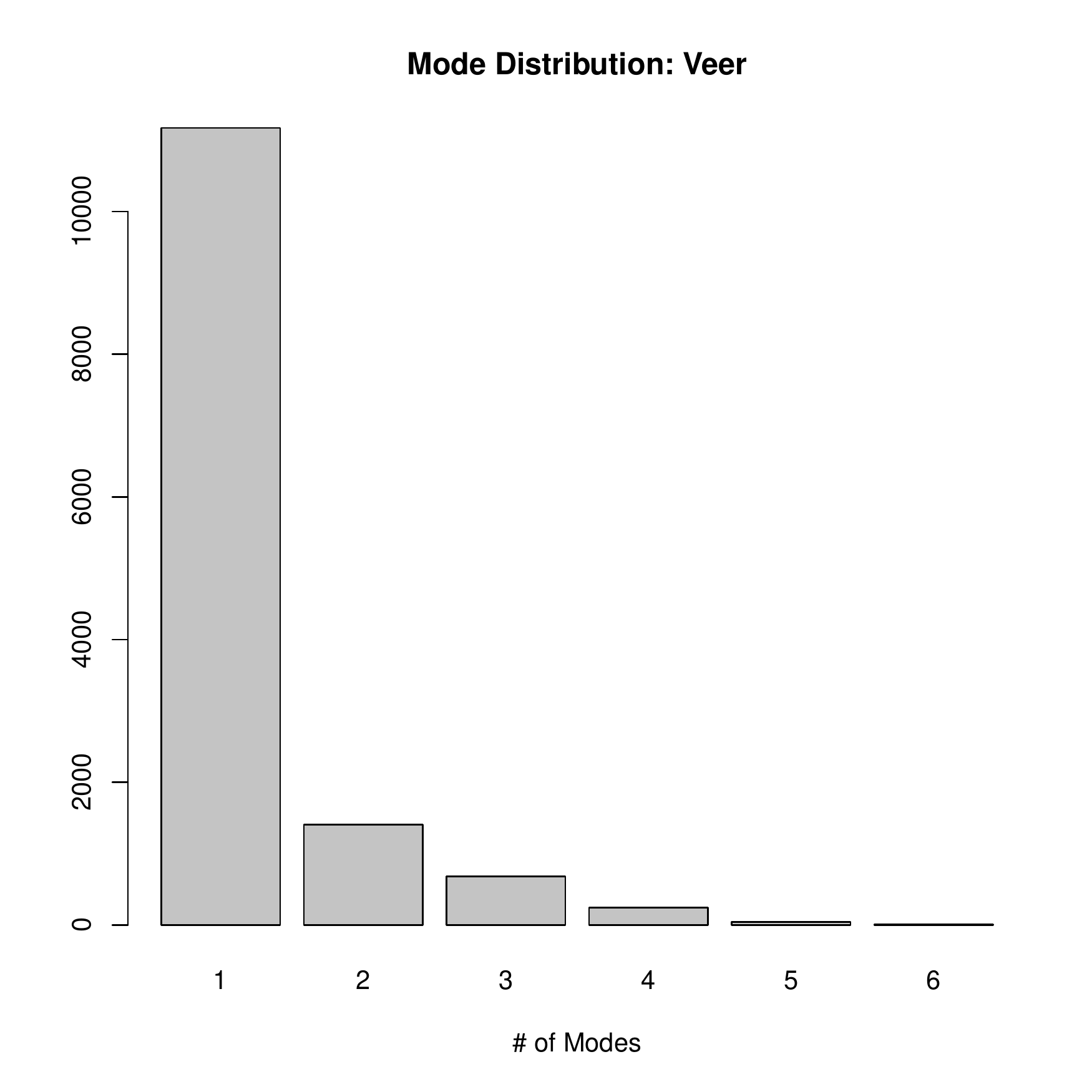}
\includegraphics[height=\textheight,width=.31\textwidth,keepaspectratio,trim=1cm 1cm 1cm 2cm]{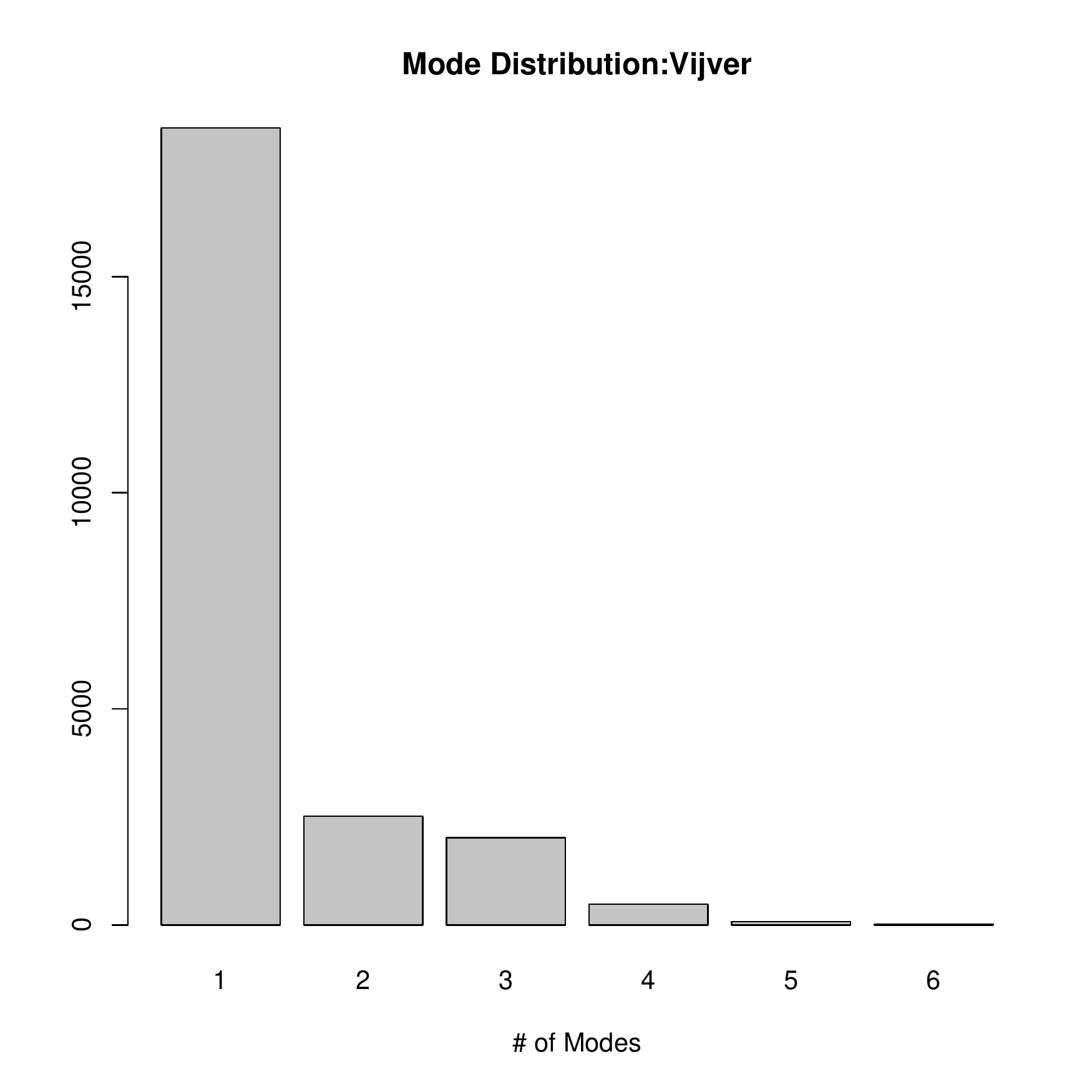}
\includegraphics[height=\textheight,width=.31\textwidth,keepaspectratio,trim=1cm 1cm 2cm 2cm]{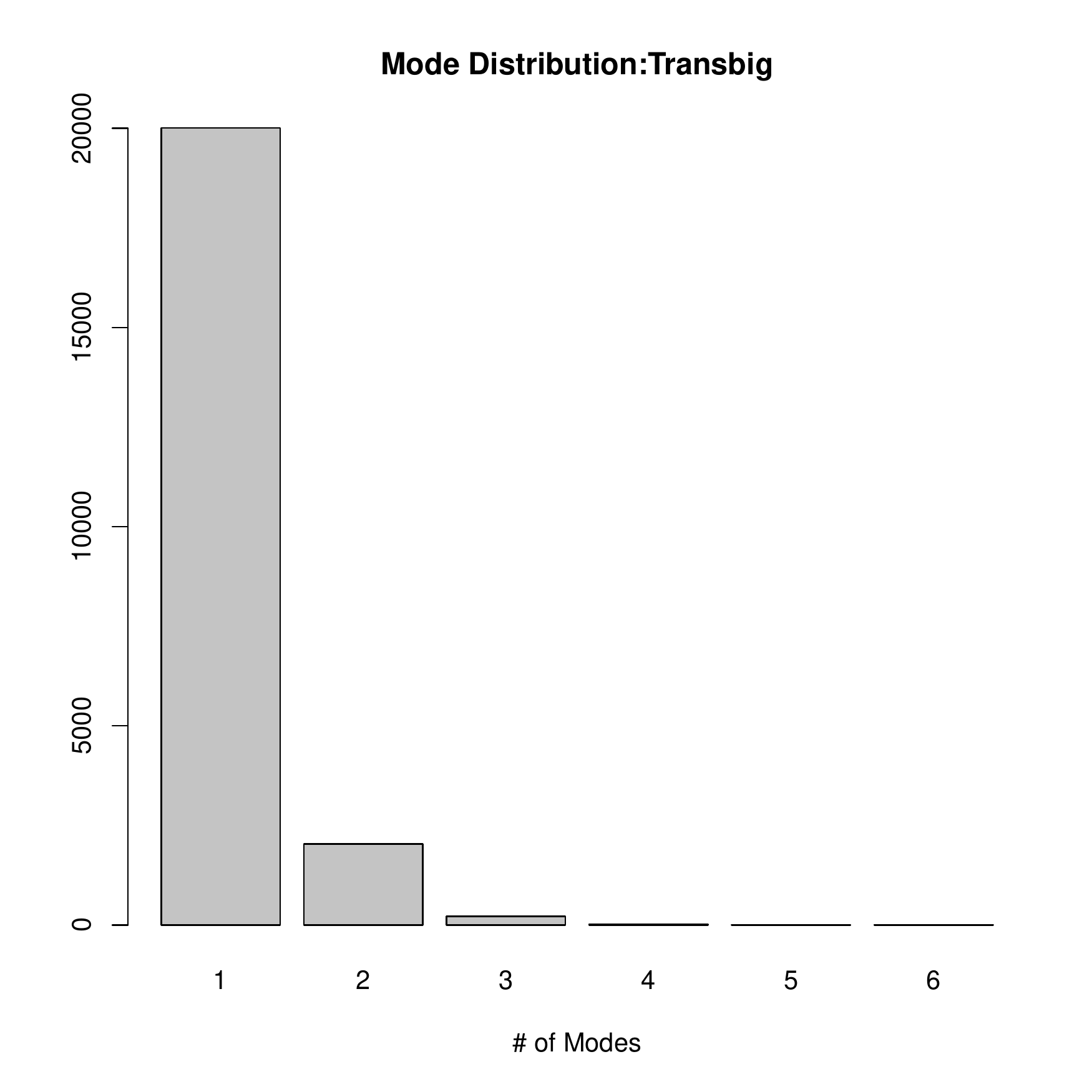}\\[4em]
\includegraphics[height=\textheight,width=.31\textwidth,keepaspectratio,trim=2cm 1cm .5cm .5cm]{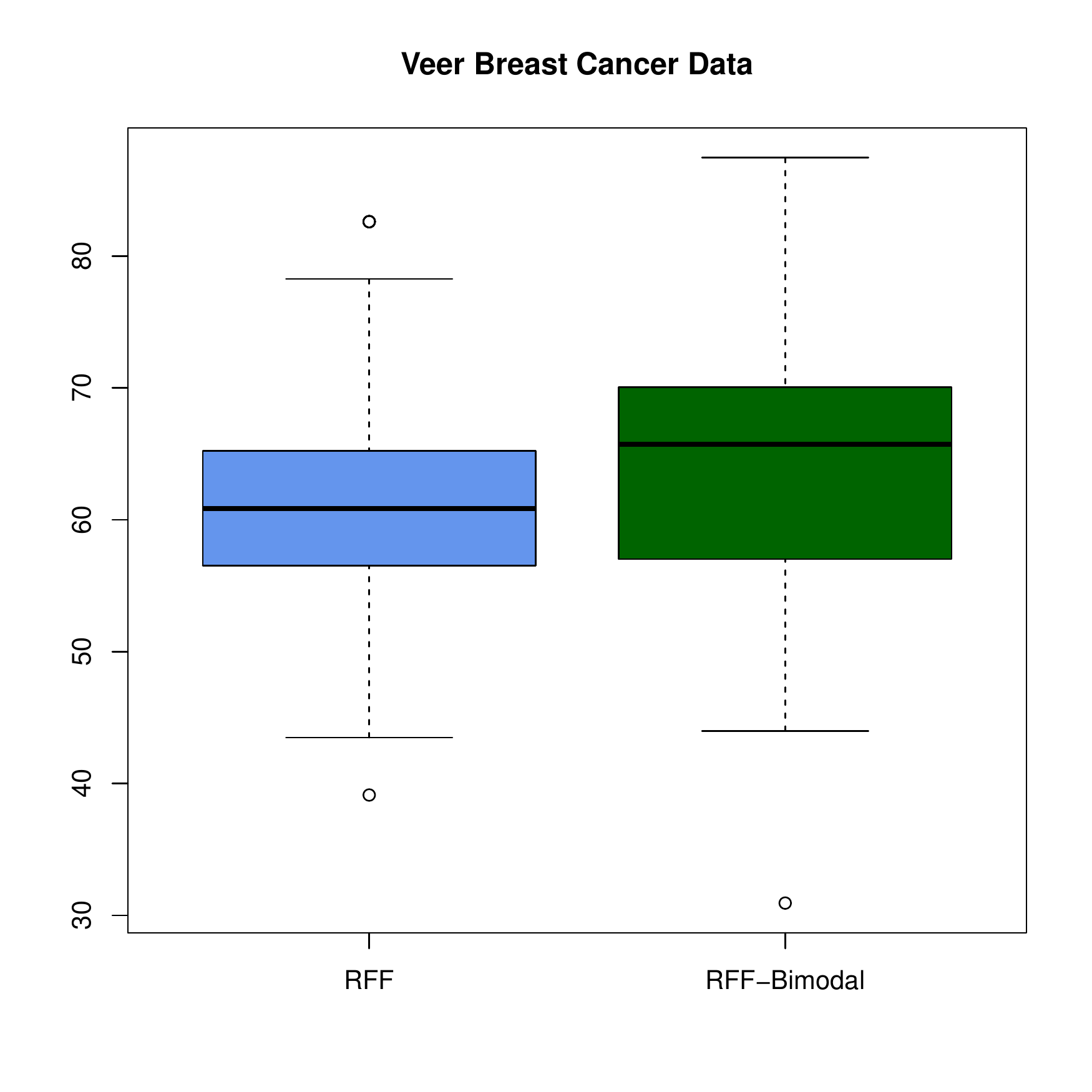}
\includegraphics[height=\textheight,width=.31\textwidth,keepaspectratio,trim=1cm 1cm 1.5cm .5cm]{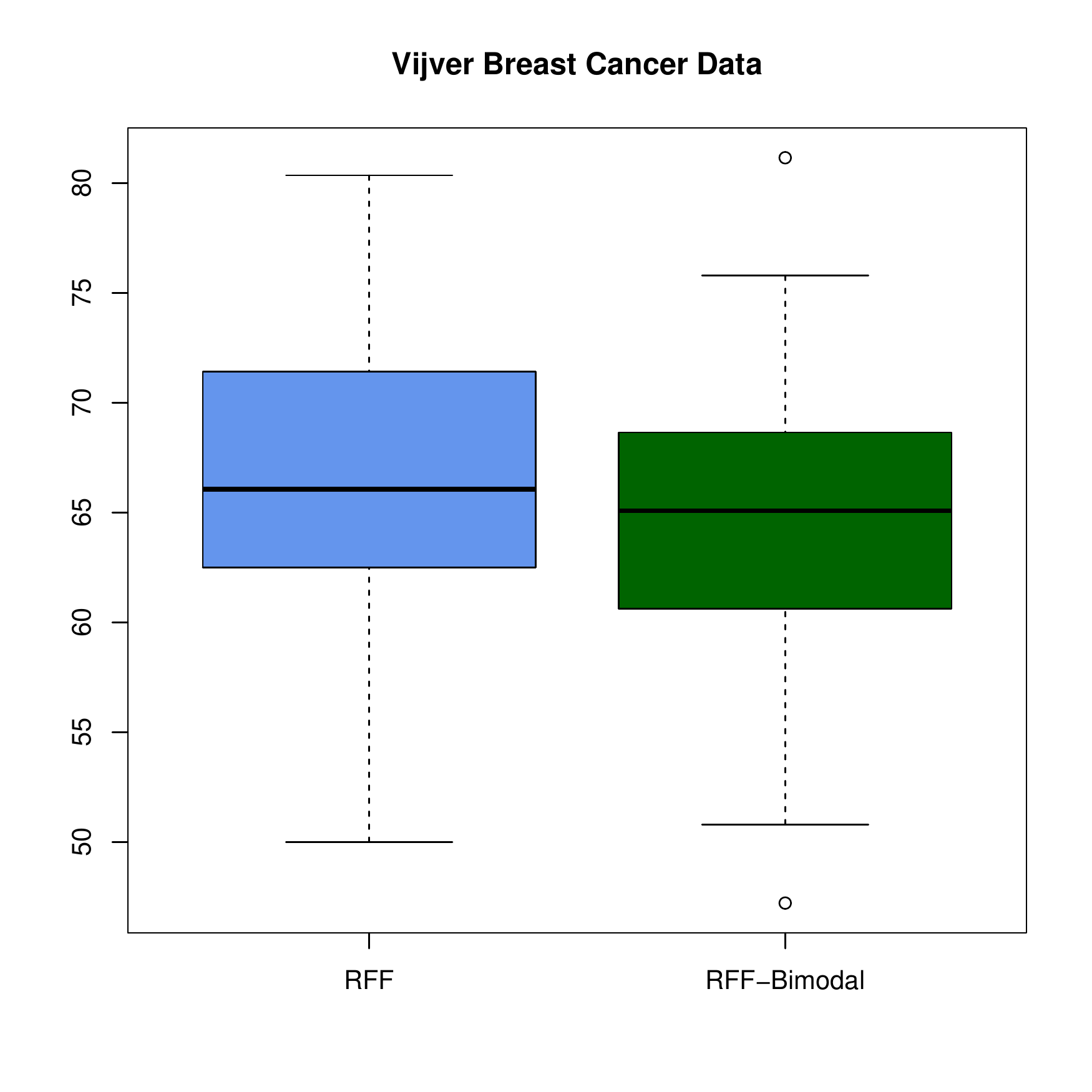}
\includegraphics[height=\textheight,width=.31\textwidth,keepaspectratio,trim=.5cm 1cm 2.5cm .5cm]{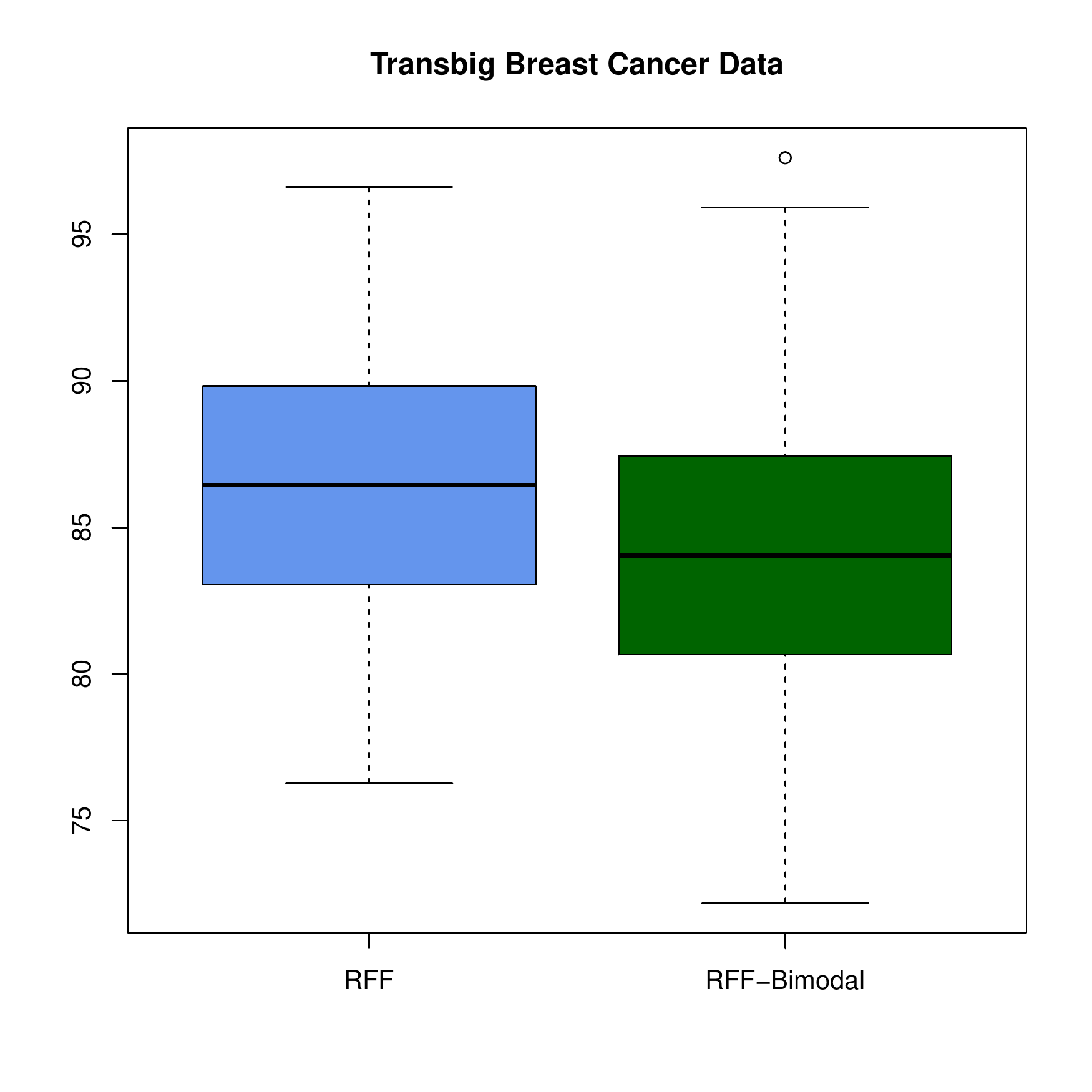}
\vskip1em
\caption{The top panel shows the bar plot which categorizes the genes according to modality for each breast cancer microarray data sets. The bottom panel compares the accuracy of disease based classification rules using Random Forest (RF). We have compared two methods for each data sets: (a) all the genes have used to build the RF model (shown in blue col boxplot), (b) only the LPMode selected multi-modal genes are used as features to build the RF model (shown in green col boxplot).} \label{fig:cancer}
\end{figure*}

To investigate the significance of the multi-modal genes for predicting cancer classes, we fit random forest \citep{breimanRF}, model-free classifier that can tackle high-dimensional covariates, under two different scenarios where the feature matrix $X$ contains (a) all the genes; or (b) only the multi-modal genes.  The necessary first step is to identify the modal structures of the genes. The huge dimensionality (thousands of gene expression distributions) makes this a challenging problem. However, \texttt{LPMode} algorithm provides a systematic (and automatic) nonparametric exploration strategy for such large-scale mode identification problems. LPMode categorizes the genes based on their modal shapes as shown in the top panel of Fig \ref{fig:cancer}.

\begin{table}[!ht]
\setlength{\tabcolsep}{12pt}
\caption{\textit{Three breast cancer microarray datasets.}} 
\vskip1em
\centering 
\begin{tabular}{ccccc} 
\hline\hline 
Datasets & Reference &\# samples ($n$) & \# genes ($p$) \\ [0.5ex] 
\hline 
Veer data &\cite{veer2002} & $78$  & $24,481$  \\
Vijver data &\cite{vijver2002} & $295$ & $22,283$\\
Transbig data &\cite{transbig2007}&$198$&$22,283$\\
\hline 
\end{tabular}
\label{tab:cg} 
\end{table}

We apply the LPMode bump-hunting based unsupervised gene selection strategy (by screening only the multi-modal genes) for supervised classification learning. We randomly sampled 30\% of the data to form a test set. This entire process was repeated $100$ times, and test set accuracy rates are shown in the bottom panel of Fig. \ref{fig:cancer}. The predictive performance is summarized in Table \ref{tab:cg2}. The surprising fact to note that based on only the multi-modal gene expression signatures (which achieves significant compression by ignoring all the unimodal genes) we get almost the \emph{same accuracy} for cancer classification.

\begin{figure*}[!thb]
\centering
\vspace{1em}
\includegraphics[height=.31\textheight,width=.31\textwidth,keepaspectratio,trim=2cm 1cm .5cm 2cm]{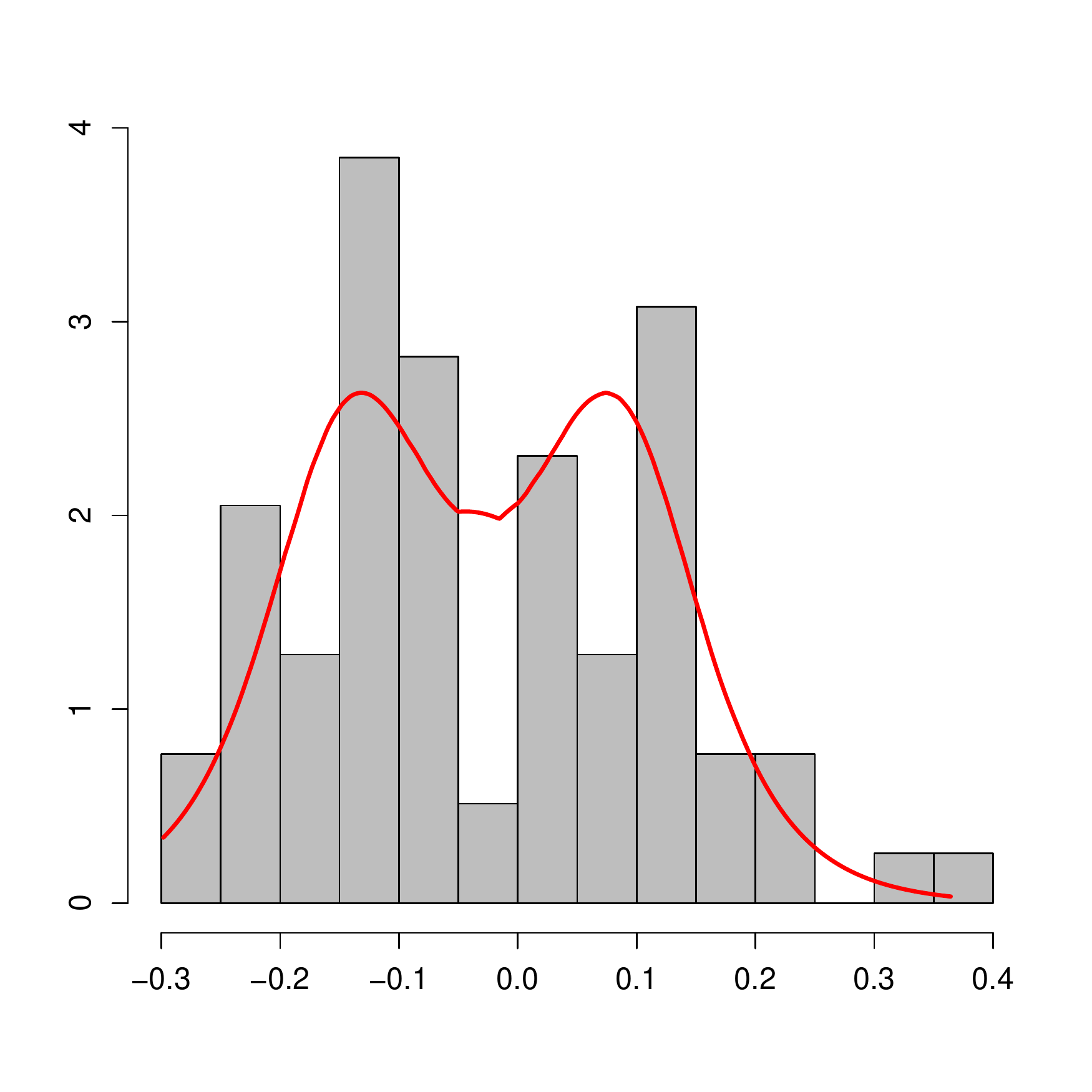}
\includegraphics[height=.31\textheight,width=.31\textwidth,keepaspectratio,trim=1cm 1cm 1cm 2cm]{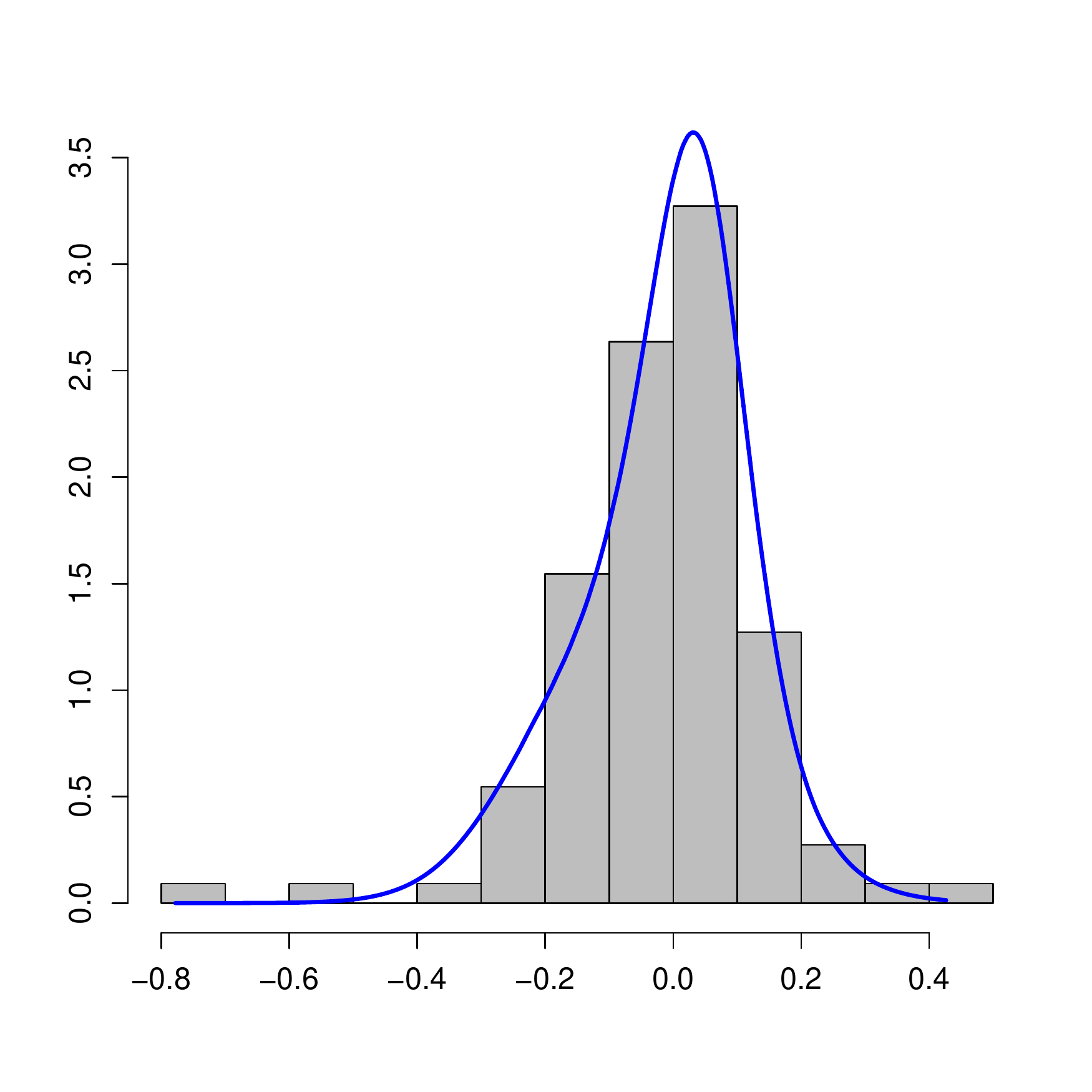}
\includegraphics[height=.31\textheight,width=.31\textwidth,keepaspectratio,trim=1cm 1cm 2cm 2cm]{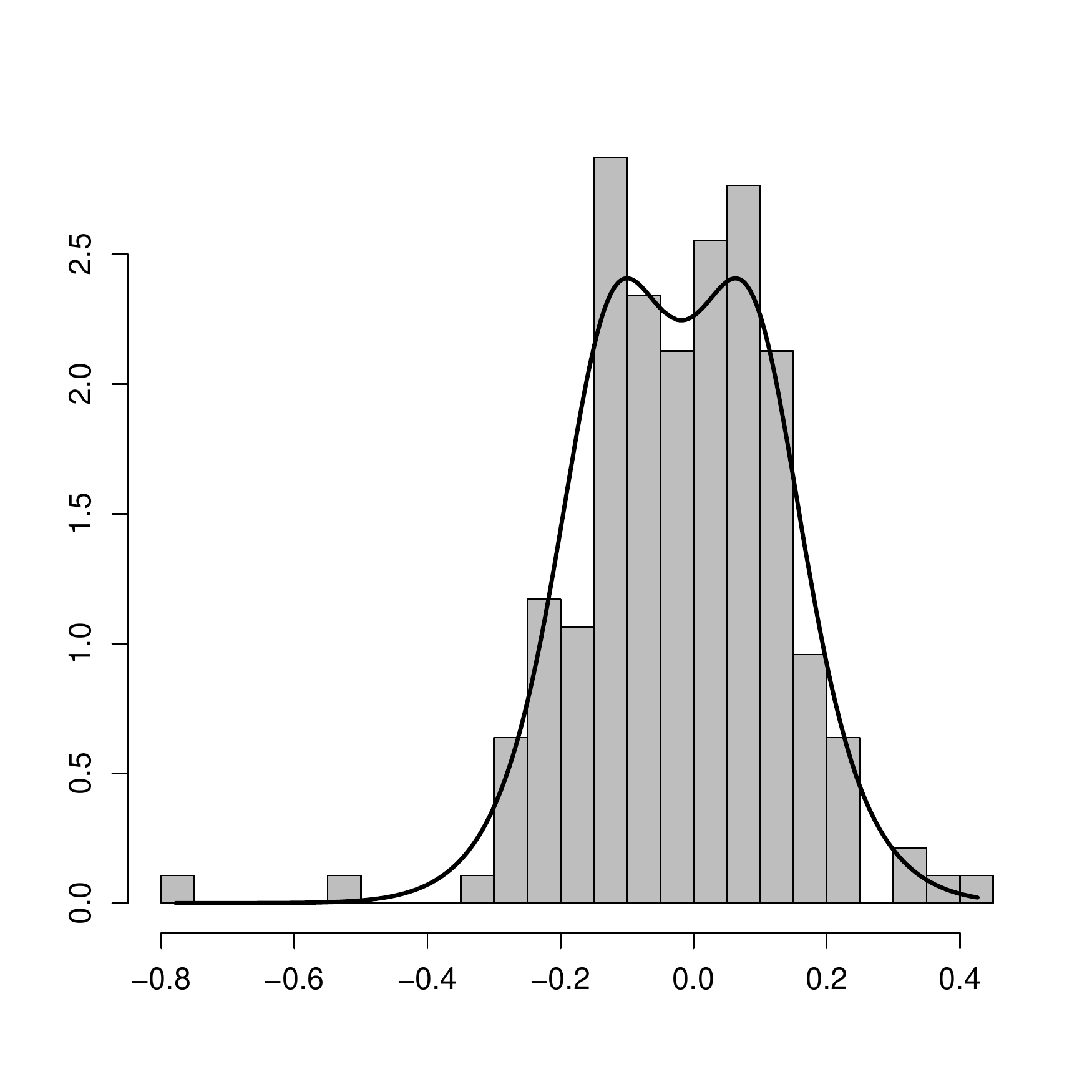}\\[1em]
\includegraphics[height=.31\textheight,width=.31\textwidth,keepaspectratio,trim=2cm 1cm .5cm .5cm]{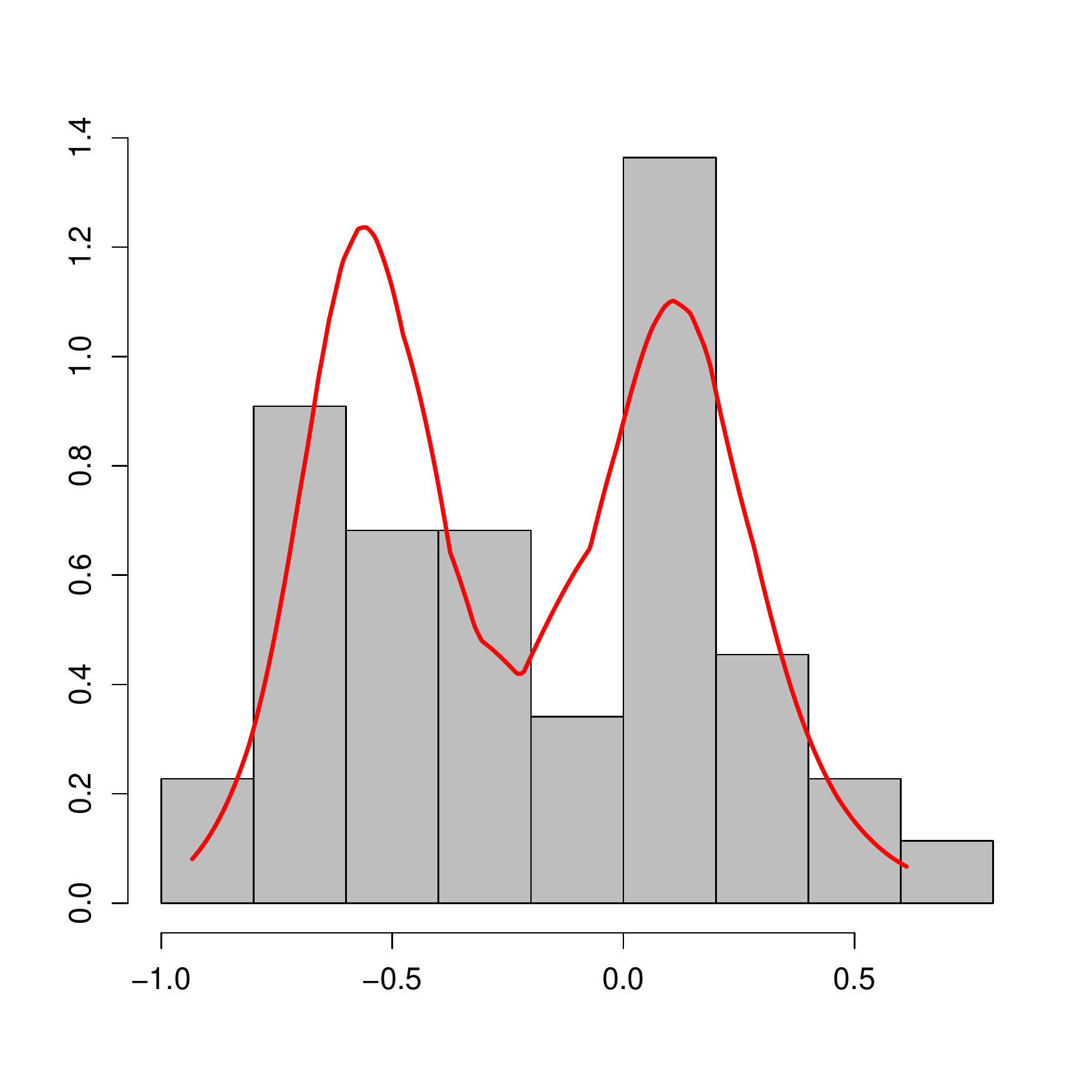}
\includegraphics[height=.31\textheight,width=.31\textwidth,keepaspectratio,trim=1cm 1cm 1.5cm .5cm]{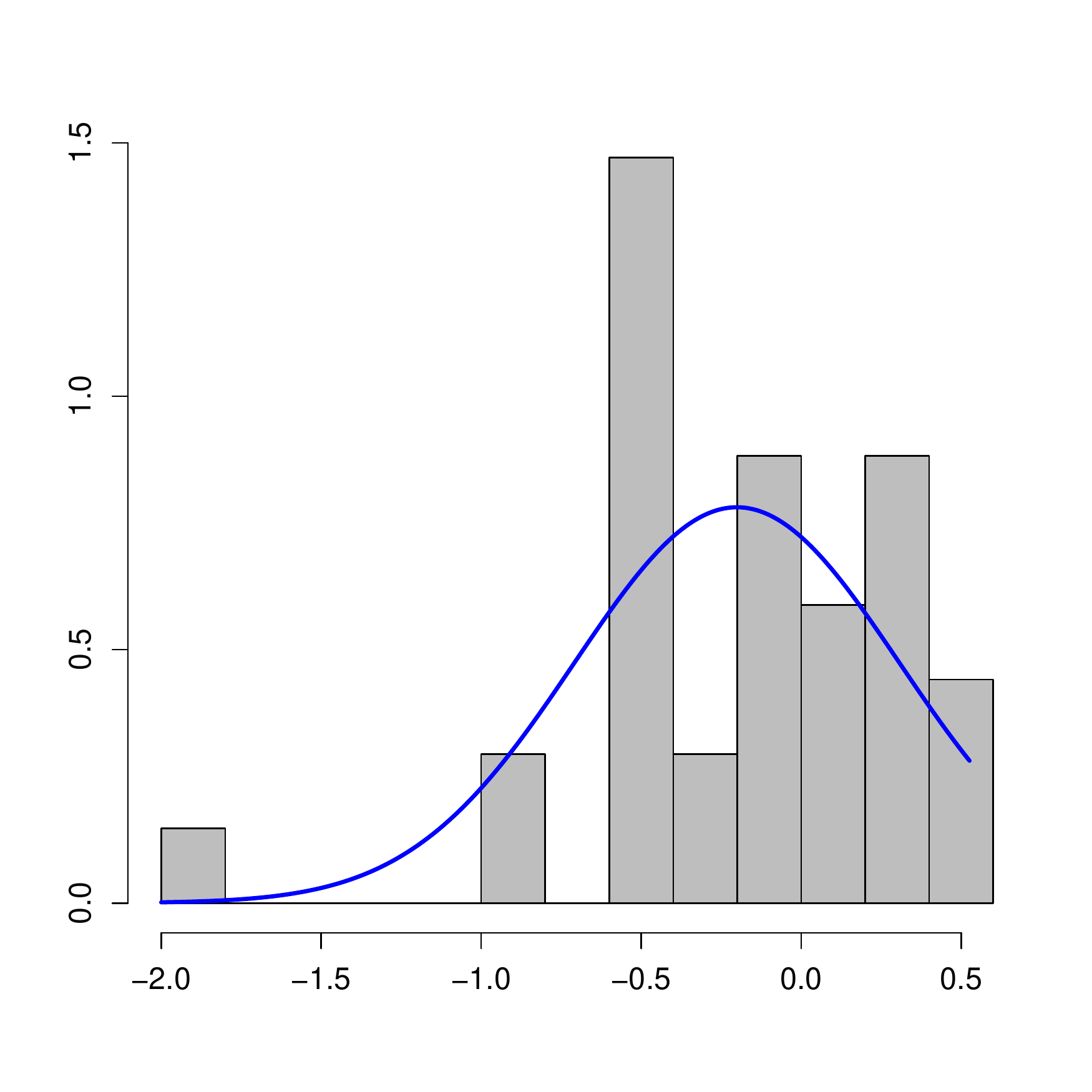}
\includegraphics[height=.31\textheight,width=.31\textwidth,keepaspectratio,trim=.5cm 1cm 2.5cm .5cm]{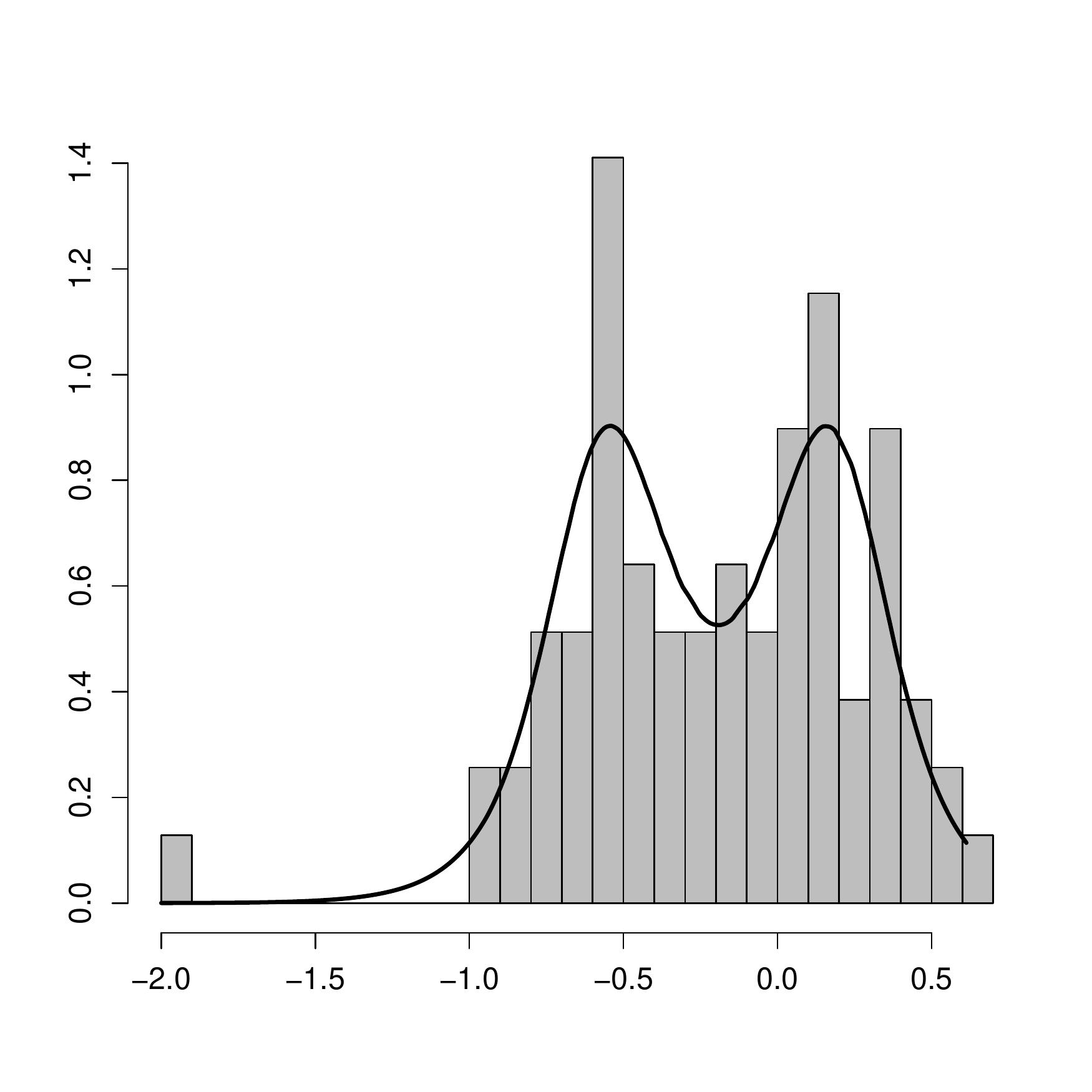}\\[1em]
\includegraphics[height=.31\textheight,width=.31\textwidth,keepaspectratio,trim=2cm 1cm .5cm .5cm]{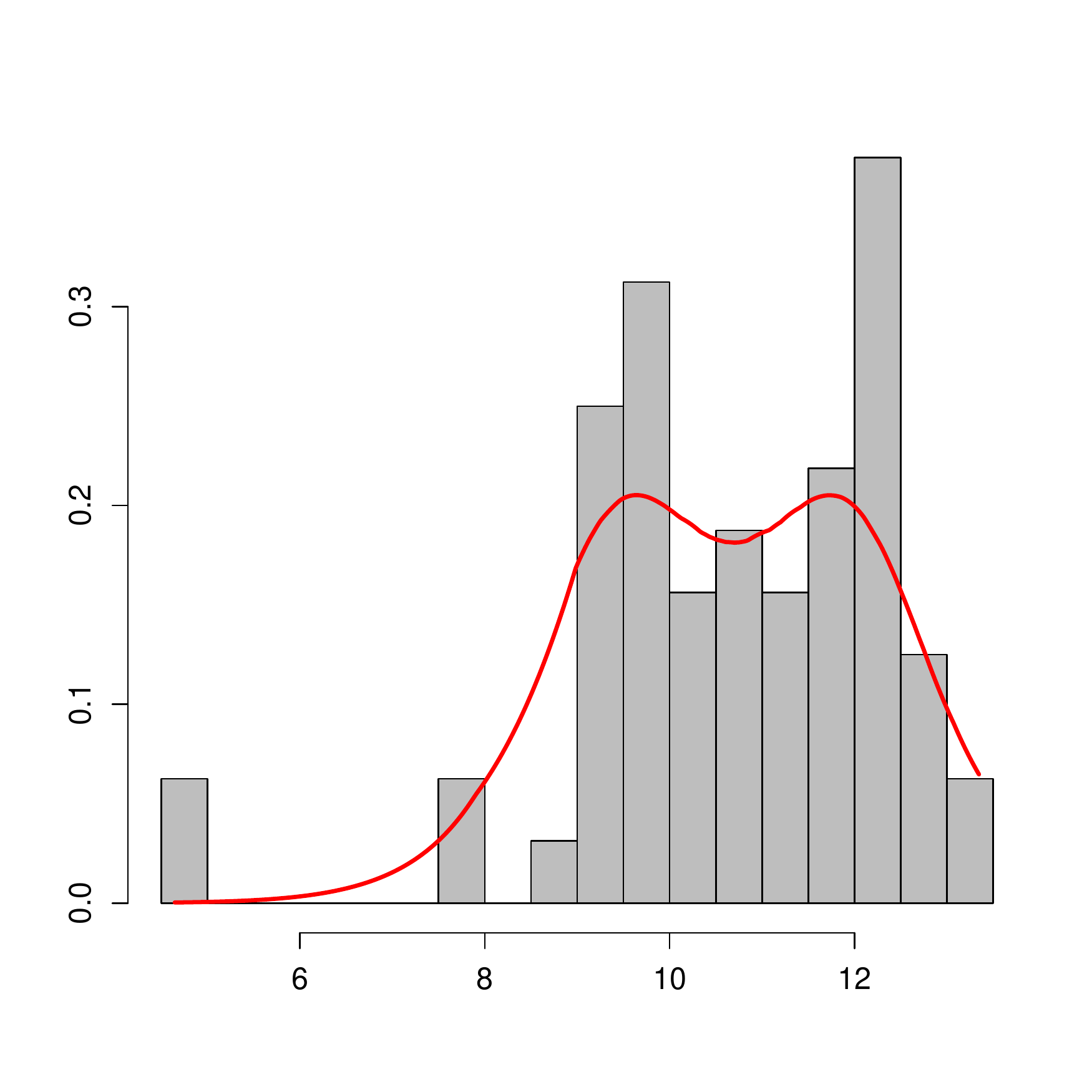}
\includegraphics[height=.31\textheight,width=.31\textwidth,keepaspectratio,trim=1cm 1cm 1.5cm .5cm]{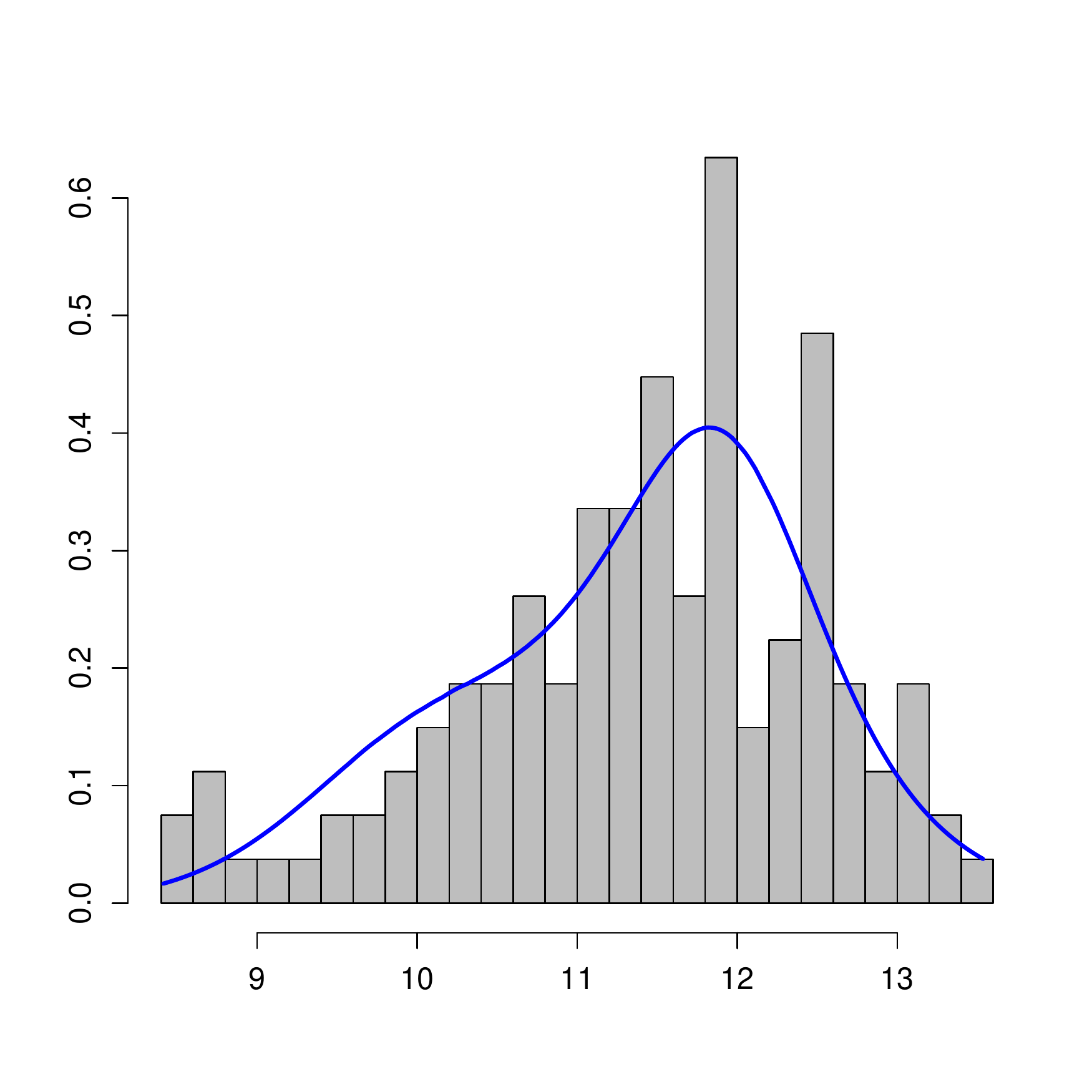}
\includegraphics[height=.31\textheight,width=.31\textwidth,keepaspectratio,trim=.5cm 1cm 2.5cm .5cm]{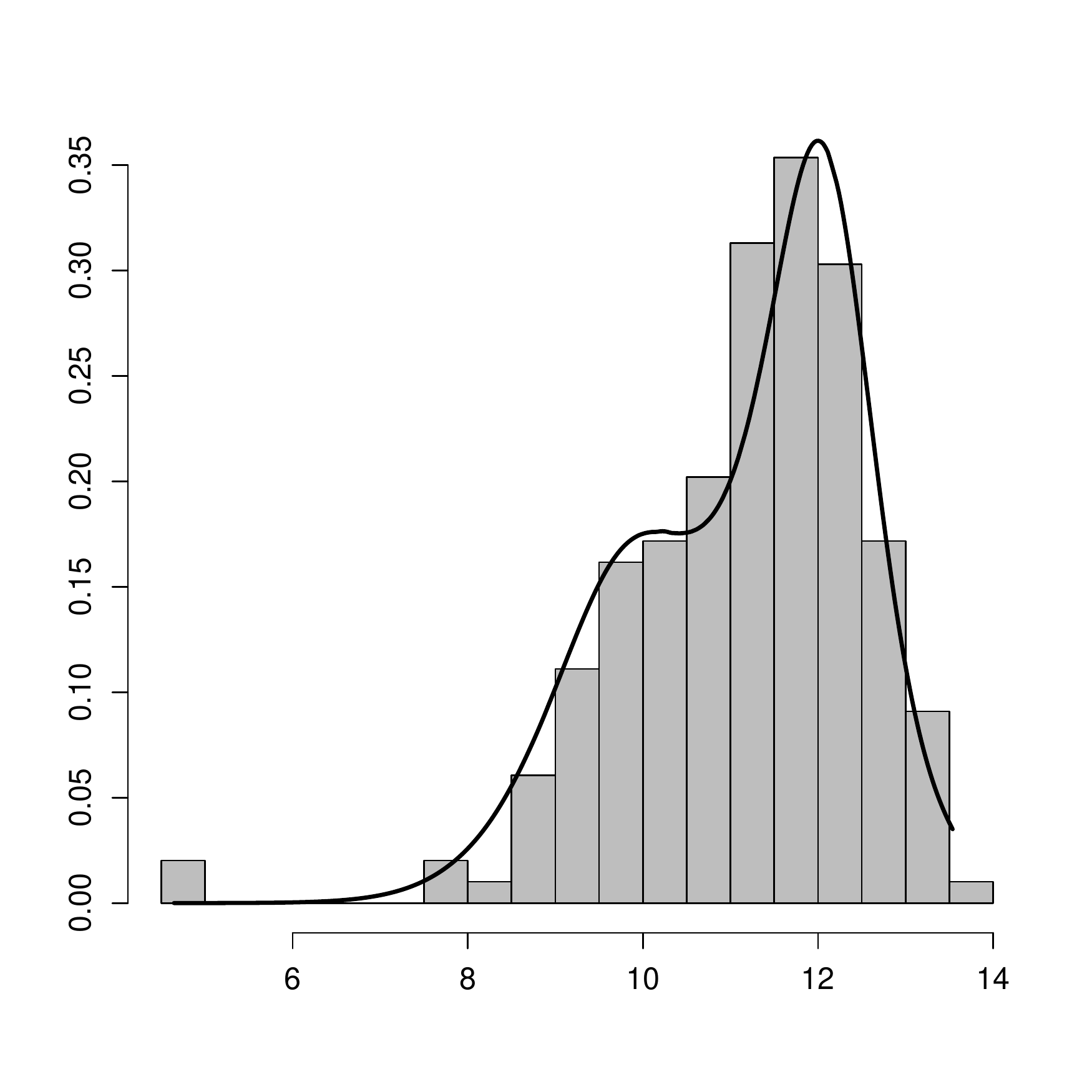}

\caption{The shapes of one selected bimodal gene from each breast cancer dataset is shown. Rows corresponds to each data set. Red and blue color respectively denotes the distribution over two classes and black indicates the pooled distribution.} \label{fig:cancer-g}
\end{figure*}

\begin{table}[!ht]
\setlength{\tabcolsep}{12pt}
\caption{\textit{Comparing the prediction accuracy of random forest (RF) classifier based on (a) all the genes, and (b) only the multimodal genes.}} 
\vskip1em
\centering 
\begin{tabular}{ccccc} 
\hline\hline 
Datasets & \% of unimodal genes & All genes & Multi-modal genes \\ [0.5ex] 
\hline 
Veer data & 82.46\% & $61.89\, (8.8)$  & $63.58\, (9.5)$  \\
Vijver data & 78.30\% &  $67.09\, (5.63)$ & $64.53\, (5.93)$  \\
Transbig data & 89.8\% &  $86.66\, (3.92)$ & $85.15\, (3.92)$  \\
\hline 
\end{tabular}
\label{tab:cg2} 
\end{table}
The distributions (over two classes) of few selected bimodal genes are shown in Fig \ref{fig:cancer-g}, which indicates the multimodal-gene predictors are highly informative for classifying tumor samples and can be considered promising candidates for disease-specific biomarker. LPMode algorithm provides a new computationally efficient tool for large-scale bump-hunting that could be useful for cancer biologists for discovering novel biomarker genes, which were previously unknown.

\subsection{Astronomy}
\subsubsection{Asteroid data}
We analyze the physical density (in grams/cm$^3$) of $26$ asteroids  \citep{marchis2006} in the main asteroid belt of our solar system. The goal is to understand the internal structure of asteroids (largely unexplored research area) and classify them into various classes based on their density characteristics.

LPMode finds three modes:

$\bullet$   $L^2$ modes are at $1.20\,(0.411),\, 2.52 \, (0.61),$ and $3.82\, (0.63)$.

$\bullet$   MaxEnt modes are at $1.364 \,( 0.54),\, 2.503 \,(0.71),$ and $3.490 \, (0.66)$.

The trimodal shape denotes the presence of $3$ types of asteroids: The low density peak corresponds to C-type (composed of  dark carbonaceous objects), the middle one S-type (composed of silicaceous  objects), and the higher-mode denotes the X-type asteroids (also known as M-type whose composition is partially known and apparently dominated by metallic iron and possibly icy and rocky composition).

Surprisingly, our pure data-driven findings are in agreement with the modern asteroid taxonomic system - both Tholen classification and more recent SMASS classification confirm the existence of three broad categories based on surface compositions. Our empirical discovery is also validated by the recent work of \cite{marchis2012} on ``Multiple asteroid systems.'' Overall \texttt{LPMode} could be a handy tool for astronomers to get more refined asteroid classification schemes as we obtain more observations in the future. We emphasize that our analysis is purely data-driven, which finds the underlying three types of asteroids without taking into account any astrophysical models of asteroids.
\subsubsection{Galaxy color data}
We analyze the rest-frame u-r color distribution (which is sensitive to star formation history) of $24,346$ galaxies with $M_r \le -18$ and $z < 0.08$, drawn from the Sloan Digital Sky Survey database \citep{fukugita1996sloan, york2000sloan, balogh2004}. For details, see \url{http://www.sdss.org/}.

\begin{figure*}[!thb]
\centering
\includegraphics[height=.4\textheight,width=.75\textwidth,keepaspectratio,trim=2cm 1cm 2cm 1.5cm]{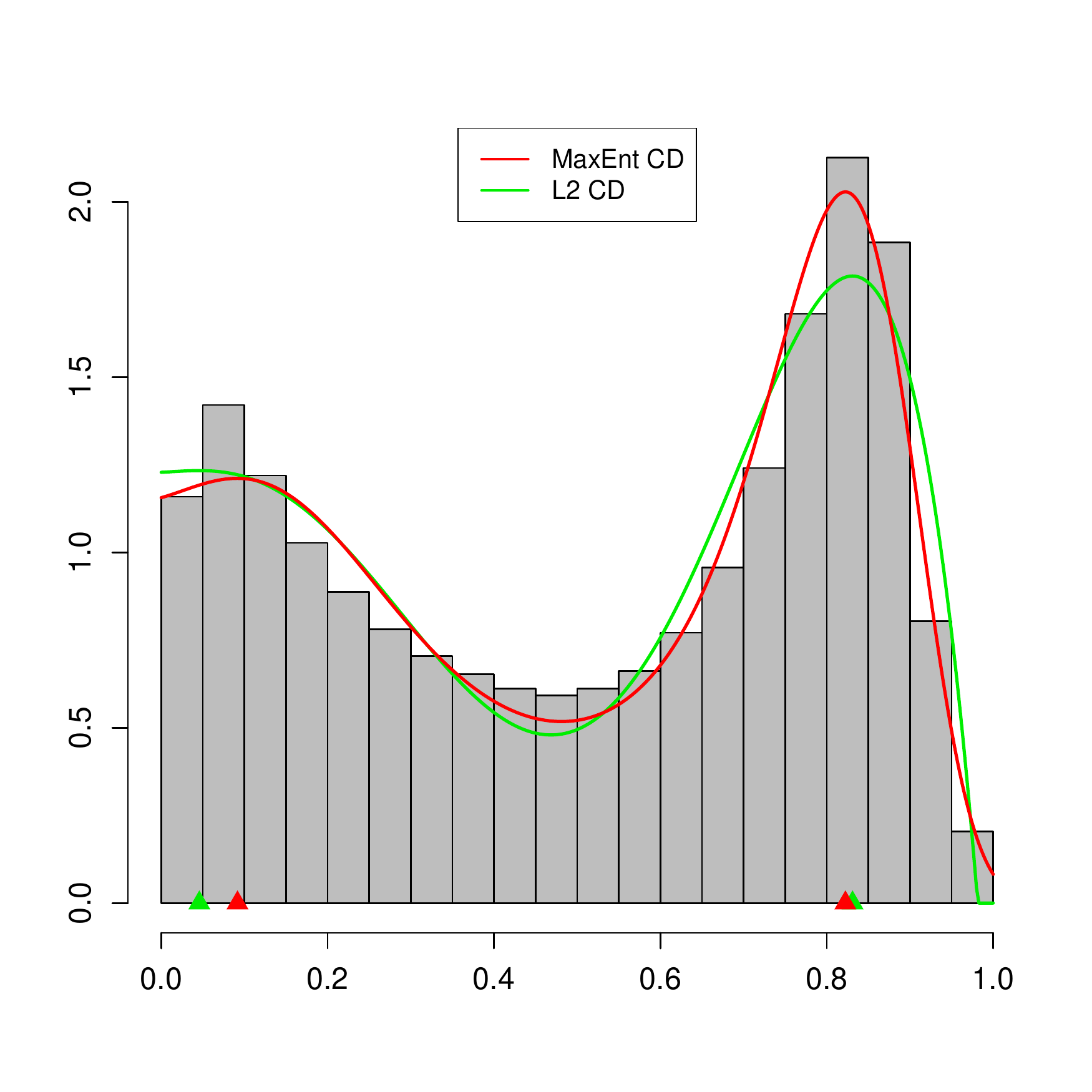}\\
\includegraphics[height=.45\textheight,width=.75\textwidth,keepaspectratio,trim=2cm 2cm 2cm .5cm]{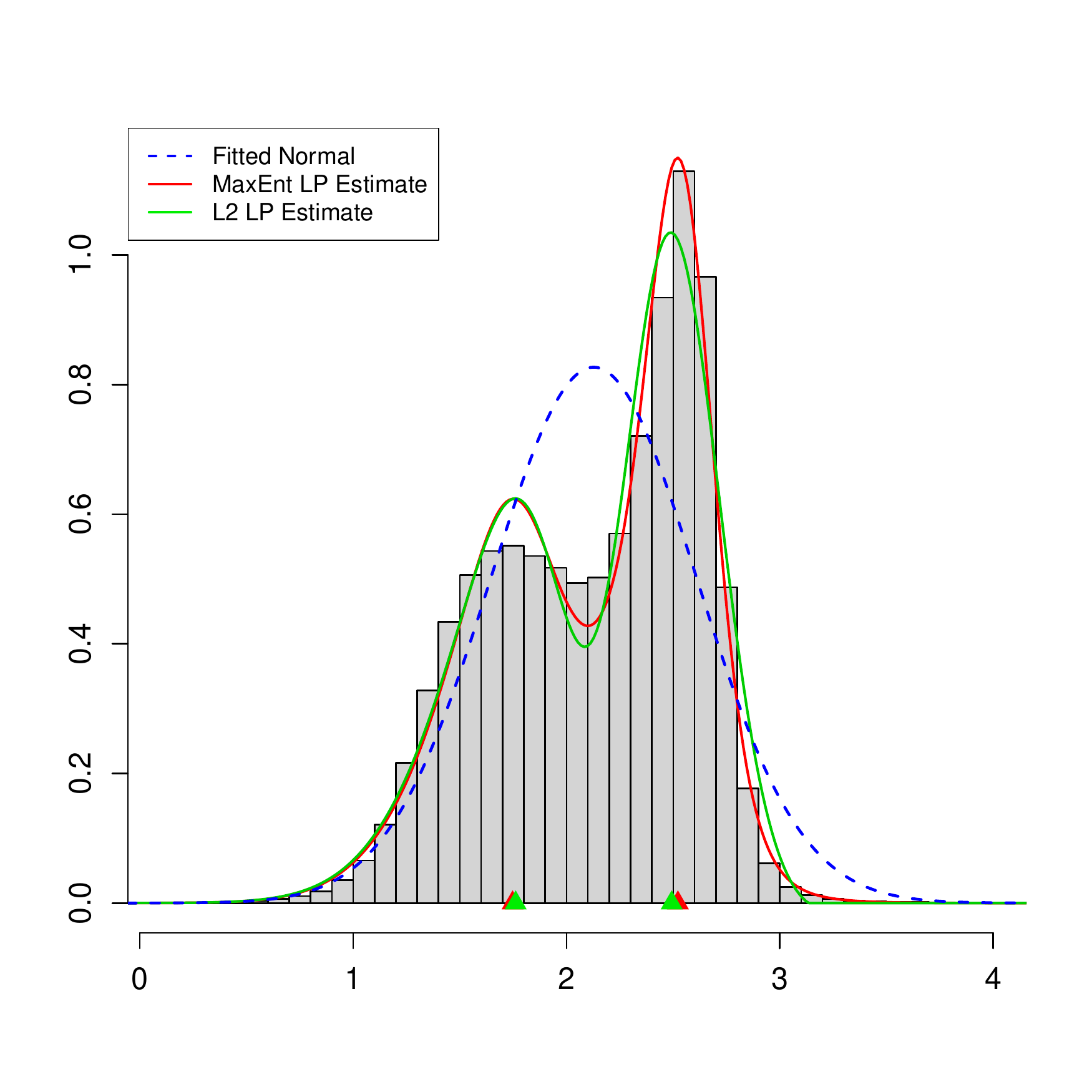}
\caption{Galaxy Color data. [Top] The $L^2$ and MaxEnt estimated comparison density. The $L^2$ modes are at $(0.0458, 0.831)$, green triangles; MaxEnt modes are $(0.092, 0.822)$, red triangles. [Bottom] The estimated densities of the observed color distributions along with the base Normal $G$. The $L^2$ modes are at $(1.761, 2.492)$ shown as green triangles; MaxEnt modes are $(1.747, 2.522)$ shown as red triangles, which respectively denote the blue and red galaxies.}
\label{fig:cd-astro}
\end{figure*}

LPMode algorithm finds two prominent modes of the galaxy color distribution. Fig \ref{fig:cd-astro} shows the LP-skew estimate of the color distribution with the mode locations and the standard errors given by
\begin{center}
$L^2$ Modes:~~~$(1.761,\, 2.492)$ with standard error $(0.047,0.057)$.

MaxEnt Modes:~~~$(1.747,\, 2.522)$ with standard error $(0.0039,0.0033)$.
\end{center}
The color bimodality conveys important clues for galaxy formation and evolution. In particular, the twin peaks represent two types of galaxy populations, which astrophysicists classify as: red (old stellar populations with no or little cold gas) and blue (young stellar populations with abundant cold gas) or, alternatively as early- and late-type, produced by two different sets of formation processes. Our purely data-driven discovery can be explained (in terms of what caused this bimodality) using galactic formation theories. See \cite{baldry2004} for further details.

\subsection{Analytical chemistry}
The pH of bioethanol is one of the most important parameters of the quality of biofuel, since its value establishes the grade of corrosiveness that a motor vehicle can withstand.  Here we consider $n=78$ measurements from the proficiency testing (PT) scheme organized by Inmetro -- the Brazilian National Metrology Institute \citep{sarmanho2015}. Nineteen laboratories participated in measuring the pH in bioethanol and the participant laboratories were allowed to use their own procedure, i.e. a specific electrode for measuring pH. The different measuring procedures in analytical chemistry are the main cause of multi-modal distribution. The challenge of PT scheme providers is to access the comparability and reliability of the measurements by identifying and estimating consensus values (modes) along with the measure of variability/uncertainty (standard errors of each mode positions); see  \cite{lowthian02}, for more details on the role of bump-hunting for proficiency testing.

LP modeling starts by selecting $G$ to be Gaussian to check whether the measurements deviate from normality. LPMode finds $2$ major peaks. $L^2$ Modes are located at $5.35\,(0.084)$ and $6.56\,(0.12)$ and the MaxEnt Modes at $ 5.35\,(.09)$ and $6.65\,(0.15)$, denoting the high variability of the second mode. Our analysis strongly suggests the presence of two groups of laboratories based on the methods for pH measurements. This is further confirmed by the fact that one group of laboratories carried out the pH measurements with electrodes containing saturated LiCl as the internal filling solution, whereas the other group used electrodes with a $3.0$ mol L$^{-1}$ KCI internal filling solution.

The  Gaussian kernel density estimate (the bandwidth $h = 0.4261$ is selected using Silverman's rule of thumb) based method finds mode at $(5.34, 6.59)$ with bootstrap estimated standard errors $(0.07,0.10)$ respectively for the two modes. The BIC select Gaussian mixture model
\[\widehat f(x)=  .58\,\cN(5.32, .26) + .42\cN(6.65, .26).\]
\vspace{-2em}
\subsection{Biological science}
We investigate the activity of an enzyme in the blood involved in the metabolism of carcinogenic substances based on the data from $n=245$ individuals. The goal is to study the possible bimodality to identify the slow and fast metabolizers as a polymorphic genetic marker in the general population. This data set has been analysed by \cite{bechtel1993}, who identified a mixture of two \emph{skewed} distributions by using maximum likelihood techniques. \cite{richardson1997} used a normal mixture estimated using reversible jump MCMC to estimate the distribution of the enzymatic activity.

\begin{figure*}[!thb]
\centering
\includegraphics[height=.45\textheight,width=.75\textwidth,keepaspectratio,trim=2cm .5cm 2cm 2cm]{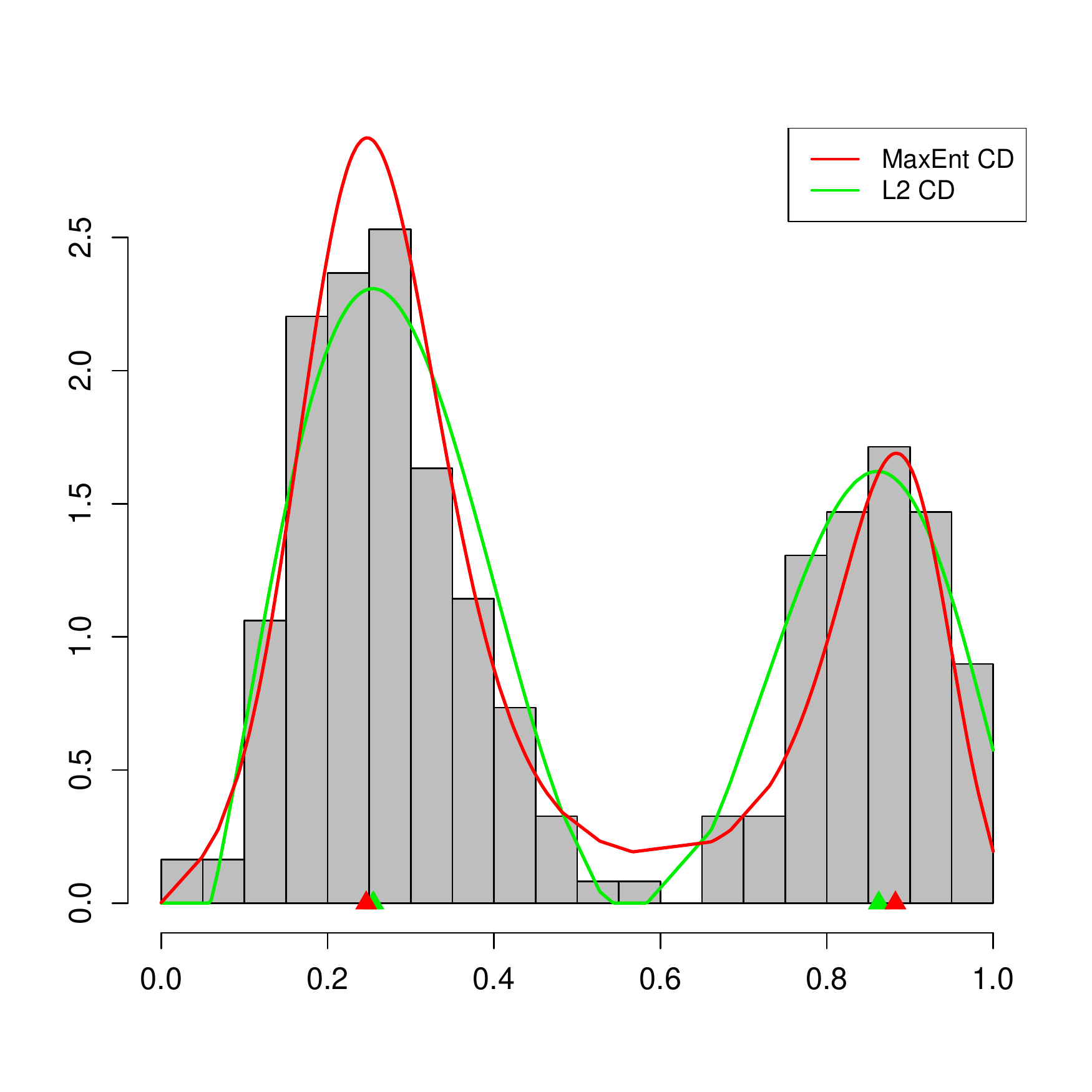}\\
\includegraphics[height=.43\textheight,width=.75\textwidth,keepaspectratio,trim=2cm 1.5cm 2cm 2cm]{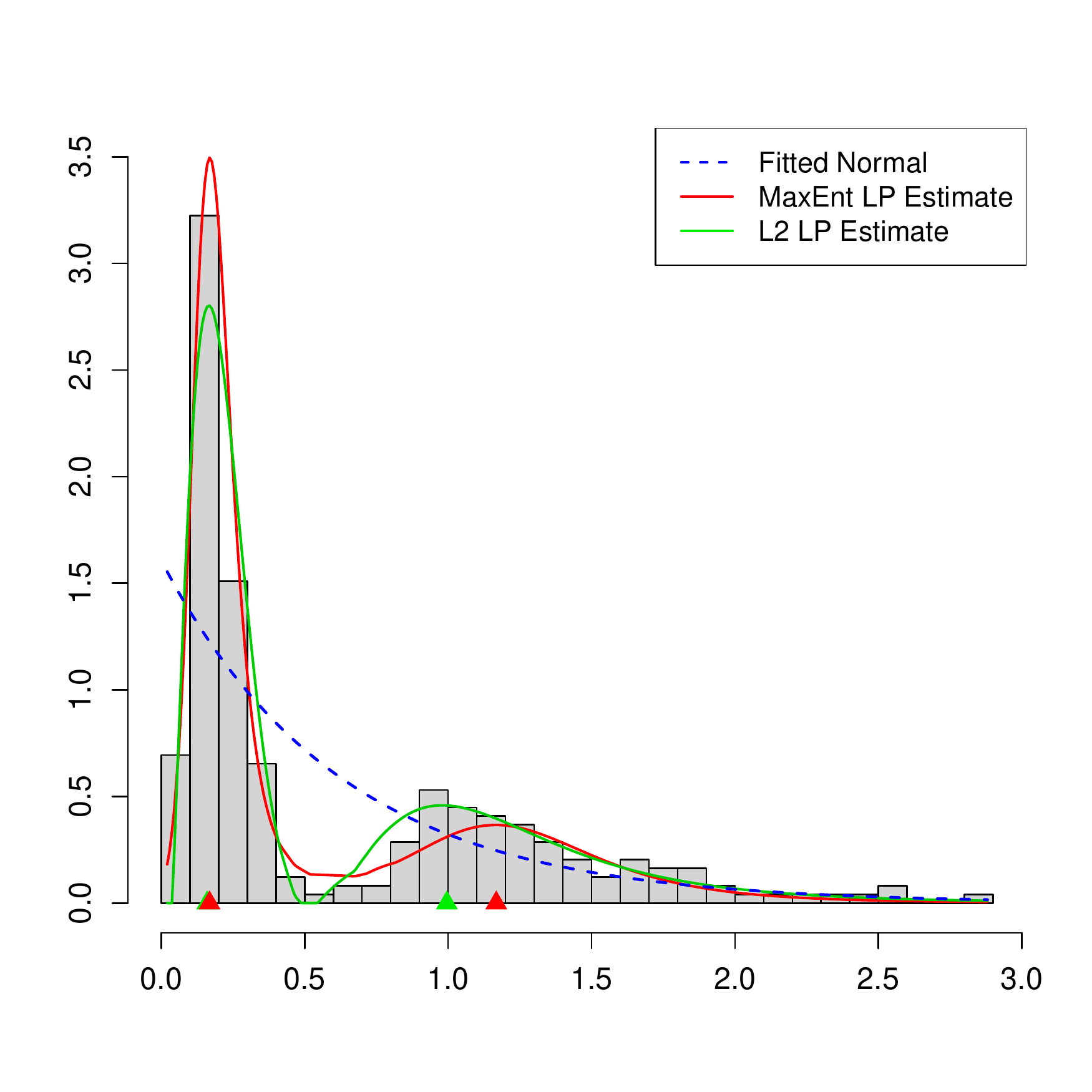}
\vspace{-1em}
\caption{Enzyme data. (A) The estimated smooth comparison density using both $L^2$ and MaxEnt methods. The $L^2$ modes are at $(0.26, 0.86)$ shown as green triangle; MaxEnt modes are $(0.25, 0.88)$ shown as red triangle; (B) The estimated densities along with the base exponential model $G$. The $L^2$ modes are at $(0.160, 0.996)$ shown as green triangle; MaxEnt modes are $(0.168, 1.168)$ shown as red triangle.}
\label{fig:cd-en}
\end{figure*}

\begin{figure*}[!thb]
\centering
\includegraphics[height=.5\textheight,width=.75\textwidth,keepaspectratio,trim=2cm 1cm 2cm 1.5cm]{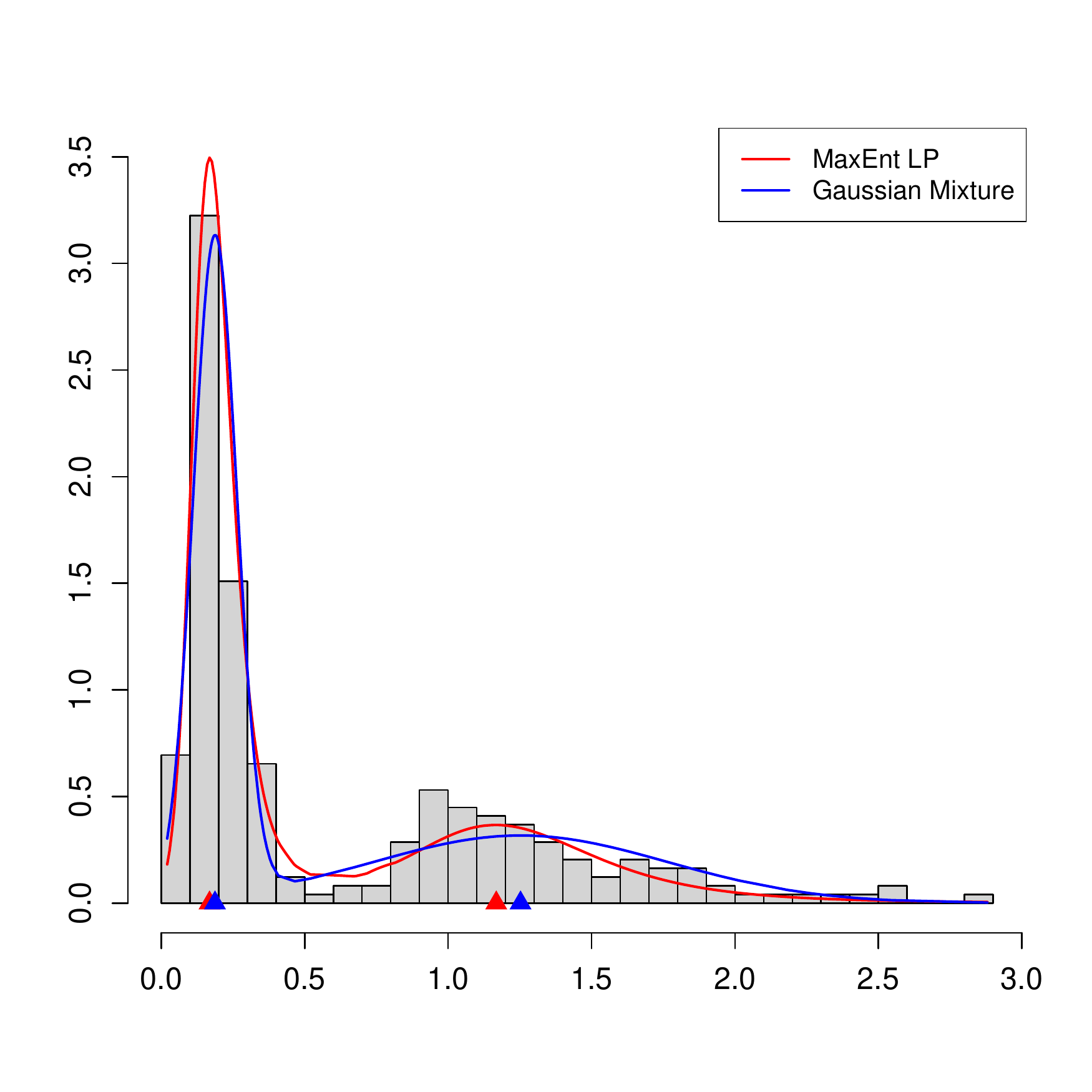}
\caption{Enzyme data. Comparing the MaxEnt LP estimate and the Gaussian mixture model density estimate. Note that the second component  of Gaussian mixture model shows inflated variance as a consequence of failing to capture the underlying skewness.}
\label{fig:mixGh-ex}
\end{figure*}

LPMode starts by selecting $G$ as exponential distribution with mean $.622$ (the MLE estimate). The estimated $L^2$ and MaxEnt comparison density is given by
\beas
d(u;G,\widetilde F) - 1 &=& 0.48\Leg_3(u) - 0.52\Leg_4(u) - 0.25\Leg_5(u) + 0.19 \Leg_6(u) \\
\log d(u;G,\widetilde F)&=& 0.64\Leg_3(u) - 0.67\Leg_4(u) - 0.30 \Leg_5(u) + 0.06\Leg_6(u) -0.43.
\eeas
The shape of the comparison density clearly suggests the presence of bimodality. The LP nonparametric skew-G density estimate $\widehat f(x) = g(x)\,\times\,d(u;G,\widetilde F)$ is shown in Fig \ref{fig:cd-en}(B). Our analysis finds two prominent subgroups of metabolizers -- slow and fast -- as a marker of genetic polymorphism in the population. Note that the two components associated with the two modes are of very different nature in terms of variability and skewness. Our approach is flexible enough to capture this satisfactorily. $L^2$ Modes are located at $0.160 \,(0.011)$ and $0.996\,(0.047)$ with 95\% C.I $(.141, 0.186)$ and $(.922, 1.091)$. The MaxEnt Modes are located at $ 0.168\,(0.014)$ and $1.168\,(0.103)$ with 95\% C.I $(0.157, 0.188)$ and $(1.084, 1.325)$. Fig \ref{fig:mixGh-ex} compares the shapes of the estimated LP skew-G density and the parametric Gaussian mixture model (GMM)
\[\widehat f(x) ~=~ .59\, \cN(.187,  .0058) + .41 \,\cN(1.25,  0.264).\]
The excess variance of the second component is caused by the underlying skewness, which GMM fails to capture by design.
\subsection{Philately}
Fig \ref{fig:ecology}(B) analyzes the distribution of the thicknesses of Hidalgo Stamp data measured by Walton Van Winkle and analyzed first by \cite{wilson1983} and then by many researchers including \cite{izenman1988}.

The estimated $L^2$ and MaxEnt comparison density is given by
\beas
d(u;G,\widetilde F) - 1 &=& - 0.11\Leg_1(u) + 0.60\Leg_3(u)  - 0.28 \Leg_4(u) + 0.24\Leg_6(u)  \\
\log d(u;G,\widetilde F)&=&  0.05 \Leg_1(u) + 0.74\Leg_3(u)  - 0.46\Leg_4(u) + 0.08\Leg_6(u)  -0.35.
\eeas
Our LPMode algorithm finds $2$ major peaks. $L^2$ Modes are located at $0.0771 \,(0.00084)$ and $0.0970\,(0.00284)$ and the MaxEnt Modes at $ 0.0761\,(0.000857)$ and $0.102\,(0.00259)$. A similar bimodality conclusion was drawn by \cite{efron1994book} using a bootstrap based kernel density estimate. \cite{wilson1983} also arrived at the same solution of bi-modality and concluded that ``the un-watermarked stamps were printed on two different papers!'' Interestingly, Wilson found the two modes are near to $0.077$ mm and $.105$ mm, which is practically identical to our finding. As the data is unduly rounded, the traditional mixture model known to find many spurious modes.
\subsection{Ecological science}
We consider the body size distribution of North American boreal forest mammals, previously investigated in a famous ecological study by \cite{holling1992}.
The important question is to check whether the distribution is different from the ``ideal'' unimodal normal distribution and in particular we would like to know if there exits multi-modality. This question has profound ecological and evolutionary implications.

\begin{figure*}[!thb]
\centering
\includegraphics[height=.4\textheight,width=.65\textwidth,keepaspectratio,trim=2cm 1cm 2cm 1.5cm]{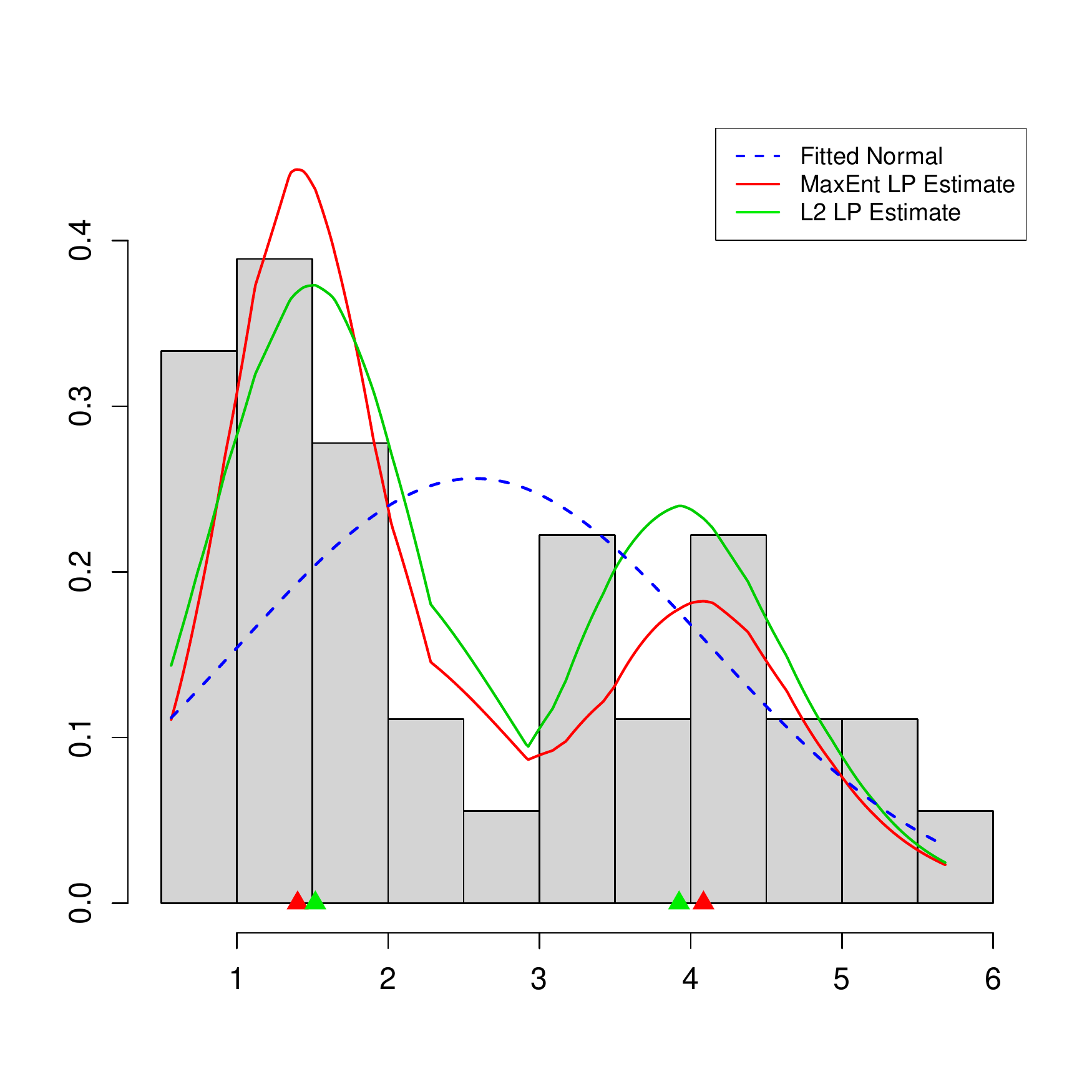}\\
\includegraphics[height=.4\textheight,width=.65\textwidth,keepaspectratio,trim=2cm 1cm 2cm 1.5cm]{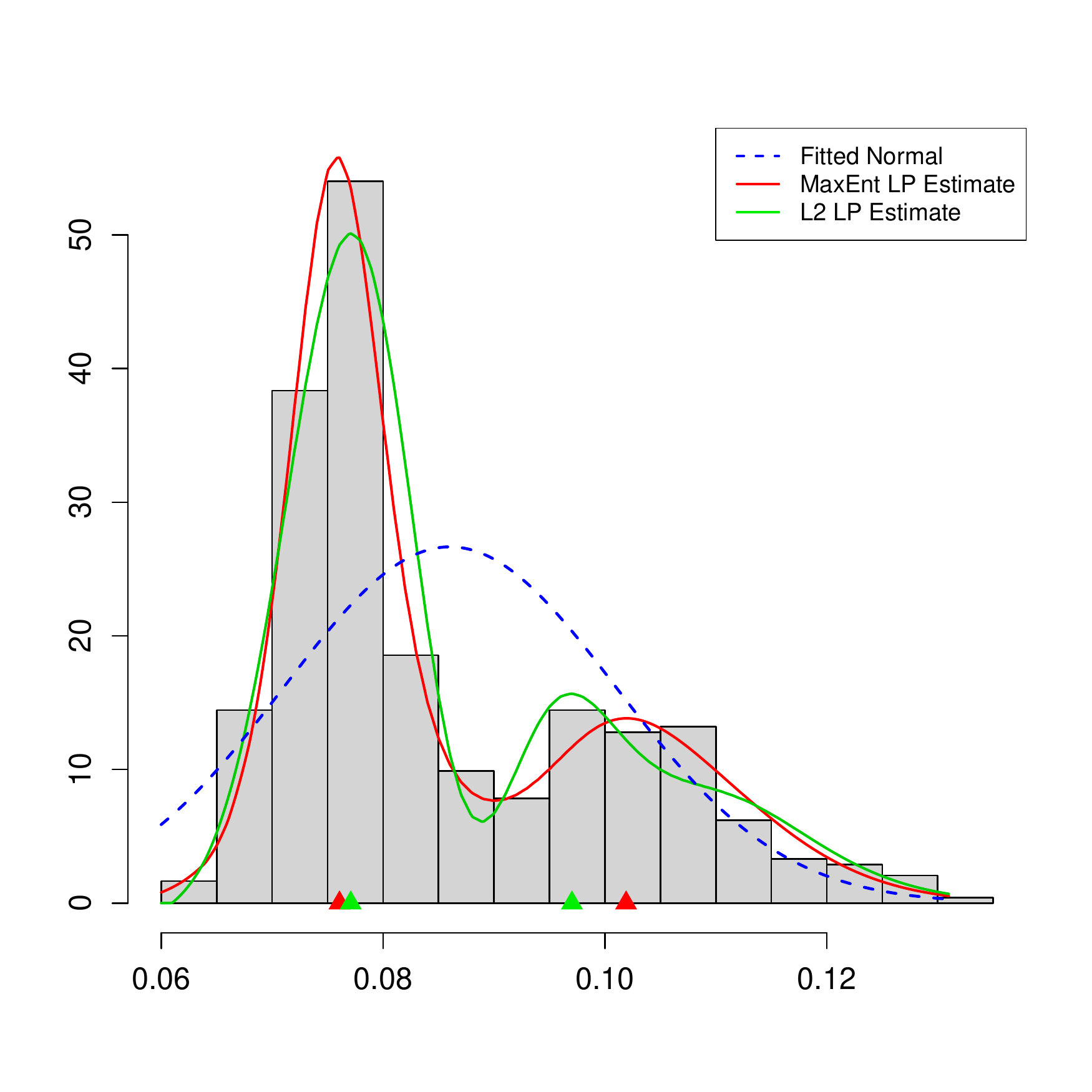}
\caption{(A) [Top] Boreal Forest Mammal Data. The estimated LP-skew estimate (along with the baseline Normal distribution). The $L^2$ modes are at $(1.52, 3.92)$ shown as green triangles; MaxEnt modes are $(1.40, 4.08)$ shown as red triangles. (B)[Bottom] Hidalgo Stamp data. The estimated LP-skew estimate (along with the baseline Normal distribution). The $L^2$ modes are at $(0.077, 0.097)$ shown as green triangles; MaxEnt modes are $(0.076, 0.102)$ shown as red triangles.}
\label{fig:ecology}
\end{figure*}

LPMode algorithm strongly reveals the presence of bi-modality as shown in Fig \ref{fig:ecology}(A). The estimated $L^2$ and MaxEnt comparison density is given by
\beas
d(u;G,\widetilde F) - 1 &=& 0.32\Leg_3(u) - 0.53\Leg_4(u)  \\
\log d(u;G,\widetilde F)&=& 0.47\Leg_3(u) -.58 \Leg_4(u)  -0.30.
\eeas

$L^2$ Modes are located at $1.52 \,(0.267)$ and $3.92\,(0.372)$; 95\% C.I $(1.17,\,  2.14)$ and $(2.95,  4.46)$. The MaxEnt Modes at $ 1.40\,(0.434)$ and $4.08\,(0.424)$; 95\% C.I $(1.15,\,  1.95)$ and $(3.07,\,  4.68)$.

Possible ecological explanations of the observed bimodal pattern is discussed in \cite{holling1992} and \cite{allen2006}. Ecophysiologists have proposed a number of completing mechanistic hypotheses (examples: biogeographical hypothesis,  textural discontinuity hypothesis, community interaction hypothesis and many more) to understand what could lead to such a pattern.
\section{Simulation studies}
\begin{figure*}[t]
\centering
\vskip1.2em
\includegraphics[height=.55\textheight,width=.8\textwidth,trim=2cm .5cm 2cm .5cm]{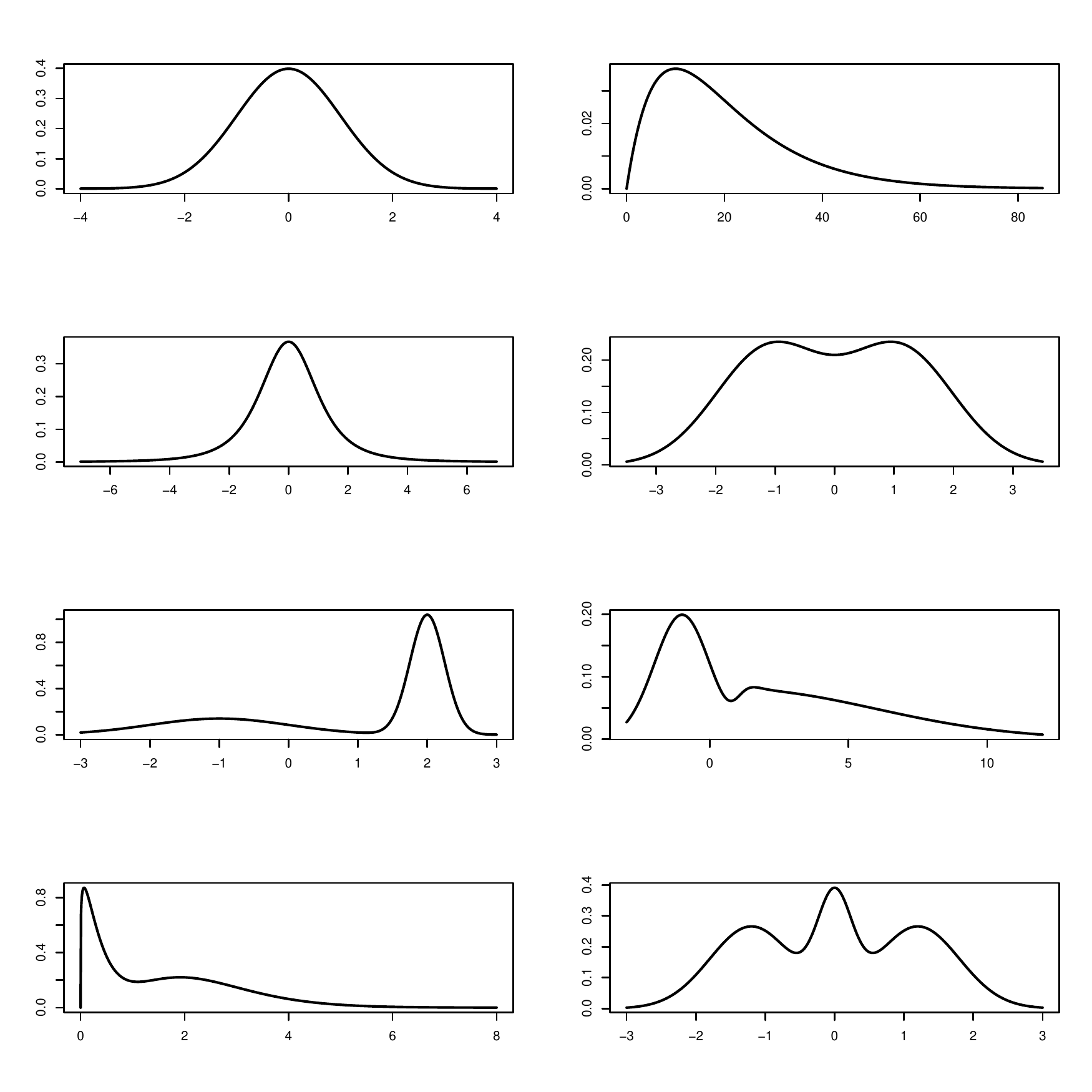}
\vspace{.5em}
\caption{Eight different density shapes with different characteristics that are considered in our simulation studies.}
\vspace{.5em}
\label{fig:den-shapes}
\end{figure*}

Extensive simulation study is conducted to compare our proposed \texttt{LPMode} algorithm with three other popular benchmark methods that practitioners routinely use for bump-hunting:
\begin{itemize}
  \item Kernel density based on Silverman's `rule of thumb'  bandwidth \citep{silverman1986book}.
  \item Kernel density based on Sheather-Jones (SJ)  bandwidth \citep{sheather1991}.
  \item Finite Gaussian mixture model  \citep{fraley2002}. We have used BIC (Bayesian Information Criterion) to select the number of mixture components throughout this comparison.
\end{itemize}
The comparison is based on simulated data from the following eight distributions with different characteristics covering a broad class of modal shapes that arise in practice. Fig \ref{fig:den-shapes} plots the densities to show the diversity of the shapes.

\begin{itemize}
  \item D1: Unimodal Gaussian:  $\cN(0,1)$.
  \item D2: Unimodal Skewed : ${\rm Gamma}(2,.1)$
  \item D3: Long-Tailed unimodal: Student's t with digress of freedom $3$.
  \item D4: Equal bimodal mixture $0.5\cN( - 1.1, 1) + 0.5\cN(1.1, 1)$.
  \item D5: Unequal bimodal mixture $0.2\cN( - 1, 1) + 0.8\cN(2, 0.25)$.
  \item D6: $1$ Component Skewed Bimodal: $.6 \cN(0,1) + .4\cSN(\xi=1,\omega=5,\alpha=15)$.
  \item D7: Skewed Bimodal: $.5 {\rm Gamma}(1,3) + .5{\rm Gamma}(5,2)$.
  \item D8: Tri-modal: $\frac{8}{20}\cN(-6/5,3/5) + \frac{8}{20}\cN(6/5,3/5) + \frac{2}{10}\cN(0,1/4)$.
\end{itemize}

Table \ref{tab:simulation} depicts the simulation results based on sample sizes of $250$, $500$ and $1000$. Each combination of distribution and sample size is replicated $1000$ times. The numbers in the table show the percentage of simulations in which the correct modality was obtained along with the standard errors of the mode distribution for each algorithm.

Table 3 implies that for unimodal Gaussian case (D1), as expected, GMM and LPMode (both $L^2$ and MaxEnt methods) performs equally well; each method correctly detects unimodality for almost 100\% of the cases for moderately large sample size. For skewed and long-tailed unimodal cases (D2 and D3) both the GMM and kernel density estimate (KDE) completely break down (the success rate is almost zero!) and produce misleading results. They tend to produce spurious bumps. LPMode successfully adapts to the modal shape and significantly outperforms both of them. Under skewed case (D2) for a moderately large sample, LPMode detects the true unimodality in 96\% of the cases, and in long-tailed case 97\% of the time. For small sample size, equal bimodal Gaussian mixture case LPMode performs best; for moderately large sample, GMM and LPMode are equally powerful; and for large sample size GMM is most efficient, which is expected. For unequal bimodal mixture KDE completely fails to capture the underlying bimodality and produces noisy bumps while GMM and LPMode show a significantly better performance. However, for skewed bimodality (D6 and D7) we can see that the GMM collapses completely and produces large number of spurious bumps. The same is true for KDE (especially SJ bandwidth selector based KDE). LPMode outperforms the other two competing methods for all sample sizes. For moderately large sample size it correctly detects the bimodality in almost 99\% of cases. For small sample tri-modal case (D8), LPmode correctly identifies the number of modes 93.6\% of the time, whereas KDE and GMM are unable to detect three modes in more than 50\% of the cases. Nevertheless for large sample size, all the methods exhibit good performance.

Our experimental results suggest that \texttt{LPMode} is the most versatile, robust and reliable mode identification algorithm compared to the other currently available state-of-the-art automatic technologies.

\begin{table}
\caption{\textit{Comparing different bump-hunting algorithms. We report \% of simulations in which the algorithm correctly identifies the modality of the distribution (out of $1000$ replications) along with the standard errors in parentheses. For each setting (row) the best performance is marked bold.}}
\vskip1em
\centering
\def\arraystretch{1.125}%
\begin{tabular}{ ccccccc }
\hline
\hline
 &    & \multicolumn{4}{c}{Methods}\\ \cmidrule{3-7}
&   &\multicolumn{2}{c}{Parzen’s kernel density} &Mixture density& \multicolumn{2}{c}{LPMode}\\ \cmidrule{3-4} \cmidrule{6-7}
Distribution  &  Size ($n$)& Silverman & SJ & &$L^2$& MaxEnt\\\hline
\multirow{3}{*}{D1} & $250$ & $61.6\,(.61)$ & $78\,(.56)$ &  $98.7\,(.09)$ &$98.7\,(.11)$&${\bf98.8}\,(.10)$\\
                    & $500$ & $60.3\,(.65)$ & $73.5\,(.53)$ &  ${\bf99.8}\,(.04)$ &$99.1\,(.09)$&$99.1\,(.09)$\\
                    & $1000$ & $55.8\,(.70)$ & $72.8\,(.53)$ & ${\bf 100}\,(0)$ &${\bf100}\,(0)$&${\bf100}\,(0)$\\ \hline

\multirow{3}{*}{D2} & $250$ & $9.60\,(.92)$ & $2.80\,(1.2)$ &  $0\,(.55)$ &${\bf94.2}\,(.30)$&$94\,(.28)$\\
                    & $500$ & $7.2\,(1.1)$ & $.89\,(1.47)$ &  $0\,(.50)$ &${\bf96.6}\,(.18)$&$96.4\,(.18)$\\
                    & $1000$ &$3.4\,(1.0)$ & $0\,(1.75)$ &  $0\,(.54)$ &${\bf96.4}\,(.20)$&$95.2\,(.21)$\\ \hline

\multirow{3}{*}{D3} & $250$ & $0\,(9.8)$ & $0\,(10.2)$ &  $1.4\,(.23)$ &$39.6\,(1.1)$&${\bf73.8}\,(.70)$\\
                    & $500$ & $0\,(12.5)$ & $0\,(12.7)$ &  $0\,(.35)$ &$43.8\,(1.0)$&${\bf88.4}\,(.47)$\\
                    & $1000$ &$0\,(14.5)$ & $0\,(15.0)$ &  $0\,(.5)$ &$ 48.2\,(.88)$&${\bf97.4},(.22)$\\ \hline

\multirow{3}{*}{D4} & $250$ & $57\,(.56)$ & $36.8\,(.59)$ &  $36\,(.48)$ &${\bf64.2}\,(.51)$&$63\,(.54)$\\
                    & $500$ & $60\,(.61)$ & $44.8\,(.60)$ &  ${\bf75.4}\,(.43)$ &$75\,(.44)$&$75.2\,(.44)$\\
                    & $1000$ & $64.6\,(.61)$ & $53.2\,(.61)$ &  ${\bf99.2}\,(.08)$ &$88.6\,(.31)$&$88.6\,(.31)$\\ \hline

\multirow{3}{*}{D5} & $250$ & $0\,(1.4)$ & $0\,(1.6)$ &  ${\bf100}\,(0)$ &$53\,(.50)$&$96\,(.20)$\\
                    & $500$ & $0\,(1.6)$ & $0\,(1.7)$ &  ${\bf100}\,(0)$ &$46.4\,(.49)$&$99.6\,(.06)$\\
                    & $1000$ & $0\,(1.5)$ & $0\,(1.8)$ &  ${\bf100}\,(0)$ &$50\,(.43)$&${\bf100}\,(0)$\\ \hline

\multirow{3}{*}{D6} & $250$ & $44.8\,(.78)$ & $.6\,(1.3)$ &  $93.8\,(.25)$ &$93\,(.26)$&${\bf95.4}\,(.21)$\\
                    & $500$ & $40.4\,(.87)$ & $0\,(1.4)$ &  $75.8\,(.42)$ &$98.4\,(.12)$&${\bf99.6}\,(.06)$\\
                    & $1000$ & $45.4\,(.82)$ & $0\,(1.6)$ &  $34.6\,(.49)$ &${\bf100}\,(0)$&${\bf100}\,(0)$\\ \hline

\multirow{3}{*}{D7} & $250$ & $45\,(.74)$ & $0\,(1.7)$ &  $6.2\,(.76)$ &${\bf87.8}\,(.34)$&$76.2\,(.44)$\\
                    & $500$ & $44.4\,(.76)$ & $0\,(3.1)$ &  $0\,(.79)$ &${\bf96.8}\,(.17)$&$93.6\,(.24)$\\
                    & $1000$ & $29.8\,(.80)$ & $0\,(3.4)$ &  $0\,(.76)$ &${\bf98.4}\,(.12)$&$97.8\,(.14)$\\ \hline

\multirow{3}{*}{D8} & $250$ & $45.6\,(.62)$ & $57.6\,(.81)$ &  $52.2\,(.62)$ &${\bf93.6}\,(.28)$&${\bf93.6}\,(.28)$\\
                    & $500$ & $73.8\,(.47)$ & $73\,(.60)$ &  $86.6\,(.35)$ &${\bf96.2}\,(.24)$&${\bf96.2}\,(.24)$\\
                    & $1000$ & $94.2\,(.24)$ & $61.2\,(.67)$ &  $99.6\,(.06)$ &${\bf99.7}\,(.09)$&${\bf99.7}\,(.09)$\\[.25em] \hline
\hline
\end{tabular}
\label{tab:simulation}
\end{table}
\section{Discussion}
One of the fundamental problem of statistical science is to detect signal from large-scale i.i.d observation $X_1,\ldots,X_n \sim F$, where $F$ is often called the signal-plus-background distribution. The task depends on the following two scenarios: \vspace{-.25em}
\begin{itemize}
  \item[(a)] First scenario, where signal hides at the tail of the probability distribution; \vspace{-.4em}
  \item[(b)] Second scenario, often encountered in practice, where signal appear as a form of bump or mode in the probability distribution.
\end{itemize}
\vspace{-.25em}
A huge research enterprise has been developed to address (a) under the banner `Large Scale Inference' \citep{efron2010book,deep16LSSD}. At the same time, problem (b) has gained renewed relevance in this 21st century with the advent of the `big data.' Despite of its practical significance and multidisciplinary utility, the progress (to develop \emph{pragmatic} mode discovery tool) has not been satisfactory since the pioneering work by \cite{parzen1962}. In this article, we intend to fill that gap by offering a new point of view to the problem of mode identification whose foundation is rooted in the modern nonparametric machineries \citep{Deep14LP}.

Our LPMode bump-hunting algorithm achieves {\bf three goals} all at once:
\vskip.15em
$\bullet$ The modes of $\widehat d(u;G,F)$ indicates the existence of `\emph{bump(s) above background}' (by choosing $G$ to be the background distribution), which is often scientifically \emph{more} interesting (as is the case with Higgs boson discovery) than the modality of the original distribution $F$. \vskip.15em
$\bullet$ It avoids the challenging problem of spurious bumps by \emph{jointly analyzing} the connector density $\widehat d(u;G,F)$ and the original density estimate $\widehat{f}(x)$, which is fundamentally different from all the existing techniques.
\vskip.15em
$\bullet$ It provides a systematic (and automatic) nonparametric exploration (not based on predetermined parametric functional form) tool that is suitable for large-scale problems.
\vskip.15em
The applicability of the algorithm is demonstrated using examples from environmental science, ecology, cancer genomics, astronomy, analytical chemistry, and econometrics.
\section*{Acknowledgments}
The author would like to express his sincere appreciation to Emanuel Parzen for several valuable comments and suggestions.

We are also thankful to Daniel Henderson and Christopher Parmeter for pointing out the vital reference which explains the tri-modal GDP per worker distribution; Gabriel F. Sarmanho for providing the data on proficiency testing (PT) of pH in bioethanol acquired by the Brazilian
National Metrology Institute; Michael Thompson for several fruitful discussions on the role of bump-hunting for the proficiency test; Franck Marchis and Dan Britt for pointing out the scientific explanation of the tri-modal structure of the asteroid data; Ivan Baldry and Karl Gebhardt for providing SDSS galaxy color data and for sharing several astronomy insights.

Finally, I would like to express my sincere thanks to the Editor and the anonymous reviewers for their in-depth comments, which have greatly improved the manuscript.
\bib

\newpage

\pagenumbering{arabic}
\renewcommand{\thepage} {A--\arabic{page}}
\begin{center}
{\large {\bf \underline{Appendix Section}  \\[1em]  Large-Scale Mode Identification and Data-Driven Sciences}}
\\[.5in]
Subhadeep Mukhopadhyay\\
Temple University, Department of Statistical Science \\ Philadelphia, Pennsylvania, 19122, U.S.A.\\
\texttt{deep@temple.edu}\\[.25in]

All the figures referenced in the main paper are shown in this this appendix section.
\end{center}
\vskip2em

\begin{figure*}[!thb]
\centering
\includegraphics[height=.65\textheight,width=.75\textwidth,keepaspectratio,trim=2cm 1cm 2cm 1.5cm]{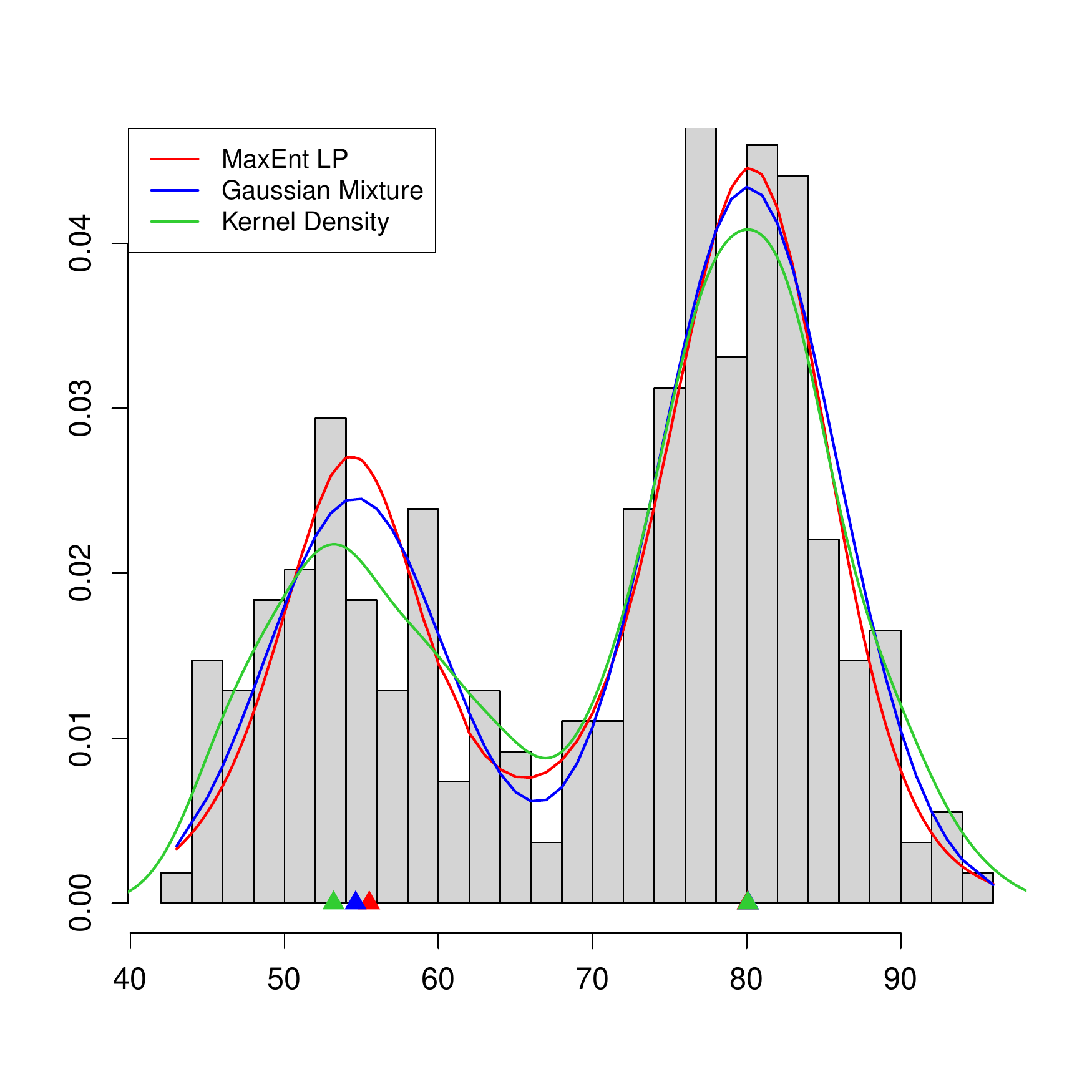}
\caption{Comparison of LP Skew-G density estimate with the Gaussian mixture model and Kernel density estimate. The mixture model and LP density show very similar modal shapes.}
\label{fig:mixGh}
\end{figure*}
\end{document}